\documentclass[preprint]{imsart}
\usepackage[utf8]{inputenc}
\usepackage[T1]{fontenc}
\usepackage{lmodern}
\usepackage{dsfont}
\usepackage{amsmath}
\usepackage{bbold}
\usepackage{amsfonts}
\usepackage{amssymb}
\usepackage{amsthm}
\usepackage{siunitx}
\usepackage{stmaryrd}

\usepackage[numbers]{natbib}
\usepackage{csquotes}
\usepackage[xindy]{glossaries}
\usepackage{algorithm}
\usepackage{algpseudocode}
\usepackage{tikz}
\usetikzlibrary{decorations.pathreplacing}
\usepackage{forest} 
\usepackage{xcolor}
\usepackage{array}
\usepackage{booktabs}
\usepackage{ltablex}
\usepackage{rotfloat}
\usepackage{multirow, makecell}
\usepackage{adjustbox}

\usepackage[indent]{parskip}
\usepackage{lscape}

\usepackage{graphicx}

\usepackage{centernot}
\usepackage[english]{babel} %
\usepackage{subcaption}
\usepackage{listings}
\usepackage{hyperref}
\usepackage{xr-hyper}

\DeclareMathOperator*{\argmax}{arg\,max}

\theoremstyle{plain}
\newtheorem{definition}{Definition}
\newtheorem{theorem}{Theorem}
\newtheorem{proposition}{Proposition}
\newtheorem{lemma}{Lemma}[theorem]
\newtheorem{corollary}{Corollary}
\theoremstyle{remarks}

\newcommand{\EE}{\mathbb{E}}
\newcommand{\PP}{\mathbb{P}}

\newcommand{\Hzero}{\mathcal{H}_0}

\newcommand{\refdistrib}{\ensuremath{\mathcal{P}_{0}}}
\newcommand{\winsize}{m}

\startlocaldefs
\numberwithin{equation}{section}
\theoremstyle{plain}
\endlocaldefs

\graphicspath{{Images/}}

\makeglossaries

\begin{document}
	\begin{frontmatter}
	\title{FDR control for Online Anomaly Detection}
	\runtitle{FDR control for Online Anomaly Detection}
	\runauthor{E. Krönert et al.}
	\author[A]{\fnms{Etienne}~\snm{Krönert}\ead[label=e1]{etienne.kronert@worldline.com}},
	\author[B]{\fnms{Alain}~\snm{Célisse}\ead[label=e2]{alain.celisse@univ-paris1.fr}}
	\and
	\author[A]{\fnms{Dalila}~\snm{Hattab}\ead[label=e3]{dalila.hattab@worldline.com}}
	\address[A]{FS Lab, Worldline, France\printead[presep={,\ }]{e1,e3}}
	\address[B]{SAMM, Paris 1 Panthéon-Sorbonne University, France\printead[presep={,\ }]{e2}}

		\begin{abstract}
		A new online multiple testing procedure is described in the context of anomaly detection, which controls the False Discovery Rate (FDR). An accurate anomaly detector must control the false positive rate at a prescribed level while keeping the false negative rate as low as possible. However in the online context, such a constraint remains highly challenging due to the usual lack of FDR control: the online framework makes it impossible to use classical multiple testing approaches such as the Benjamini-Hochberg (BH) procedure, which would require knowing the entire time series. The developed strategy relies on exploiting the local control of the “modified FDR” (mFDR) criterion. It turns out that the local control of mFDR enables global control of the FDR over the full series up to additional modifications of the multiple testing procedures. An important ingredient in this control is the cardinality of the calibration dataset used to compute the empirical p-values. A dedicated strategy for tuning this parameter is designed for achieving the prescribed FDR control over the entire time series. The good statistical performance of the full strategy is analyzed by theoretical guarantees. Its practical behavior is assessed by several simulation experiments which support our conclusions.
		\end{abstract}
		
	\end{frontmatter}
	\bigskip\bigskip

	\section{Introduction} 
	\subsection{Context}
	By monitoring indicators in real time to check the system health, online anomaly detection aims at raising an alarm when abnormal patterns are detected \cite{ahmed2016survey, li2022intelligent}. This contrasts with offline anomaly detection, which can only analyze historical data retrospectively. One motivation for automatic anomaly detection is to reduce the workload of operations teams by allowing them to prioritize their efforts where necessary. This is typically accomplished through the use of statistical and machine learning models \cite{chandolaAnomalyDetectionSurvey2009, braei2020anomaly, blazquez-garciaReviewOutlierAnomaly2020}. However, a poorly calibrated anomaly detector leads to alarm fatigue. An overwhelming number of alarms desensitizes the people tasked with responding to them, leading to missed or ignored alarms or delayed responses \cite{cvach2012monitor, blum2010alarms}. One of the reasons for alarm fatigue is the high number of false positives which take time to manage \cite{solet2012managing, lewandowska2023determining}. The main goal of the present work is to design a new (theoretically grounded) strategy allowing to control the number of false positives when performing online anomaly detection.
	
	\subsection{Related works}\label{sec:rw}
	Due to the lack and cost of labeled data, anomaly detection is most often formulated in the context of unsupervised learning \cite{chandolaAnomalyDetectionSurvey2009, blazquez-garciaReviewOutlierAnomaly2020, ruff2021unifying}. The unlabeled training set is used to learn atypicity scores that are difficult to interpret, such as those based on likelihood or distance \cite{chandolaAnomalyDetectionSurvey2009}. Furthermore, due to the lack of labels, test data is rarely available to evaluate an anomaly detector. This makes empirical evaluation of anomaly detectors difficult \cite{wu2021current}.
	Conformal Anomaly Detection (CAD) \cite{LaxhammarConformalanomalydetection2014} is a method based on Conformal Prediction \cite{AngelopoulosGentleIntroductionConformal2022}. The goal of CAD is to give a probabilistic interpretation of the atypicity score using a calibration set and a conformal $p$-value estimator. CAD allows control of the expected number of false positives within a time period \cite{LaxhammarConformalanomalydetection2014}. However, it does not provide control over the false discovery rate (FDR), i.e., the proportion of false positives among all detections, which is the focus of the present work.
	The Benjamini-Hochberg (BH) procedure \cite{benjamini1995, Benjaminicontrolfalsediscovery2001} is a multiple testing procedure that controls the proportion of false positives among rejections, the False Discovery Rate (FDR). BH has been used to control FDR for anomaly detection in the offline context \cite{bates2023testing, MarandonMachinelearningmeets2022}.
	In online anomaly detection, however, the decision to classify a new observation as an anomaly must be made instantaneously. Therefore, BH cannot be applied as it requires the entire time series.
The \cite{javanmard2018, ramdas2018, xu2022dynamic} introduce, in a different context than anomaly detection, an online multiple testing procedure based on alpha investing. The $p$-value is compared to an adaptive threshold which depends on previous decisions. However, this method is difficult to apply to conformal $p$-values due to its low power. In \cite{WangOnlineFDRControlled2019} the author suggests using the principle of local FDR. At each observation, a decision is taken depending on the estimation of the local FDR. Unfortunately, this approach relies on a Gaussian assumption that limits its generality.
Approaches based on online breakpoint detection are particularly well suited for collective anomaly detection, thanks to methods such as CUSUM \cite{page1954continuous, aue2024state}, which can accumulate the atypicity scores of several consecutive observations, thus increasing the power of the test. There are numerous theoretical results concerning this method about the average run length \cite{wu2007inference} or the probability of false negatives and false positives \cite{kirch2022sequential}.  However, throughout this paper, anomalies are assumed to be punctual. In this case, online breakpoint detectors do not improve detection power. For this reason, these detectors are not studied in this paper. 

	Controlling false positives for online anomaly detection remains a difficult task. 
	The core of the paper is a theoretical analysis which aims to control FDR for online anomaly detection in the case where the observations are iid. Although this is a strong assumption, this is the only situation for which there are theoretical results is the case of offline multiple testing \cite{MarandonMachinelearningmeets2022}.
	In particular two challenges arise with online anomaly detection. First, the true $p$-values are unknown and need to be estimated. Second, the decisions are made in an online context, whereas most multiple testing methods are done in the offline context. 
	The main contributions are to address these challenges. More precisely it is established that it is possible to design online anomaly detectors controlling FDR of the time series.
	\begin{itemize}
		\item This paper investigates the relationship between FDR and the cardinality of the calibration set used to estimate $p$-values. To ensure FDR control, a calibration set cardinality tuning method is proposed.
		\item This paper describes a data-driven threshold for anomaly detection based on multiple testing in order to control the False Discovery Rate (FDR) of the entire time series. It explains how global control of the time series FDR can be obtained from local control of a modified version of the FDR criterion. Then, a modified version of the Benjamini-Hochberg procedure is suggested to achieve local control of the modified FDR.
	\end{itemize}

    \subsection{Description of the paper}
   
   First, an anomaly detector controlling the global FDR is introduced in Section~\ref{sec:fdr-online-anomaly-detection}. Second, Section~\ref{sec:fdr_emp} deals with conditions on empirical $p$-values to ensure local control of FDR at a desired level $\alpha$. Third, algorithms that allow global control of the FDR time series from local control of the modified FDR are developed in Section~\ref{sec:golbal}. Then, the anomaly detector is empirically studied in Section~\ref{sec.FDR.mFDR.empirical}. Finally, it is evaluated against a competitor from the literature in Section~\ref{sec:empirical-simulation}.
   
	\section{FDR control for Online Anomaly Detection}\label{sec:fdr-online-anomaly-detection}
	
	\subsection{Statistical framework}\label{sec:problem-settings}
	\label{sec:model-ad}
	
	Let $(\Omega, \mathcal F,\mathbb{P})$ be a probability space, with $\Omega$ the set of all possible outcomes, $\mathcal F$ a $\sigma$ algebra on $\Omega$, and $\mathbb P$ a probability measure on $\mathcal F$. Assume a sequential realization of the random variables $(X_t)_{1 \leq t \leq T}$, where $X_t$ takes values in a set $\mathcal X$ for all $t$. $T\in \mathbb{N}\cup\lbrace \infty \rbrace$ is the size of the time series. Let $\mathcal P_0$ be a probability distribution, called the reference distribution, on the space $\mathcal X$.   
	For each instant $t$, the observation $X_t$ is called ``normal'' if $X_t \sim \mathcal P_0$. Otherwise $X_t$ is an ``anomaly''. The aim of an anomaly detector is to find all anomalies among the new observations along the time series $(X_t)_{t\geq 1}$. In \emph{online} anomaly detection, for each time $t\geq1$, a decision about the status of $X_t$ is made based on \emph{past observations}: $(X_s)_{1\leq s \leq t}$. This contrasts with the offline context, where the complete series is known when a decision is made.
	
	Since the present goal is to use FDR, a natural strategy is to rephrase the online anomaly detection problem as a multiple testing problem:
	at  each step $1 \leq t\leq T$, a statistical test is performed on the hypotheses:
	$$\mathcal H_{0t}, ``X_t \text{ is not an anomaly }"  \qquad \mbox{against}\qquad \mathcal H_{1t} ``X_t \text{ is an anomaly}".$$ 
	Each statistical test is performed by comparing an estimation of the $p$-value under $\mathcal H_{0t}$, noted $\hat p_t$, with a threshold $\hat \varepsilon_t$. Details about how to calculate these quantities are given in the next section. If the $p$-value $\hat p_t$ is below the threshold $\hat \varepsilon_t$, then $\mathcal H_{0t}$ is rejected, $X_t$ is detected as an anomaly. In the online context, $\hat p_t$ and $\hat\varepsilon_t$ are computed sequentially when $X_t$ is observed without knowledge of the next observations.
	A classical criterion controlling the proportion of type I errors (False Positives) in the whole time series is FDR \citep{benjamini1995}. For a given data-driven threshold $\hat\varepsilon$ and a set of \emph{estimated} $p$-values $\hat p = (\hat p_t)_{t\geq 1}$, the FDR criterion of the sequence from 1 to $T$ is given by
	\begin{align*}
		&FDR_1^T(\hat \varepsilon, \hat p) = \mathbb E[FDP_1^T(\hat\varepsilon, \hat p)],\\
		& \mbox{with}\quad  FDP_1^T(\hat\varepsilon, \hat p) = \frac{\sum_{t \in \mathcal H_0}\mathbb{1}[\hat p_t \leq \hat \varepsilon_t]}{\sum_{t=1}^T\mathbb{1}[\hat p_t \leq\hat\varepsilon_t]},
	\end{align*}
	with the convention that $0/0=0$.
In the above expression, $FDP_1^T$ denotes the \emph{False Discovery Proportion} (FDP) of the time series from 1 to $T$. Also $\mathcal H_0 = \lbrace t\in \mathbb N^*| \mathcal H_{0t} \mbox{ is true}\rbrace$ is called the set of null hypotheses. 
Let us emphasize that the anomalies satisfy $\mathbb 1 [\hat p_t \leq \hat\varepsilon_t] = 1$.
The main objective of the present work is to define a data-driven sequence  $\hat \varepsilon:$ $t\mapsto \hat\varepsilon_t$  such that, for a given control level $\alpha\in[0,1]$, under weak assumptions on the sequence $\hat p: t \mapsto \hat p_t$,
\begin{equation}\label{eq:fdrcontr}
		FDR_1^T(\hat\varepsilon, \hat p) \leq \alpha
\end{equation}
Note that the $FDR_1^T$ must be controlled on the whole time series, even if decisions are made locally.
The control is said exact when ``$\leq$'' is replaced with ``$=$''. Such a control would imply that for a level $\alpha = 0.1$, at most 10\% of the detected anomalies along the entire time series are false positives. 
The detection power of the anomaly detector is measured by means of the \emph{False Negative Rate} defined, for the sequence from 1 to $T$, by
\begin{align}
	&FNR_1^T(\hat \varepsilon, \hat p) = \mathbb E[FNP_1^T(\hat\varepsilon, \hat p)],\\
	& \mbox{with}\quad  FNP_1^T(\hat\varepsilon, \hat p) = \frac{\sum_{t \in \mathcal H_1}\mathbb{1}[\hat p_t > \hat \varepsilon_t]}{|\mathcal H_1|},
\end{align}\label{equ:fnr}
where $FNP_1^T$ denotes the \emph{False Negative Proportion} (FNP) of the sequence from 1 to $T$ and $\mathcal H_1 = \lbrace t\in \mathbb N^*| \mathcal H_{1t} \mbox{ is true}\rbrace$ is the set of alternative hypotheses. 
	
The Benjamini-Hochberg procedure, introduced in \cite{benjamini1995} and described in Definition~\ref{bhdef}, allows the control of the at a desired level $\alpha$, for a family of $m$ $p$-values.

\begin{figure}[H]
	\centering
	\includegraphics[width=0.3\linewidth]{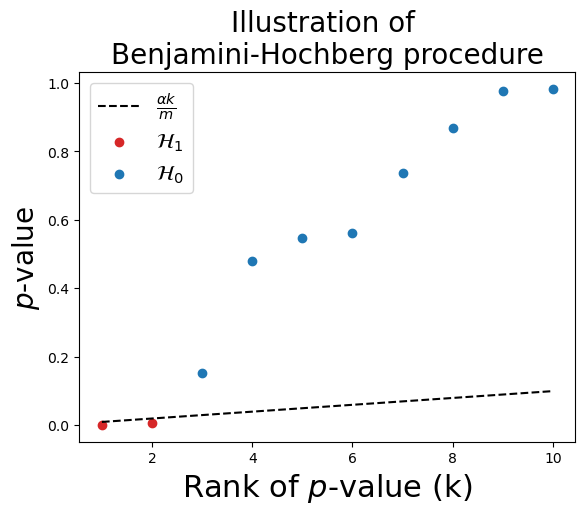}
	\caption{Illustration of the Benjamini-Hochberg procedure. The $p$-values are sorted in ascending order. The threshold is the largest $p$-value that is lower than $\alpha k/m$, where $k$ is the rank of the $p$-value.}
	\label{fig:illustrationbh1}
\end{figure}

\begin{definition}[Benjamini-Hochberg \cite{benjamini1995}]\label{bhdef}
	Let $\winsize$ be an integer and $\alpha\in[0,1]$.  Let $(p_i)_{1\leq i\leq \winsize}\in [0,1]^\winsize$ be a family of $p$-values.
	The Benjamini-Hochberg (BH) procedure, denoted by $BH_\alpha$, is given by
	\begin{itemize}
		\item a data-driven threshold: 
		\begin{align*}
			\varepsilon_{BH_\alpha} = max\{\frac{\alpha k}{\winsize}; p_{(k)}\leq \frac{\alpha k}{\winsize}, k\in \llbracket 1,\winsize\rrbracket\},
		\end{align*}
		\item a set of rejected hypotheses:
		\begin{align*}
			BH_\alpha\left((p_i; i\in  \llbracket 1,\winsize\rrbracket)\right) = \left\lbrace i; p_i\leq  \varepsilon_{BH_\alpha}, i \in \llbracket 1,\winsize\rrbracket\right\rbrace.
		\end{align*}
	\end{itemize}
\end{definition}
The intuition behind this procedure consists in drawing the ordered statistics $i \mapsto p_{(i)}$ (Figure~\ref{fig:illustrationbh1}) with $p_{(1)}\leq \dots \leq p_{(n)}$ and the straight line $i \mapsto \frac{\alpha i}{m}$. Then the BH procedure will reject all hypotheses corresponding to $p$-values smaller than the last crossing point between the straight line and the ordered $p$-values curve.
	
However, a crucial remark at this stage is that controlling FDR on the full time series is a very challenging task in the current online context for at least two reasons:
\begin{itemize}
	\item The main existing approaches to control FDR are described in an ``offline'' framework where the whole series is observed first, and decisions are taken afterwards \citep{benjamini1995,MarandonMachinelearningmeets2022}. This makes these approaches useless in the present context.	
	
	\item The existing approaches designed in the online context \citep{javanmard2018,xu2022dynamic} are difficult to parameterize and hard to apply with empirical $p$-values. Realistic scenarios usually exclude the knowledge of the true probability distribution of the test statistic, leading to the approximation or estimation of the corresponding $p$-values in practice.
\end{itemize}

The new anomaly detector introduced in the next section provides solutions to these problems and allows global control of FDR.

\subsection{Description of the method}\label{sec:descrition.method}

 Like many other anomaly detectors, our detector is based on the notions of atypicity score, $p$-value, and threshold. The novelty of our approach lies in the use of a data-driven threshold, which allows to control FDR at a desired $\alpha$ level. This threshold is calculated by applying the Benjamini-Hochberg procedure to a subseries of length $m$ with a carefully chosen $\alpha'$ level. At the same time, the size of the calibration set must be correctly specified. 
 More precisely, the new anomaly detector described in Algorithm~\ref{alg:fdrcontroad} relies on:
\begin{enumerate}
	\item \textbf{Atypicity score}:  A score $a:\mathcal X \rightarrow \mathbb R$ is a function reflecting the atypicity of an observation $X_t$. To be more specific, the further $\mathcal P_{X_t}$ is from $\mathcal P_0$, the larger $a(X_t)$ is expected to be. It is often implemented using a non-conformity measure (NCM) \cite{LaxhammarConformalanomalydetection2014}, which measures how different a point $X_t$ is from a training set $\mathcal X^{train}$, $a(X_t) = \overline a(X_t, \mathcal X^{train})$.
	\item \textbf{$p$-value}: It is the probability of observing $a(X)$ higher than $a(X_t)$ if $X \sim \mathcal P_0$. It is estimated using the empirical $p$-value, by the following equation:
	\begin{align}
		\hat p_e(s_t, \mathcal S^{cal}_t)= \frac{1}{|\mathcal S_t^{cal}|}\sum_{s\in \mathcal S_t^{cal}}\mathbb{1}[s_t>s]\label{eq:empv}
	\end{align}
    Where $S_t^{cal} = \lbrace a(X_{u_1}),\ldots, a(X_{u_n})\rbrace$ is the calibration set with $n$ data points. 
   The $p$-value enables an interpretable criterion measuring how unlikely $X_t \sim \mathcal{P}_0$ is. 
   The empirical $p$-value is chosen because it is agnostic to the true and unknown distribution, which is not the case for the Gaussian $p$-value estimator. 
   The performance of the BH procedure applied to empirical $p$-values depends strongly on the cardinality of the calibration set ($|\mathcal S_t^{cal}|=n$).
   Section~\ref{sec:FDRemp} investigates how to optimally choose this number.
	\item \textbf{Detection threshold}: $\varepsilon_t \in [0,1]$, it discriminates observations considered as abnormal from others. The observations considered as anomalies are $X_t$ whose (estimated) $p$-value is smaller than the threshold $\varepsilon_t$. To control FDR of the entire time series, a data-driven threshold is computed using the $m$ most recent $p$-values. This detection threshold is computed using a multiple testing procedure inspired by Benjamini-Hochberg (BH) and described in Section~\ref{sec.modified.BH}. This procedure requires the calculation of an $\alpha'$ value estimated from a training set or using heuristics. See Appendix~\ref{sec.heuristic.arguments} for more details.
\end{enumerate} 

\begin{algorithm}[!h]
	\begin{algorithmic}[1]
		\Require $T$ length of the time series, $(X_t)_{1\leq t\leq T}$ time series, $\alpha$ desired FDR, $m$ subseries length, $\nu$ integer to tune the calibration set cardinality
		\Require Either $(Z_t)$ an historical dataset or $\pi$ the proportion of anomalies 
		\If{historical dataset}
		\State $\alpha' \gets \argmax_{\tilde\alpha}\left(\frac{\hat \mu_{R^{**}_{\tilde\alpha}}}{\hat \mu_{R_{\tilde\alpha}}}\tilde \alpha \leq \alpha\right)$ \Comment{Estimate $\alpha'$ with training set}
		\Else
		\State $\alpha' \gets \frac{\alpha}{1+\frac{1-\alpha}{m\pi}}$ \Comment{Estimate $\alpha'$ with heuristics}
		\EndIf 
		\State $n\gets \nu\cdot\frac{m}{\alpha'}-1$ \Comment{Get the calibration set cardinality}
		\For{$t$ in $[1,T]$}
		\State $s_t \gets a(X_t)$       
		\State $\hat p_t \gets \hat p_e(s_t, \mathcal S^{cal}_t)$ \Comment{Compute empirical $p$-value}
		\State $\hat \varepsilon_t \gets \hat \varepsilon_{BH_{\alpha'}}(\hat p_{t-m}, \ldots,\hat p_t)$  \Comment{Get the threshold using Benjamini-Hochberg}
		\If{$\hat p_t<\hat \varepsilon_t$}   \Comment{Retrieve anomalies}
		\State $d_t = 1$
		\Else
		\State $d_t = 0$
		\EndIf
		\EndFor
		\\{\bf Output:} $(d_{t})_{t=1}^T$ boolean list that represent the detected anomalies.
	\end{algorithmic}
	\caption{FDR Control Online Anomaly Detection}
	\label{alg:fdrcontroad}
\end{algorithm}

In the next sections, the design choices of Algorithm~\ref{alg:fdrcontroad} are specified and justified by a theoretical study of the detector. Section~\ref{sec:fdr_emp} studies the local control of FDR when BH is applied to subseries of $m$ empirical $p$-values. This allows to specify the choice of the calibration set cardinality used in Algorithm~\ref{alg:fdrcontroad}. Then, in Section~\ref{sec:golbal}, the FDR control on the complete time series is obtained from the local control of some modified FDR. This allows to specify the procedure for selecting $\alpha'$ in Algorithm~\ref{alg:fdrcontroad}.
	
	\section{Local control of FDR with Empirical $p$-values}
	\label{sec:fdr_emp}
	
	The goal here is to describe a strategy achieving the desired FDR control for a  time series of length $m$ when using empirical $p$-values. 
	A motivating example is first introduced for emphasizing the issue in Section~\ref{sec:motiv-example}. Then a theoretical understanding is provided along Section~\ref{sec:FDRemp} which results in a new solution which applies to independent empirical $p$-values. An extension to some dependency setups is then discussed in Section~\ref{sec:extniid}. 
	
	\subsection{Motivating example}
	\label{sec:motiv-example}
	The purpose here is to further explore the effect of the calibration set cardinality on the actual FDR control when using empirical $p$-values. This gives us more insight on how to find mathematical solutions. 
\begin{figure}[h!] 
	\centering
	\includegraphics[width=0.95\textwidth]{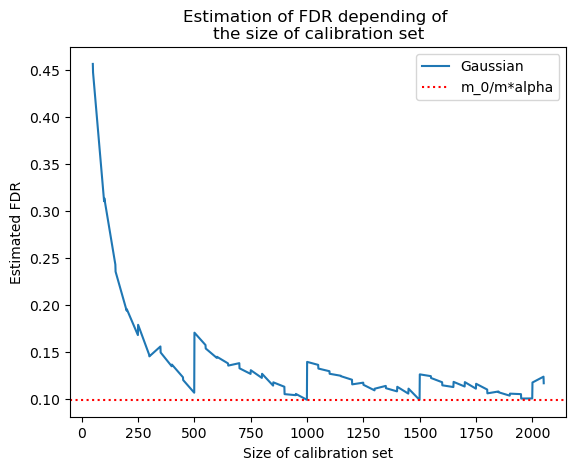}
	\caption[]{}
	\label{fig:motivexp}
\end{figure}
Let us start by generating observations using two distributions.
The reference distribution is $\mathcal P_0 = \mathcal N (0,1)$ and the alternative distribution is $\mathcal P_1 = \mathcal N (4, 10^{-4})$. The anomalies are located in the right tail of the reference distribution.
The length $m$ of the signal is $m=100$. The number of observations under $\mathcal P_0$ is $m_0=99$.
The experiments have been repeated $K=10^4$ times. 
Figure~\ref{fig:motivexp} displays the actual value of FDR as a function of the cardinality $n$ of the calibration set $\lbrace X_1, \ldots, X_n \rbrace$ used to compute the empirical $p$-values (see Eq.~\ref{eq:empv}). 
One clearly see that FDR is not uniformly controlled at level $m_0/m \alpha$. However there exist particular values of $n$ for which this level of control is nevertheless achieved.
As long as $n$ has become large enough ($n\geq 500$), repeated picks can be observed with a decreasing height as $n$ grows.

	\subsection{FDR control: main results for \emph{i.i.d.} $p$-values}\label{sec:FDRemp}
	
The present section aims first to explain the shape of the curve shown in Figure~\ref{fig:motivexp}. This will help to specify conditions on the calibration set cardinality to get FDR control for subseries of length $m$.

	\subsubsection{Compute FDR when BH is applied to empirical $p$-values}

The main focus is now given to independent $p$-values. In what follows, the classical proof (see Section~1.1 in the Supplementary Materials) of the FDR control by the BH-procedure is revisited, leading to the next result. Its main merit is to provide the mathematical expression of the plain blue curve observed in Figure~\ref{fig:motivexp}. 
	\begin{theorem}
		\label{thm:fdr-bh-emp}
		Let $n$ be the cardinality of the calibration set and $\winsize$ be that of the set of tested hypotheses where $\lbrace X_1,\ldots, X_m\rbrace$ denotes a set of random variables.
	    Let $m_0\leq m$ be the cardinality of the random variables from the reference distribution $\mathcal P_0$.
		Let the empirical $p$-value be denoted, for any $i \in \llbracket 1,\winsize\rrbracket$, by $\hat p_i = \hat p_e(X_i, \lbrace Z_{i,1},\ldots, Z_{i,n}\rbrace)$,
		where the calibration set is $\lbrace Z_{i,1},\ldots, Z_{i,n}\rbrace$ and each $Z_{i,j} \sim \mathcal{P}_0$.
		Let  the random variable $R(i)$ be the number of detections, when $BH_\alpha$ is applied on the family ${p'}_1^m$, defined by $p_i'=0$ and $p_j'=p_j$ for $j\neq i$, for some $i$ in $\mathcal H_0$.
Then, for every $\alpha \in ]0,1]$, the FDR value over the sequence from 1 to $m$ is given by
		\begin{align*}
		FDR_1^\winsize(\hat \varepsilon_{{BH}_{\alpha}}, \hat p)=\winsize_0\sum_{k=1}^\winsize \frac{\frac{\lfloor \frac{\alpha k n}{\winsize}\rfloor+1}{n+1}}{k}\mathbb{P}(R(i)=k),
		\end{align*}
		where $\hat \varepsilon_{{BH}_{\alpha}}$ denotes the $BH_\alpha$ threshold from Definition~\ref{bhdef} when the BH-procedure is applied to the empirical $p$-values $\hat p=(\hat p_i; 1\leq i\leq m)$.
	\end{theorem}
    The proof is delayed to Appendix~\ref{sec:proof-thm-empfdr}.
    In general it is not possible to compute the exact value of the $FDR$ without knowing the distribution of the random variables $R(i)$. This is in contrast with the case of true $p$-values where $\PP (p_i \leq \frac{\alpha k}{m}) =\frac{\alpha k}{m} $, whereas with empirical $p$-values $\PP (\hat p_i \leq \frac{\alpha k}{m}) = \frac{\lfloor \frac{\alpha k n}{m}\rfloor + 1}{n+1}$, which prevents any simplification of the final bound.
	Nevertheless, this value still suggests a solution to circumvent this difficulty: requiring conditions on $\alpha$, $m$, and $n$ such that $\frac{\lfloor \frac{\alpha k n}{m}\rfloor + 1}{n+1}=\frac{\alpha k}{m}$, for all $k$.  
	This is precisely the purpose of the next Corollary~\ref{th:empFDR}.

	\subsubsection{Tuning of the calibration set cardinality}
	\label{sec:tuning-calibration-cardinality}
\begin{corollary}\label{th:empFDR}
Under the same notations and assumptions as Theorem~\ref{thm:fdr-bh-emp}, the next two results hold true. 
\begin{enumerate}
	\item Assume that there exists an integer $1\leq \nu$ such that $\frac{\nu m}{\alpha}$ is an integer. If the cardinality $n$ of the calibration set satisfies $n = n_\nu-1 = \nu \winsize/\alpha-1$, then
	\begin{align*}
	FDR_1^\winsize(\hat \varepsilon_{{BH}_{\alpha}}, \hat p) =\frac{\winsize_0\alpha}{\winsize}.
	\end{align*}
	
	\item For every $\alpha \in ]0,1]$, assume that the cardinality of the calibration set satisfies $n=n_\nu-1=\lceil\frac{\nu\winsize}{\alpha}\rceil-1$, for any integer $\nu\geq 1$. Then,
	\begin{align*}
	\frac{n}{(n+1)} \frac{\winsize_0\alpha}{\winsize} \leq FDR_1^\winsize(\hat \varepsilon_{{BH}_{\alpha}}, \hat p) \leq \frac{\winsize_0\alpha}{\winsize}.
	\end{align*}
	
\end{enumerate}

\end{corollary}
The proof is moved to Appendix~\ref{proof:empFDR}.
The first statement in Corollary~\ref{th:empFDR} establishes that it is possible to recover the desired control of FDR at the exact prescribed level $\alpha$, provided that the cardinality of the calibration set is equal to $n = \nu \winsize/\alpha-1$. The (mild) restriction on the values of $\alpha$ reflects that the empirical $p$ values do not satisfy the super uniformity property.
By contrast, the second statement gives the desired control at the level $\alpha m_0/m$ by means of lower and upper bounds. In particular, the lower bound tells us that the FDR value cannot be smaller than the desired level $\alpha m_0/m$ up to a multiplicative factor equal to $1-1/n$, which goes to 1 as $n$ grows. For instance, with $\alpha = 0.1$ and $m=100$, $n_\nu=1000$ would result in $FDR_1^\winsize(\hat \varepsilon_{{BH}_{\alpha}}, \hat p) \cdot m/(m_0\alpha) \in [0.999, 1]$. This small lack of control is the price of allowing any value of $\alpha\in]0,1]$.
It is also important to remember that in anomaly detection, abnormal events are expected to be rare. As a consequence, $\frac{m_0}{m}$ is close to $1$ and the actual $FDR$ is close to the desired $\alpha$. However, in situations where $m_0/m$ could deviate too far from 1, it would be helpful to include an estimator of $m_0/m$.

\subsection{Extension to dependent $p$-values}
\label{sec:extniid}

In Section~\ref{sec:FDRemp}, Corollary~\ref{th:empFDR} states that FDR is controlled at a prescribed level $\alpha$ with empirical $p$-values, as long as the calibration set cardinality is correctly tuned, even if the super uniformity property is not satisfied. This result was established in the case where the calibration sets of the different $p$-values are independent. The purpose of the present section is to extend these results to dependent $p$-values. 

A classical result for BH, established in \cite{Benjaminicontrolfalsediscovery2001}, proves that FDR is upper bounded by $\alpha m_0/m$ provided the $p$-value family satisfies the PRDS and super uniformity properties. The PRDS property is a form of positive dependence between $p$-values where all pairwise $p$-value correlations are positive. It results that a small $p$-value for a given observation makes other $p$-values for all considered observations simultaneously small as well, and vice-versa \citep{bates2023testing}. One important example of $p$-values family that are PRDS are conformal $p$-values sharing a unique calibration set. Furthermore, another important result from the literature gives that FDR can also be lower bounded in the case where the calibration set is the same for all (conformal) $p$-values \cite{MarandonMachinelearningmeets2022}. 
It turns out that all these results can be extended to empirical $p$-value estimators (see Eq~\ref{eq:empv}) when the calibration set cardinality is the one discussed in Corollary~\ref{th:empFDR}.
Furthermore, in the context of online anomaly detection, moving windows are classically used to capture and process the incoming data. In results, the calibration sets of the $p$-value family will overlap. Thus, FDR control is also studied in the case of overlapping calibration sets. In the rest of this section, the results obtained for these different cases are presented.

\paragraph{PRDS $p$-values}
For this extension, the concept of positive regression dependency on each one from a subset called PRDS \cite{Benjaminicontrolfalsediscovery2001} is recalled. 

\begin{definition}[PRDS property]\label{def:prds}
	A family of $p$-values $\hat p_1^m$ is PRDS on a set $I_0\subset \lbrace 1,\ldots, m \rbrace$, if for any $i \in I_0$ and any increasing set $A$, the probability $\mathbb P[\hat p_1^m \in A | \hat p_i =u]$ is increasing in $u$.
\end{definition}

The classical result in \cite{Benjaminicontrolfalsediscovery2001} can be extended to empirical $p$-values with the same choice of calibration set cardinality as in Corollary~\ref{th:empFDR}. It ensures that FDR is upper bounded by $\frac{m_0}{m}\alpha$ in the case where the $p$-value family is PRDS. 

\begin{corollary}[Corollary of Theorem 1.2 in \cite{Benjaminicontrolfalsediscovery2001}]
	Suppose the family of $p$-values $\hat p_1^m$ is PRDS on the set $\mathcal H_0$ of true null hypotheses, and suppose that  $\hat p_1^m$ respects super uniformity on all thresholds that can result from BH 
	\begin{align*}
		\forall i \in \mathcal{H}_0,\forall k \in \llbracket 1,m\rrbracket,\qquad\mathbb P(\hat p_i < \frac{\alpha k}{m}) \leq  \frac{\alpha k}{m},
	\end{align*}

	Then, the $FDR$ is upper-bounded by $\alpha$
	\begin{align*}
		FDR(\hat \varepsilon_{BH}, \hat p) \leq \frac{m_0\alpha}{m}.
	\end{align*}

\end{corollary}
The proof is not reproduced here since it is almost the same as the one of Theorem 1.2 in \cite{Benjaminicontrolfalsediscovery2001}.

\paragraph{Unique calibration set}

Moreover, in the case where the calibration set is the same for all $p$-values, FDR can also be lower bounded.
This result extends the one from in \cite{MarandonMachinelearningmeets2022}, to empirical $p$-values using a calibration set with cardinality tuned as proposed in Corollary~\ref{th:empFDR}. 

\begin{corollary}[Corollary of Theorem 3.4 in \cite{MarandonMachinelearningmeets2022}]
	Assuming the following conditions:
	Let $n$ be the cardinality of the calibration set, $\winsize$ the cardinality of the active set and $\winsize_0$ the number of normal observations.
	Let $\refdistrib$ be the reference distribution.
	Let $Z_{i}$ for $i$ in $\llbracket 1,\winsize\rrbracket$ be independent random variables, following $\refdistrib$.
	Let $X_i$ for $i$ in $\llbracket 1,\winsize\rrbracket$ be independent random variables and independents of $(Z_j)$. There are exactly $\winsize_0$ random variables following the $\refdistrib$ distribution.
	Let $a$ be a scoring function.
	For all $i$ in $\llbracket 1, \winsize\rrbracket$, let $\hat p_i$ be the empirical $p$-values associated with the random variables $X_i$ and computed as follows, $\hat p_i  = p_e(a(X_i), \lbrace a(Z_{1}),...,a(Z_{ \winsize})\rbrace, a)$.
	
	If the cardinality of the calibration set is a multiple of $n=n_\nu=\nu\winsize/\alpha-1$, then FDR using $\hat{BH}_\alpha$ on $(\hat p_i)_{1\leq i\leq \winsize}$ is equal to $\frac{m_0 \alpha}{m}$:
	$$FDR_1^\winsize(\hat \varepsilon_{{BH}_{\alpha}}, \hat p) =\frac{\winsize_0\alpha}{\winsize}$$ 
\end{corollary}
The proof is not provided in this paper since it is too similar to the one of Theorem 3.4 in \cite{MarandonMachinelearningmeets2022}.

\paragraph{Overlapping calibration set}

In the context of online anomaly detection, the calibration sets of the $p$-value family will often overlap, since moving windows are classically used to capture and process the incoming data.
To have a perfect control of FDR, an upper and lower bound is needed. According to Proposition~\ref{prop:prds-overcal}, $p$-values with overlapping calibration sets are PRDS.

\begin{proposition}[PRDS property for overlapping calibration sets]
	\label{prop:prds-overcal}
Let $X_i$ for $i$ in $\llbracket 1,\winsize\rrbracket$ be independent random variables. There are exactly $\winsize_0$ random variables that follow the $\refdistrib$ distribution and thus belong to $\mathcal H_0$.
Let $\textbf{Z}$ be the random vector that combines all calibration sets, all elements of $\textbf{Z}$ are generated from $\mathcal P_0$. The set of $n$ indices defining the elements of the calibration set related to $\hat p_i$ in $\textbf{Z}$ is noted as $\mathcal{D}_i$. The calibration set related to $X_1$ is noted $\textbf{Z}_{\mathcal D_1}=(\textbf{Z}_{i_1},...,\textbf{Z}_{i_n})$.
For all $i$ in $\llbracket 1, m \rrbracket$: $\hat p_i = p_e(X_i, \textbf{Z}_{\mathcal{D}_i})$.

Under these conditions, the set of $p$-values is PRDS on $\mathcal H_0$.
\end{proposition}

The proof of the proposition is in delayed to  Appendix~\ref{sec:proof-prds}.
Since such $p$-values are PRDS, it gives an upper bound control of FDR using Corollary~\ref{cor:overcal}.

\begin{corollary}[PRDS property for overlapping calibration sets]
	\label{cor:overcal}
	Under the same conditions as Proposition~\ref{prop:prds-overcal} and the following condition on calibration set cardinality: $\exists \nu \geq 1, n = \nu \frac{m}{\alpha}-1$:
	
	$$FDR(\hat \varepsilon_{BH}, \hat p) \leq  \frac{m_0 \alpha}{m}.$$ 
\end{corollary}
For perfect FDR control, a lower bound is required in addition to the upper bound proven above.
However, no theoretical results exist to compute the lower bound. Indeed, the existing proof in \cite{MarandonMachinelearningmeets2022} did not extend to overlapping calibration sets. 
Therefore, the experiment in Appendix~\ref{sec:experiment-overlapping-calibrationset} suggests to establish this lower bound empirically.  Table~\ref{table:fdr-dependence-calibration-permtest2} shows the $p$-value associated with hypothesis $\mathcal{H}_0$ ``FDR is identical whatever the overlap proportion between calibration sets''. Please note that the $p$-values displayed in this table should not be confused with $\hat p_t$ which is associated with the $H_{0t}$ hypothesis ``$X_t$ is generated by $\mathcal P_0$''.
All tested hypotheses have a $p$-values greater than the (Bonferroni) threshold equal to $0.00625$. According to this experiment, there are no significant differences in the resulting FDR between the different proportions of overlap in the calibration sets. 
This would suggest that considering overlapping calibration sets should not worsen the control of false positives and negatives too much. 
\begin{table}[H]
	\caption{$p$-values resulting from permutation tests \label{table:fdr-dependence-calibration-permtest2}}
	\begin{tabular}{l|rr|rr|rr|rr}
		\toprule
		n &  249  &  250  &  499  &  500  &  749  &  750  &  999  &  1000 \\
		\midrule
		$p$-value of the test &0.300& 0.0326& 0.572& 0.313& 0.588& 0.435& 0.735& 0.690\\
		\bottomrule
	\end{tabular}
\end{table}

	\section{Global control of FDR  over the full time series}
	\label{sec:golbal}	\label{sec:locglofdr}
\begin{figure}[h!]
   	\centering
   	\includegraphics[width=0.6\linewidth]{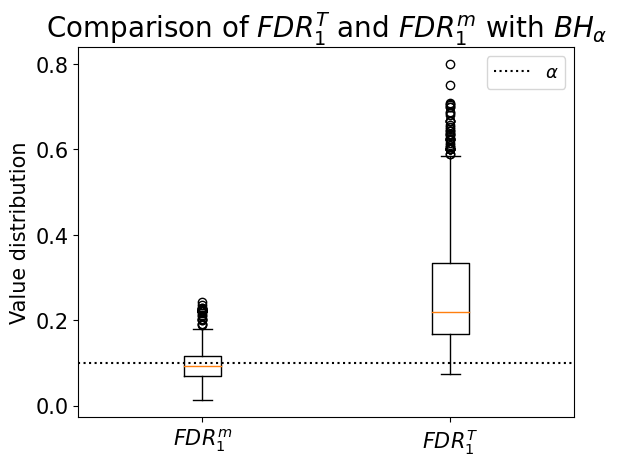}
   	\caption{Comparison of the calculation of FDR computed locally on a subseries and FDR computed globally on the entire time series using the Benjamini-Hochberg procedure applied to a subseries. }
   	\label{fig:locvsglo}
   \end{figure}

	In the previous Section~\ref{sec:fdr_emp}, it was seen that the BH procedure provides control on FDR of subseries of length $m$ if the calibration set cardinality is carefully tuned.
	While working with streaming time series data, the anomaly detection problem requires to control the FDR value of the full time series to ensure that the global false detection rate (FDR) remains under control at the end of the iterative process. The final criterion to control is then the global FDR criterion given by
	\begin{align*}
		FDR_1^\infty(\hat\varepsilon, \hat p) ,
	\end{align*}
	where $\hat \varepsilon = (\hat \varepsilon_t)_{t\geq 1}$ denotes a sequence of data-driven thresholds, and $ \hat p$ stands for a sequence of empirical $p$-values (see Section~\ref{sec:fdr_emp} for further details).
   However, the control of FDR of subseries does not imply control of the global FDR of the entire time series. This fact is illustrated in Figure~\ref{fig:locvsglo}. This figure is obtained by splitting a series of length 1000 into 10 subseries of length 100. The BH procedure is then applied to each subseries. FDP is calculated for each subseries and for each complete time series. The results are displayed in two boxplots. The left boxplot shows that $BH_\alpha$ provides the desired control at the level $\alpha=10\%$ for FDR computed over the subseries of length $m$. However, the right boxplot clearly deviates from $\alpha$, meaning that the actual FDR value for the full time series of length $1\,000$ is much larger than $\alpha$ (more than 20\% on average), leading to more false positives at the level of the full time series. 

    This implies that controlling FDR on subseries is not a good criterion. However, in contrast to the global objective, anomaly detection still requires a decision at each time step, i.e., for each new observation, without knowing what the next ones will be.
    This justifies the need for another local criterion that is used to make a decision at each iteration, leading to the sequence of data-driven thresholds $\hat \varepsilon = (\hat \varepsilon_t)_{t\geq 1}$ computed over subseries.
    An additional difficulty arises from the connection that must be made between this local criterion and FDR of the full time series.
	
	To this end, Section~\ref{sec:locmfdr} explores the connection between FDR of the full time series and the so-called modified-FDR (mFDR) of the subseries. In particular, it turns out that controlling the mFDR value for all subseries of a given length $m$ yields the desired FDR control for the full time series. Section~\ref{sec.modified.BH} then explains how the classical BH-procedure can be modified to get the mFDR control for subseries of length $m$.

\subsection{How mFDR can help in controlling FDR of the full time series?}
	\label{sec:locmfdr}

	\paragraph{Mixture model and time series}
	In this section one assumes that the time series is generated from a mixture process between a reference distribution $\mathcal P_0$ and an alternative distribution $\mathcal P_1$. 
	The anomaly locations are supposed to be independent and generated by a Bernoulli distribution. 
	This is a common assumption usually made in the literature \cite{huber1992robust, staerman2022functional} for simplicity. 
	\begin{definition}[Time series process with anomalies]\label{def:tsa}
		Let $\pi \in [0,1]$ be the anomaly proportion and $\mathcal P_0$ and $\mathcal P_1$ be two probability distributions on the observation domain $\mathcal X$. $\mathcal P_0$ is the reference distribution and $\mathcal P_1$ denotes the alternative distribution.		
		The generation process of a time series containing anomalies $(A_t)_{t\geq 0}$ is given, for each $t\geq 0 $, by
		\begin{itemize}
			\item $A_t \sim B(\pi)$ (Bernoulli distribution)
			\item if $A_{t}=0$, then $X_t \sim \mathcal{P}_0$.
			\item if $A_t=1$, then $X_t \sim \mathcal{P}_1$.
		\end{itemize}
	Moreover, given the above scheme, $(X_t)_{t\geq 0}$ is a random process with independent and identically distributed random variables $X_t \sim (1-\pi)\mathcal P_0 + \pi\mathcal P_1$.
	\end{definition}
	This definition details how anomalies are generated. In particular, it assumes that anomalies are independent from each other. It should be noted that this does not prevent us from observing a sequence of successive anomalies along the time series. However, this scheme differs substantially from the case analyzed by \cite{gretton2012kernel}, where specific patterns of successive anomalies are searched for.

	\paragraph{Threshold on subseries}
	%
	
	Let us start with a subseries of length $m$ in which each observation is summarized by its corresponding empirical $p$-value, and let us assume that there exists a function $f_\winsize :$ $[0,1]^\winsize \to [0,1] $ that maps a set of $m$ empirical $p$-values onto a real-valued variable. This random variable corresponds to the data-driven threshold. This function $f_\winsize$ is called a \emph{local threshold} function because it returns a threshold when applied to a subset of length $m$.
	Given the notation above, the threshold sequence $\hat \varepsilon_o=(\hat \varepsilon_{o,t})_{t}$ can be defined as follows.
	
     \textbf{Threshold on (overlapping) subseries:}
		$\hat \varepsilon_o:$ $t \mapsto  \hat\varepsilon_{o,t}$ is given by
		\begin{align}\label{def.threshold.overlapping}
		\forall t\geq m, \qquad\hat \varepsilon_{o, t}  =  f_{\winsize}(\hat p_{t-\winsize+1}, \dots ,\hat p_t).
		\end{align}

Figure~\ref{tikz:winb} illustrates this data-driven threshold.   Because two successive subseries differ from each other by two observations, the thresholds are different at each time step.
	\begin{figure}[h!]
		\centering\fbox{
			\begin{tikzpicture}[scale=0.75]
			\draw (1, 0) node[right] {$\hat p_1, ~\dots, ~\hat p_{t-2m}, ~\dots,~\hat p_{t-m-1}, ~\hat p_{t-m},~ \hat p_{t-m+1}, ~\dots,~\hat p_{t-1}, ~\hat p_{t}$};
			
			\draw[black, thick] (7.7,0.5) rectangle (12.1,-0.5);
			\draw[black, thick] (6.6,0.4) rectangle (11.2,-0.6);
			\draw[black, thick] (2.6,0.5) rectangle (6.3,-0.5);
			
			\draw[black, thick, ->] (11.9,-0.5) -- (11.9,-1.7);
			\draw[black, thick, ->] (10.7,-0.6) -- (10.7,-1.7);
			\draw[black, thick, ->] (5.2,-0.5) -- (5.2,-1.7);
			
			\draw (11.9, -1) node[left] {$f_m$};
			\draw (10.7, -1) node[left] {$f_m$};
			\draw (5.2, -1) node[left] {$f_m$};
			
			\draw (1, -2) node[right] {$\hat \varepsilon_{o,1}, \dots, \hat \varepsilon_{o,t-2m}, \dots, \hat \varepsilon_{o,t-m-1}, \hat \varepsilon_{o,t-m}, \hat \varepsilon_{o,t-m+1}  \dots, \hat \varepsilon_{o,t-1}, \hat \varepsilon_{o,t}$};		
			\end{tikzpicture}}
	\caption{Illustration of a threshold on (overlapping) subseries. }\label{tikz:winb}
\end{figure}
It shows that the variables $\hat \varepsilon_{o,t}$ and $\hat\varepsilon_{o,t-1}$ are dependent because they share $m-1$ common observations. But for any $t$, $\hat \varepsilon_{o,t}$ is independent from $\hat \varepsilon_{o,t-m-1}$. This can be reformulated as $\hat \varepsilon_{o,t_1}$, $\hat \varepsilon_{o,t_2}$ are independent if and only if $|t_1-t_2|>m$.
The local threshold function is said to be permutation invariant if its values does not depends on the  $p$-value order. For any $\rho$ permutation of $\llbracket 1,m\rrbracket$,
\begin{align}
 f_m(p_{\rho(1)}, \ldots, p_{\rho(m)}) = f_m(p_{1}, \ldots, p_{m}).\label{eq:perm-inv}
\end{align}

In Section~1.2 of the Supplementary Materials, a different scheme for defining a data-driven threshold from subseries is presented. The time series is divided into disjoint subseries and the same threshold is applied to all points belonging to that subseries. The term ``overlapping subseries'' is used to distinguish it from the disjoint case.
In the present online anomaly detection context, considering the overlapping case sounds more convenient since the detection threshold can be updated at each time step (as soon as a new observation is given), which makes the anomaly detector more versatile. However, disjoint are easier to study.

	\paragraph{FDR control with threshold on (overlapping) subseries}
    \label{label:fdrcntr-overlapping}
    
	As illustrated in Figure~\ref{fig:locvsglo}, controlling FDR on each subseries of length $m$ (locally) is not equivalent to controlling FDR (globally) on the full time series. 
	However in online anomaly detection, a decision has to be made at each time step regarding the potential anomalous status of each new observation. 
	This requires a criterion to be controlled locally (on subseries) in such a way that the resulting global FDR value (the one of the full time series) can be proven to be controlled at the desired level $\alpha$.
	This requirement for a local criterion	justifies the introduction of the modified FDR criterion, denoted by mFDR \cite{xu2022dynamic, foster2008alpha}, which is defined as follows.
	\begin{definition}[mFDR]\label{def.mFDR} With the previous notations, the mFDR expression of the subseries from $t-m+1$ to $t$ is given by
		\begin{align*}
			mFDR_{t-m+1}^t(\hat\varepsilon, \hat p) = \frac{\mathbb E \left[\sum_{u \in \mathcal H_0, t-m+1 \leq u \leq t}\mathbb{1}[\hat p_u \leq \hat \varepsilon_u] \right] }{\mathbb E \left[\sum_{u=t-m+1}^t\mathbb{1}[\hat p_u \leq \hat\varepsilon_u] \right] } ,
		\end{align*}
		where $\hat\varepsilon=(\hat\varepsilon_u)_{t-m+1\leq u\leq t}$ denotes a sequence of thresholds, $\hat p$ is a sequence of empirical $p$-values evaluated at each observation of the subseries from $t-m+1$ to $t$.
	\end{definition}
	Mathematically the difference between mFDR and FDR is that the expectation is no longer on the ratio but independently on the numerator and the denominator. 
	The main interest for mFDR is clarified by Theorem~\ref{thm:adosw}, which establishes that the control of the latter at the $\alpha$ level provides a global control of FDR at the same level under simple conditions.

%
\begin{theorem}[Global FDP control using local threshold] \label{thm:adosw}
	Let $\hat p  = (\hat p_t)_{t\geq 1}$ be the $p$-value random process, for a time series that follows the scheme detailed in Definition~\ref{def:tsa}. Let $\hat{\mathbf{P}}=(\hat {\mathbf{P}}_{t,k})_{t\geq 1,1\leq k\leq m}$ a process of $p$-values vectors, such that $\hat{\mathbf{P}}_{t,m}=\hat p_t$. 
	Assume that $\hat \varepsilon_o:t \mapsto  \varepsilon_{o,t}$ is given by $\hat \varepsilon_{o,t} = f_m(\hat{\mathbf{P}}_t) $, for any $ t\geq 1$, with $f_m :$ $[0,1]^\winsize \to [0,1] $ permutation invariant (see Eq.~\eqref{eq:perm-inv}).
	Let us also assume that there exists $n$ such that $|t_1 - t_2| > n$ implies that $\hat p_{t_1} $ and $\hat p_{t_2}$ are independent, and $|t_1 - t_2| > n+m$ implies that $\hat P_{t_1} $ and $\hat P_{t_2}$ are independent. For all $t$, $\hat{\mathbf{P}}_t = (\hat{\mathbf{P}}_{t,1}, \ldots, \hat{\mathbf{P}}_{t,m})$ is exchangeable.
	Then, the global FDR (and FDP) value of the full (infinite) time series is equal to the local mFDR value of the any subseries of length $\winsize$ computed at time $t\in\mathbb{N}^*$. More specifically
	$$FDP_1^\infty(\hat \varepsilon_{o}, \hat p) = FDR_1^\infty(\hat \varepsilon_{o}, \hat p) = mFDR_{t-\winsize+1}^t(\hat \varepsilon_{o}, \hat{\mathbf{P}}_t) = mFDR_{1}^m(\hat \varepsilon_{o},\hat{\mathbf{P}}_m)$$
\end{theorem}
%
The proof is postponed to Appendix~\ref{sec:proof-thm-adosw}.
An important consequence is that the desired control for FDR of the full (infinite) time series at level $\alpha$ can be achieved if one can control the successive mFDR of all (shifted) subseries of length $m$ at level $\alpha$. This point is not obvious at all and constitutes the main concern of Section~\ref{sec.modified.BH} where a new multiple testing procedure is designed to yield the desired control of the mFDR criterion.
The main limitation of Theorem~\ref{thm:adosw} is the requirement that $f_m$ has to be permutation invariant. Note that this property holds for the BH procedure, for example.
Let us also mention that the empirical $p$-values computed as $\hat p_t = \hat p\mbox{-value}(a(X_t), \lbrace a(X_{t-n}),\ldots,a(X_{t-1})\rbrace )$ and vector $p$-values defined as $\hat{\mathbf{P}}_t=(\hat p_{t-m+1},\ldots, \hat p_{t})$ actually satisfy the requirements of Theorem~\ref{thm:adosw} regarding the independence and the stationarity. 
	
	\subsection{Modified BH-procedure and mFDR control}
	\label{sec.modified.BH}
	As shown in Section~\ref{sec:locmfdr}, controlling the FDR value of the full time series is equivalent to controlling mFDR of subseries of length $m$. The strategy then consists in controlling the mFDR criterion of all successive subseries of length $m$ along the full time series at level $\alpha$. 
	The main challenge addressed in the present section is to design a new multiple testing procedure that controls the local mFDR criterion at a prescribed level $\alpha$.

In Section~\ref{sec.mFDR.BH}, it is shown that applying $BH_\alpha$ to a subseries of size $m$ does not result in mFDR being controlled to the desired level $\alpha$. However, mFDR is upper bounded by a quantity that depends on the level of control $\alpha$ used in the BH procedure. This leads to a strategy of using the BH procedure, but with a different, well-chosen $\alpha'$, to allow control of mFDR at the desired level $\alpha$. This so-called modified BH-procedure is introduced in Section~\ref{sec.modifiedBH}.
	
	\subsubsection{mFDR control with the BH-procedure} \label{sec.mFDR.BH}
	The next Proposition~\ref{thm:mfr-bh} establishes the actual mFDR level achieved by the BH-procedure.
	\begin{proposition}
		\label{thm:mfr-bh}
		Let $(X_i)_1^\winsize$ satisfy the requirements detailed by Definition~\ref{def:tsa}, with an abnormality proportion $\pi$. $m_0$ is the random variable representing the number of data points generated by $\mathcal P_0$. Let $p=(p_1,\ldots, p_m)$ be the associated $p$-values. Let $\alpha$ belong to $[0,1]$. Suppose there are integers $\nu$ and $n$ such that $n = \nu \frac{m}{\alpha}-1$.
		Let $R_{\alpha, 1}^m$ be the random variable representing the number of rejections when $BH_{\alpha}$ is applied to $p$.
		Suppose one of the following three statements is true:
		\begin{enumerate}
			\item  $(p_1,\ldots, p_m)$ are true $p$-values that is, for any $1\leq i \leq m$, $p_i = \mathbb P_{X\sim \mathcal P_0}(a(X) \geq a(X_i))$.
			\item  $(p_1,\ldots, p_m)$ are empirical $p$-values with independent calibration sets of cardinality $n$, $p_i = \hat p_e(a(X_i), \lbrace a(Z_{i,1}), \ldots, a(Z_{i,n}))$. With $Z_{i,j}$ are iid random variables generated by $\mathcal P_0$.
			
			 In these two cases, let choose $i$ in $\mathcal H_0$, the sequence $(p_j')_{1\leq j\leq m}$ is defined by $p'_i=0$ and $p_j'=p_j$. Let $R^{*,m}_{\alpha,1}$ define the number of rejections when applying $BH_\alpha$ on $p'$
		\end{enumerate}

	    \begin{enumerate}
	    	 \setcounter{enumi}{2}
	    	\item  $(p_1,\ldots, p_m)$ are empirical $p$-values with a unique calibration set of cardinality $n$, $p_i = \hat p_e(a(X_i), \lbrace a(Z_{1}), \ldots, a(Z_{n}))$. With $Z_{j}$ are iid random variables generated by $\mathcal P_0$.
	    	
	    	Let choose $i$ in $\mathcal H_0$, the sequence $(p_j')_{1\leq j\leq m}$ is defined by $p'_i=0$ and $p_j'=p_j-\frac{1}{n}\mathbb 1[p_j<p_i]$. Let $R^{*,m}_{\alpha, 1}$ define the number of rejections when applying $BH_\alpha$ on $p'$.
	    \end{enumerate}

		Then, applying $BH_\alpha$ to the $p$-values $(p_i)_{1\leq i\leq \winsize}$ leads to 
		\begin{align*}
		mFDR_1^m(p)= \alpha \frac{\mathbb E\left[ \frac{|\mathcal H_0|}{m} R_{1,\alpha}^{*, m}\right]}{\mathbb{E} R_{1,\alpha}^m} \leq \alpha \frac{\mathbb E R_{1,\alpha}^{*, m}}{\mathbb{E} R_{1,\alpha}^m},
		\end{align*}
	
	    Furthermore, if $\mathbb E[R^{*,m}_{\alpha,1}|m_0]$ is decreasing:
	    \begin{align*}
	    	mFDR_1^m(p) \leq \alpha (1-\pi) \frac{\mathbb E R_{1,\alpha}^{*, m}}{\mathbb{E} R_{1,\alpha}^m},
	    \end{align*}
	\end{proposition}
The proof is moved to Appendix~\ref{sec:proof-thm-mfr-bh}. The upper bound is generally not too large since, the proportion of normal data $\frac{|\mathcal H_0|}{m}$ is close to 1. The decreasing property of $E[R^{*,m}_{\alpha,1}|m_0]$ is verified as soon as the data points generated by $\mathcal P_1$ have a higher probability of being detected as anomalies than those generated by $\mathcal P_0$. This is verified whenever the atypicity score is chosen correctly.
Then, to get control of mFDR, it is sufficient to find $\alpha'$ such that $\frac{E R^*_{\alpha'}}{\mathbb{E} R_{\alpha'}}=\alpha$, and then apply BH with level $\alpha'$ to get the data driven threshold $\hat\varepsilon_{BH_{\alpha'}}$.
This method is called ``modified Benjamini-Hochberg'' and is described in Definition~\ref{def:mbh}. There is no theoretical result in the case where the $p$-values $p_1, \ldots, p_m$ are PRDS. In particular, there are no theoretical results if the calibration sets overlap as described in Section~\ref{sec:extniid}. 
	
\subsubsection{Modified BH}\label{sec.modifiedBH}
From the previous Section~\ref{sec.mFDR.BH}, it is now possible to suggest and analyze the new modified BH-procedure (mBH in the sequel).
\begin{definition}[Modified BH-procedure (mBH)]\label{def:mbh}
Let $\winsize$ be an integer and $\alpha\in[0,1]$. 
Let us introduce the level $\alpha' = \alpha'(\alpha)$ an estimate of $\argmax_{\tilde\alpha}\left\lbrace\frac{\mathbb{E}[R_{1,\tilde\alpha}^{*,m}]}{\mathbb{E}[R_{1,\tilde\alpha}^m]}\tilde\alpha\leq\alpha\right\rbrace$, which can be given by some procedure.
Then the modified BH-procedure, denoted by $mBH_\alpha$, is given for all true $p$-values $(p_1,\ldots,p_m) \in [0,1]^m$ by,
\begin{align*}
mBH_\alpha(p_1,\ldots,p_m) = BH_{\alpha'}(p_1,\ldots,p_m).
\end{align*}		
The related $mBH_\alpha$ threshold at level $\alpha$ is defined as
\begin{align*}
\varepsilon_{mBH_\alpha} =  \varepsilon_{BH_{\alpha'}},
\end{align*} 
when computed with true $p$-values, and 
$\hat \varepsilon_{mBH_\alpha} =  \hat\varepsilon_{BH_{\alpha'}}$ when used with empirical $p$-values.
	\end{definition} 
The above definition defines the $mBH_\alpha$ in terms of the BH-procedure by simply changing the level of control $\alpha^\prime$.
There are several ways to estimate $\alpha'$. It can be done using heuristics, which give a formula to directly calculate $\alpha'$ as a function of $\alpha$, $m$ and $\pi$. However, the heuristics rely on assumptions that are difficult to verify and on the proportion of anomalies, which is often unknown. Another solution is to estimate $\mathbb{E}[R_{1,\tilde\alpha}^{*,m}]$ and $\mathbb{E}[R_{1,\tilde\alpha}^m]$ on a grid of $\tilde\alpha$ using a training set. These two methods are described in more detail in Appendix~\ref{sec.evaluate.alphaprim}.

Combining the results of Theorem~\ref{thm:adosw} and Proposition~\ref{thm:mfr-bh}, the procedure $mBH$ allows to get the control of FDR of the complete series at the desired level $\alpha$. This property is described in Corollary~\ref{cor:mbh.contr.fdrg}.

\begin{corollary}[Control of FDR using mBH]
	\label{cor:mbh.contr.fdrg}
	Under the same notations and assumptions as Theorem~\ref{thm:adosw}. Let $m$ and $n$ be two integers.
	Suppose one of the following statements if true:
	\begin{enumerate}
	\item  For all $t$ $p$-values of $\hat{\textbf{P}}_t$ are true $p$-values.
	\item  For all $t$ $p$-values of $\hat{\textbf{P}}_t$ are are empirical $p$-values with independent calibration sets of cardinality $n$.
	
	Let $i$ be a true null hypothesis in $\hat{\textbf{P}}_t$, the sequence $\hat{\textbf{P}}_t'$ is defined by $\hat{\textbf{P}}_{t,i}'=0$ and $\hat{\textbf{P}}_{t,j}'=\hat{\textbf{P}}_{t,j}$ for $j\neq i$. Let $R^{*,m}_{\alpha,1}$ be the number of rejections when applying $BH_\alpha$ to $\hat{\textbf{P}}_{t}'$.

	\item  For all $t$, $p$-values of $\hat{\textbf{P}}_t$ are are empirical $p$-values with a unique calibration set of cardinality $n$.
	
	Let $i$ be a true null hypothesis in $\hat{\textbf{P}}_t$, the sequence $\hat{\textbf{P}}_t'$ is defined by $\hat{\textbf{P}}_{t,i}'=0$ and $\hat{\textbf{P}}_{t,j}'=\hat{\textbf{P}}_{t,j}-\frac{1}{n}\mathbb 1[\hat{\textbf{P}}_{t,j}<\hat{\textbf{P}}_{t,i}]$ for $j\neq i$. Let $R^{*,m}_{\alpha, 1}$ be the number of rejections when applying $BH_\alpha$ to $\hat{\textbf{P}}_{t}'$.
    \end{enumerate}
	
    Suppose there are $\nu$ and $\alpha'$ such that:
	\begin{align*}
		\frac{\mathbb{E}R_{1, \alpha'}^{*,m}}{\mathbb{E} R_{1,\alpha'}^m}\alpha'= \alpha 
	\end{align*}    	

	Then, FDR of the entire time series can be controlled at level $\alpha$ by using $\hat \varepsilon_{BH_{\alpha'}, t} = f_m(\hat{\mathbf{P}}_t)$
	\begin{align}
		FDR_{1}^{\infty}(\hat \varepsilon_{BH_{\alpha'}}, \hat p) \leq (1-\pi) \alpha
	\end{align}
%
\end{corollary}
The proof is moved to Appendix~\ref{sec:proof-cor-mbh.contr.fdrg}.
Corollary~\ref{cor:mbh.contr.fdrg} describes the properties that the sequence of $p$-values $\hat p$ and the sequence of vector $p$-values $\hat{\mathbf P}$ should satisfy in order to get the control on FDR.
Theorem~\ref{thm:mbh-fdr} gives practical ways to compute $\hat p$ and $\hat{\mathbf P}$ that satisfy these requirements. 
It gives the ingredient to build an anomaly detector controlling FDR at a desired level $\alpha$.
Our Algorithm~\ref{alg:fdrcontroad} implements this result.

\begin{theorem}[Global FDR control using $mBH_\alpha$] \label{thm:mbh-fdr}
 	Let $(X_t)$ be a mixture process introduced in Definition~\ref{def:tsa}.
   	Let $\alpha\in[0,1]$ be the desired FDR level for the full time series. Let $m$ and $n$ be integers.
If one of the three statements is true:
\begin{enumerate}
	\item $\forall t, \hat p_t = 1-\mathbb P_{X\sim\refdistrib}(a(X)>a(X_t))$
	\item $\forall t, \hat p_t = \hat p_e(a(X_t), \lbrace a(Z_{t,1}),\ldots, a(Z_{t,n})\rbrace)$ with $Z_{t,i} \sim \refdistrib$ 
	
	In this two cases, let $\hat{\mathbf{P}}_t=(\hat p_{t-m+1},\ldots ,\hat p_{t})$.
	Let choose $i$ in $\mathcal H_0\cap\llbracket t-m+1,t\rrbracket$, the sequence $\hat{\textbf{P}}_t'$ is defined by $\hat{\textbf{P}}_{t,i}'=0$ and $\hat{\textbf{P}}_{t,j}'=\hat{\textbf{P}}_{t,j}$ for $j\neq i$. Let $R^{*,m}_{\alpha,1}$ be the number of rejections when applying $BH_\alpha$ to $\hat{\textbf{P}}_{t}'$.
\end{enumerate}

\begin{enumerate}
    \setcounter{enumi}{2}
  		\item $\forall k \in \llbracket 1, m\rrbracket, (\hat{\textbf{P}}_t)_{t,k} = \hat p_e(a(X_{t-m+k}), \mathcal{S}^{cal})$, with the calibration set \begin{align*}\mathcal{S}^{cal}&=\lbrace (1-A_{t-n+1-m})a(X_{t-n+1-m}) + A_{t-n+1-m}a(Z_{t,1}),\ldots,\\
  		&\quad (1-A_{t-m})a(X_{t-m}) + A_{t-m}a(Z_{t,n})\rbrace\end{align*}
  	 with $Z_{t,i} \sim \refdistrib$ 
  		and $\hat p_t = \hat{\textbf{P}}_{t,m}$.
  		
  		Let choose $i$ in $\mathcal H_0\cap\llbracket t-m+1,t\rrbracket$, the sequence $\hat{\textbf{P}}_t'$ is defined by $\hat{\textbf{P}}_{t,i}'=0$ and $\hat{\textbf{P}}_{t,j}'=\hat{\textbf{P}}_{t,j}-\frac{1}{n}\mathbb 1[\hat{\textbf{P}}_{t,j}<\hat{\textbf{P}}_{t,i}]$ for $j\neq i$. Let $R^{*,m}_{\alpha, 1}$ define the number of rejections when applying $BH_\alpha$ to $\hat{\textbf{P}}_{t}'$.   
  	\end{enumerate}

 Suppose there are $\alpha'$ and $\nu$ such that:
 \begin{align*}
 	\frac{\mathbb{E}R_{1, \alpha'}^{*,m}}{\mathbb{E} R_{1,\alpha'}^m}\alpha'\leq \alpha  \quad \mbox{ and } \quad n = \nu m/\alpha' - 1.
 \end{align*}    	   
   
 Then, with $\hat \varepsilon_{BH_{\alpha'}, t} = \hat \varepsilon_{BH_{\alpha'}}(\hat{\textbf{P}})_t$:
 \begin{align*}
 	FDR_1^\infty(\hat \varepsilon_{BH_{\alpha'}}, \hat p)\leq (1-\pi) \alpha .
 \end{align*}

 %
   \end{theorem}
The proof is deferred to Appendix~\ref{sec:proof-thm-mbh-fdr}.
The main merit of Theorem~\ref{thm:mbh-fdr} is to establish the actual level of control for the global FDR of the full time series depending on the type of empirical $p$-value used in the anomaly detection process.
The last type of empirical $p$-values is (almost) the one used in practice in the present work. More specifically, Section~\ref{sec:practical-pvalue-estimators} describes empirical $p$-values based on a ``Sliding Calibration Set''.

	\section{Empirical results}
		\label{sec.FDR.mFDR.empirical}
		In this section, properties related to the anomaly detector are empirically evaluated to illustrate and complete the various theoretical results set out in the previous sections.
	First, from Corollary~\ref{th:empFDR}, it is known that when BH is applied to empirical $p$-value subseries, there are specific values for the calibration set cardinality at which FDR is controlled to the desired value $\alpha$. In Section~\ref{subsec:emprslt}, to illustrate this property and to further study the impact of cardinality, FDR and FNR are calculated as a function of calibration set cardinality in several scenarios.
	Then, according to Proposition~\ref{thm:mfr-bh}, the mBH procedure (unlike the BH procedure) allows to control mFDR of subseries. 
	However, this hides the fact that the mBH procedure requires the estimation of a parameter $\alpha'$. In case this parameter is estimated by the heuristics described in Appendix~\ref{sec.heuristic.arguments}, it is necessary to check the assumptions \ref{assume.heuristic} and \ref{assume.power}. These assumptions are difficult to ensure in practice.
	In Section~\ref{sec:emp-diswin}, the evaluation is done on simulated data where the level of atypicity of the anomalies ( affecting the power of the detector) varies from one sample to another. Different scenarios are tested to check if the mFDR control holds.
	Finally, Theorem~\ref{thm:adosw} gives asymptotic FDR control over the full time series. But there is no result about the speed of convergence, which is necessary when applied to finite time series. In Section~\ref{sec:fdr.conv}, FDR for the full time series is calculated in different situations, as a function of the time series size. It is possible to determine when the entire series reaches control of FDR.
	
		\subsection{Assessing the local control of FDR with empirical $p$-values}\label{subsec:emprslt}
	
	The purpose of the present section is to compute the actual FDR
value when empirical $p$-values are used instead of true ones.
The question raised here is whether FDR of the subseries is really controlled at a given level $\alpha$.
The empirical results must be compared with the theoretical FDR
expression established in Theorem~\ref{thm:fdr-bh-emp}.
		
	\paragraph{Experiment description}\label{sec:empbh-experiment-description}
	Two scenarios have been considered to explore how much the thickness of the distribution tails can affect the results.
	\begin{enumerate}
		\item {\bf Thin tails:}
		
		The reference probability distribution is $\mathcal
		P_0=\mathcal N (0,1)$ for normal observations and $\mathcal
		P_1=\delta_{\Delta_{\mathcal N}}$ for anomalies, where $\Delta_{\mathcal
			N} \in\mathbb{R}$ is a parameter encoding the strength of the shift.
		Here $\delta_{\Delta_{\mathcal N}}$ denotes the Dirac measure such that
		$\delta_{\Delta_{\mathcal N}}(z)= 1$ if $z=\Delta_{\mathcal N}$ and 0
		otherwise.		
		\item  {\bf Thick tails:}
		
		$\mathcal P_0= \mathcal{T}(5)$ is a Student probability distribution with 5 degrees of freedom and $\mathcal
		P_1=\delta_{\Delta_{t}}$ denotes the alternative distribution
		of anomalies, where $\Delta_{\mathcal N} \in\mathbb{R}$ is a parameter
		encoding the strength of the shift.
	\end{enumerate}
	Regarding the value of the shift strength in Scenarios 1 and 2, two
	values of $\Delta_{\mathcal N}$ have been considered $3.5$ and $4$.
	The values of $\Delta_{\mathcal{T}}$ have been chosen such that
	$$\mathbb P_{X \sim \mathcal N (0,1) }(X>\Delta_{\mathcal N }) = 
	\mathbb P_{X \sim  \mathcal{T}(5) }(X>\Delta_t)$$
	for each choice of $\Delta_{\mathcal N }$. This avoids any bias in
	the comparison of the detection power of the considered strategy
	depending on the ongoing scenario.
	
	Different cardinalities have been considered for the calibration set
	following the mathematical expression
	\begin{align*}
		n \in \{k\cdot10, k\in \llbracket 1, 200\rrbracket \} \cup \{\ell \cdot10-1, \ell \in \llbracket 1,
		200\rrbracket \}.
	\end{align*}
	In particular, all integers between 10 and 2\,000 are explored with a step size equal to 10, as well as all integers between 9 and 1\,999 with a step size of 10. This choice is justified by the particular expression of FDR value provided by Corollary~\ref{th:empFDR}.
	All the $n$ elements of the calibration set are generated from the reference distribution that is, $\lbrace Z_{1},\dots,Z_{n}\rbrace \sim \mathcal P_0$. Concerning the $m$ observations corresponding to the tested hypotheses $\lbrace X_1,\dots,X_m\rbrace$, the $m_0$ normal observations are generated according to $\mathcal{P}_0$ and the $m_1$ anomalies from $\mathcal{P}_1$. Here $m=100$ and $m_0=99$.
	Each simulation condition has been repeated $B=10^4$ times.
	For each repetition $1\leq b\leq B$, the observations are indexed by $b$ such that $X_{b,j} \sim \mathcal P_1$ for each $j \in \llbracket 1,m_1\rrbracket$, and $X_{b,j} \sim \mathcal P_0$ for $j \in \llbracket m_1+1,m\rrbracket$.
	In the present scenarios, the anomalies are all in the right tail of the reference probability distribution. Therefore, the empirical $p$-values are computed according to Eq.~\ref{eq:empv} with the scoring function $a(x)=x$. 
	For each repetition $1\leq b\leq B$,
	\begin{align*}
		\forall 1\leq j\leq \winsize, \quad  \hat p_{b,j} =
		\hat p_e(X_{b,j}, \lbrace Z_{b,1},\dots,Z_{b,n}\rbrace).
	\end{align*}
	After computing the empirical $p$-values, the $BH_\alpha$ procedure (see Definition~\ref{bhdef}) is applied to this sequence. Then FDP and FNP of the sequence are calculated. The results are averaged over all sequences to get an estimation of FDR and FNR.
		
	\paragraph{Results and analysis}
	\label{sec:empbh-experiment-results}

	\begin{figure}[h!]
		\centering
		\begin{subfigure}[b]{.45\linewidth} 
			\centering
			\includegraphics[width=0.7\linewidth]{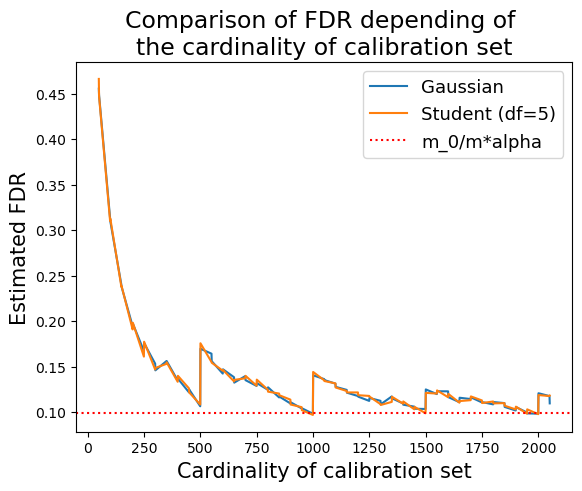}
			\caption[]{FDR depending on $n$ with 4 sigmas anomalies}
			\label{fig:fdrgauss1}
		\end{subfigure}
		\begin{subfigure}[b]{.45\linewidth} %
			\centering
			\includegraphics[width=0.7\linewidth]{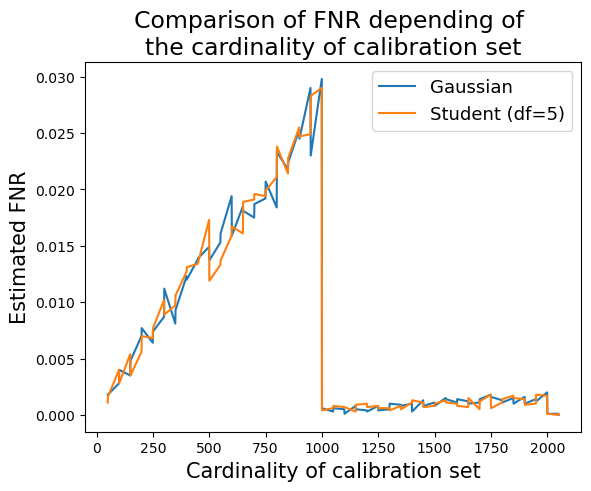}
			\caption[]{FNR depending on $n$ with 4 sigmas anomalies}
			\label{fig:fnrgauss1}
		\end{subfigure}
		\begin{subfigure}[b]{.45\linewidth} %
			\centering
			\includegraphics[width=0.7\linewidth]{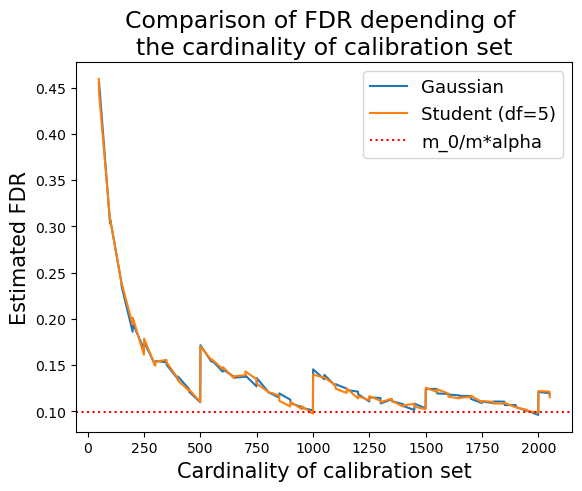}
			\caption[]{FDR depending on $n$ with 3.5 sigmas anomalies}
			\label{fig:fdrgauss2}
		\end{subfigure}
		\begin{subfigure}[b]{.45\linewidth} %
			\centering
			\includegraphics[width=0.7\linewidth]{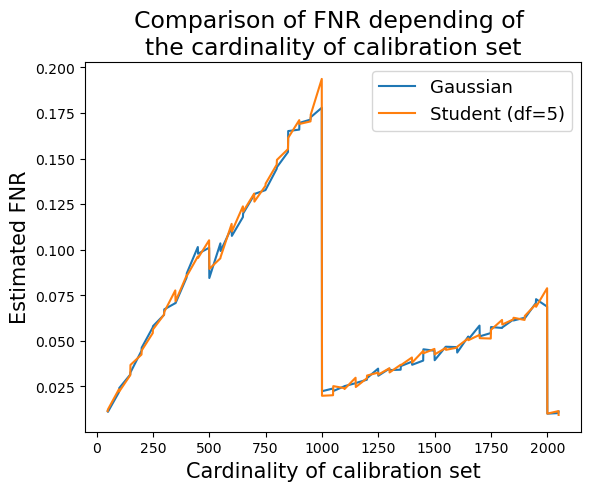}
			\caption[]{FNR depending on $n$ with 3.5 sigmas anomalies}
			\label{fig:fnrgauss2}
		\end{subfigure}
		\begin{subfigure}[b]{.45\linewidth} %
			\centering
			\includegraphics[width=0.7\linewidth]{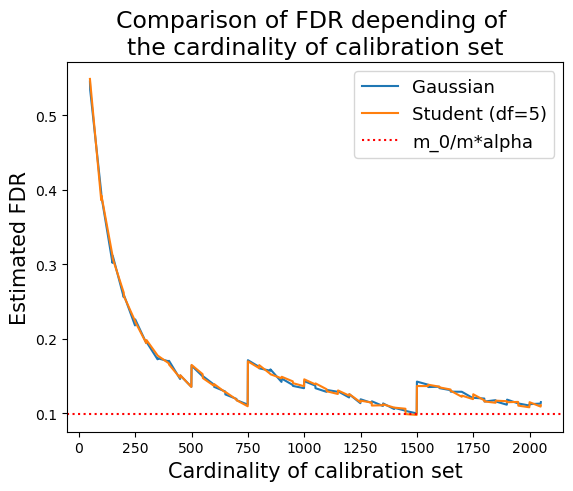}
			\caption[]{FDR depending on $n$ with 4 sigmas anomalies and
				$m=150$}
			\label{fig:fdrgauss3}
		\end{subfigure}
		\begin{subfigure}[b]{.45\linewidth} %
			\centering
			\includegraphics[width=0.7\linewidth]{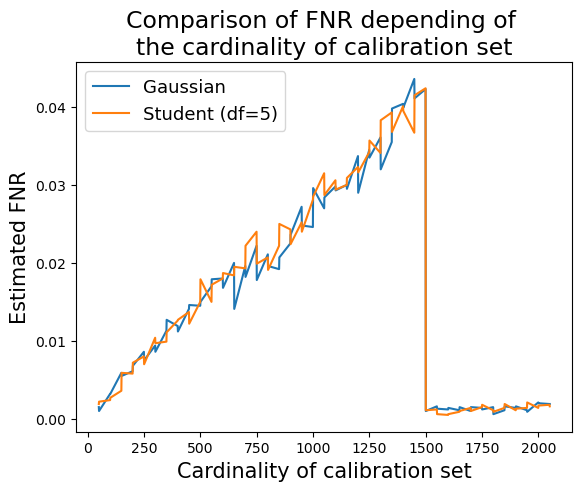}
			\caption[]{FNR depending on $n$ with 4 sigmas anomalies and
				$m=150$}
			\label{fig:fnrgauss3}
		\end{subfigure}
		\caption[]{Effect of calibration set cardinality and abnormality
			score on the FDR control and FNR}
		\label{fig:fdnrgauss}
	\end{figure}
	Figure~\ref{fig:fdnrgauss} displays FDR (left panel) and FNR (right panel) as a function of the calibration set cardinality for the two scenarios (Gaussian and Student) described in the previous paragraph. The blue (respectively orange) curve corresponds to the Gaussian (resp. Student) reference distribution.
	The horizontal line is the desired FDR control level $\alpha=0.1$.
	Figures~\ref{fig:fdrgauss1} and~\ref{fig:fnrgauss1} are obtained with $\Delta_{\mathcal N} =4$, while Figures~\ref{fig:fdrgauss2} and~\ref{fig:fnrgauss2} are obtained with $\Delta_{\mathcal N}=3.5$.\newline 
	\null\quad According to these plots, the behavior of both FDR and FNR does not exhibit a strong dependence with respect to the reference probability distribution. The results are very close for both the Gaussian and Student distributions. As shown in Figures~\ref{fig:fdrgauss1} and~\ref{fig:fdrgauss2}, FDR control is achieved at the prescribed level for particular values of the calibration set cardinality. These values coincide with the ones given by Theorem~\ref{th:empFDR}, which are of the form $\nu\alpha/m-1=\nu 10^{3}-1$ for an integer $\nu$. Notice that the FNR curve rises sharply from 1 to $n=999$. This reflects that although the FDR value becomes (close to) optimal as $n$ increases from 1 to $n=999$, the proportion of false negatives increases at the same time, leading to suboptimal statistical performance (because of too many false negatives). 
	Fortunately, a larger cardinality $n$ of the calibration set, e.g. $n=1999$, would greatly improve the results at the cost of a larger calibration set, which also increases the computational cost.
	
	 The FDR value seems not to depend on the strength of the distribution shift $\Delta$, as illustrated by Figures~\ref{fig:fdrgauss1} and~\ref{fig:fdrgauss2}. Let us mention that the focus of the expectation does not say anything about the probability distribution of FDP, which can be influenced by the strength of the shift.
	 In contrast, the comparison of Figures~\ref{fig:fnrgauss1} and~\ref{fig:fnrgauss2} clearly shows the impact of the shift strength on the FNR value. As the shift strength becomes lower, anomalies become more difficult to detect, inflating FNR.

	According to Figure~\ref{fig:fdrgauss1}, FDR can be controlled for particular values of the calibration set cardinalities that are not predicted by Corollary~\ref{th:empFDR}. For example, $n=1499$ controls FDR at $0.1$, while FDR is controlled for $n=999$ and $n=1999$ according to Corollary~\ref{th:empFDR}. 
Furthermore, complementary experiments (summarized by Figure~5 in Section~3.1 of the Supplementary Materials) illustrate that these particular values of $n$ depend on the number of anomalies $m_1$, in contrast to the values derived from Corollary~\ref{th:empFDR}. 
Using Theorem~\ref{thm:fdr-bh-emp}, their existence can be explained by the specific distribution of the number of detections. For example, Figure~5d shows a high probability of detecting $3$ anomalies. More generally, assuming that there exists $k^*\in \llbracket 1,m \rrbracket$ such that $\mathbb{P}(R(i)=k^*)\approx 1$, Theorem~\ref{thm:fdr-bh-emp} justifies that
	\begin{align*}
		FDR_1^m(\hat\varepsilon_{BH_\alpha}, \hat p) &= \frac{n}{n+1} \cdot \alpha \frac{m_0}{m} + \frac{m_0}{n+1}\sum_{k=1}^m \frac{(1-q_{n,k})}{k} \mathbb{P}(R(i)=k)\\
		&\approx \frac{n}{n+1} \cdot \alpha \frac{m_0}{m} + \frac{m_0}{n+1}\frac{(1-q_{n,k^*})}{k^*}. 
	\end{align*}
	Then the proof detailed in Appendix~\ref{proof:empFDR} yields that $1-q_{n,k^*}= \frac{\alpha}{m (n+1)} $ can be reached for all $\nu \geq 1$, such that $n=\nu\frac{m}{\alpha k^*}-1$.
	This allows to conclude that $FDR_1^m(\hat\varepsilon_{BH_\alpha}, \hat p) \approx \frac{m_0 \alpha}{m}$.

\begin{figure}[!h]  
		\begin{subfigure}{0.45\linewidth}
			\centering
			\includegraphics[width=0.7\linewidth]{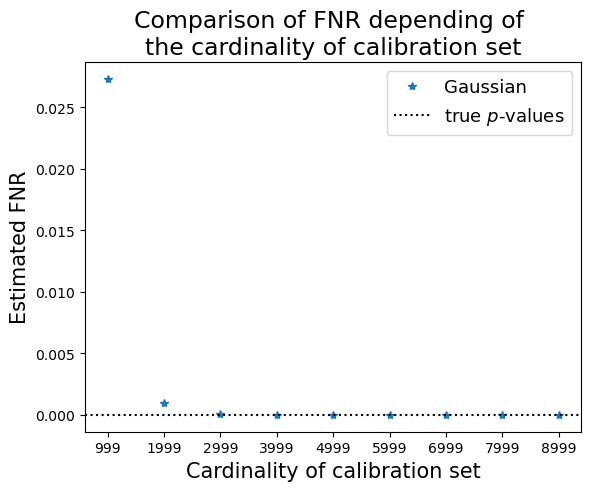}
			\caption{$\Delta=4$, $m_1=1$}
			\label{fig:fnr-case1}
		\end{subfigure}
		\begin{subfigure}{0.45\linewidth}
			\centering
			\includegraphics[width=0.7\linewidth]{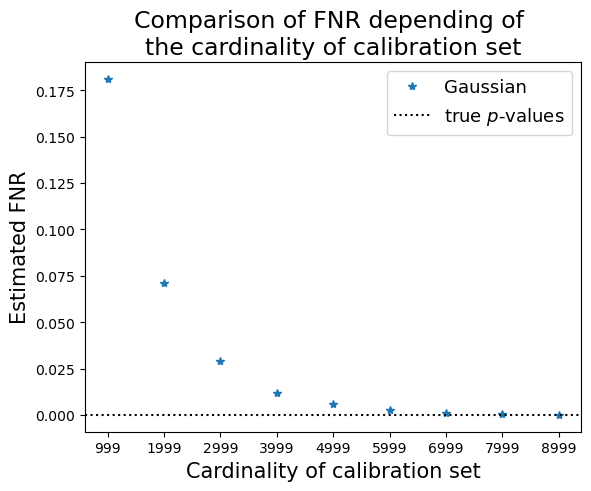}
			\caption{$\Delta=3.5$, $m_1=1$}
			\label{fig:fnr-case2}
		\end{subfigure}
		\begin{subfigure}{0.45\linewidth}
			\centering
			\includegraphics[width=0.7\linewidth]{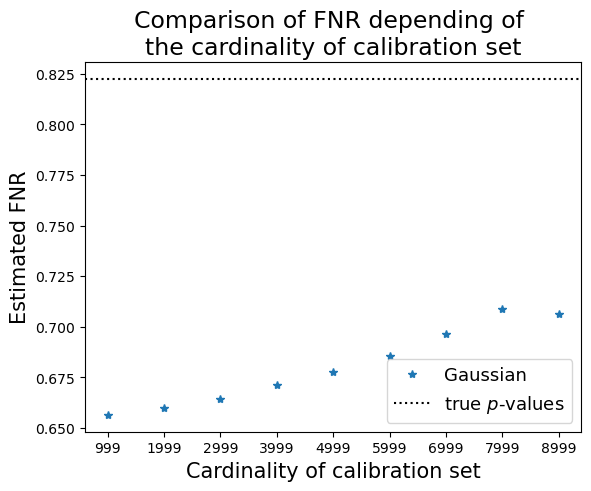}
			\caption{$\Delta=3$, $m_1=1$}
			\label{fig:fnr-case3}
		\end{subfigure}
		\begin{subfigure}{0.45\linewidth}
			\centering
			\includegraphics[width=0.7\linewidth]{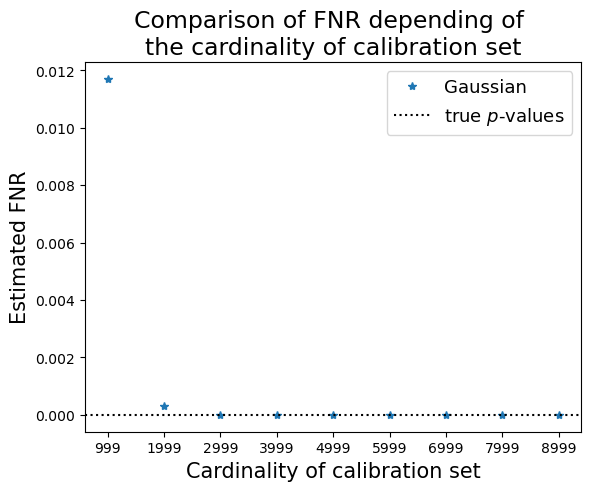}
			\caption{$\Delta=3$, $m_1=5$}
			\label{fig:fnr-case4}
		\end{subfigure}
		\caption{\label{fig:fnr-n} FNR as a function of the calibration set cardinality constrained to belong to $\mathcal{N}$.}
	\end{figure}

	\paragraph{How to choose the right cardinality of the calibration set?}
	Intuitively, an optimal choice of the cardinality $n$ of the calibration set should enable the FDR control while minimizing the number of false negatives and avoiding excessive computation time. 
	To achieve this objective, the first part of Corollary~\ref{th:empFDR} explains that  $n$ must be chosen from the set $\mathcal{N} = \lbrace \nu m/\alpha - 1,\quad  \nu \geq 1\rbrace$. 
	Using the simulation scenarios described in the paragraph ``Experiment description'', FNR is estimated for different values $n$ of the calibration set cardinality when $n \in \mathcal{N}$. The results are visualized in Figure~\ref{fig:fnr-n} where the FNR value is represented as a function of $n$. 
	For all the considered scenarios (Fig.~\ref{fig:fnr-case1}, \ref{fig:fnr-case2}, \ref{fig:fnr-case3}, \ref{fig:fnr-case4}), the FNR value converges to the value reached with true $p$-values (horizontal dashed line) as $n$ grows. From Figures~\ref{fig:fnr-case1} and~\ref{fig:fnr-case2}, the convergence speed depends on the ``difficulty'' of the problem.
	
	 In practice, however, the lack of labeled data prevents the computation of the actual FNR value, making it difficult to choose the optimal value of $n$.
To address this challenge, our suggestion is to choose the largest possible value of $n$ that does not exceed the computation time limit. This would result in a value of $n$ that minimizes the FNR criterion while satisfying the computational constraints. However, following this suggestion does not prevent us from computational drawbacks, as shown in Figure~\ref{fig:fnr-case1}, where the optimal FNR value is reached for $n=3999$, while choosing a larger $n$ does not bring any gain (but still increases the computational cost). 
	
	\subsection{Control of mFDR on subseries}
	\label{sec:emp-diswin}
	\paragraph{Experiment description}
	Let say the parameter $\alpha'$ is chosen using the heuristic strategy provided in Appendix~\ref{sec.heuristic.arguments}, it comes from Corollary~\ref{cor:mbh.contr.fdrg}, the $mBH_{\alpha}$ procedure controls the $mFDR_1^m$ only if the \ref{assume.power} assumption is satisfied. Since the power of the anomaly detector depends on how easy it is to detect anomalies, the level of atypicity $\delta$ is introduced. To quantify the atypicity of a data point $X_t$, the true $p$-value is computed as $p_t = \PP_{X \sim \refdistrib}(X>X_t)$, and the atypicity level is defined as the inverse of the $p$-value: $\delta_t = 1/p_t$. The atypicity level is preferred to the $p$-values because it is easier to plot on the x-axis of the graph when the $p$-value is small. To evaluate the effect of power, for each sample, all anomalies have their atypicity level lower bounded by a given parameter $\delta$.  Therefore, it is possible to observe the effect of a variation in the level of atypicity on $mFDR_1^m$, $FDR_1^m$, and $FNR_1^m$.
	For a given scenario - i.e. a proportion of anomalies $\pi$, a level of atypicity $\delta$, and a desired level of mFDR noted $\alpha$ - the actual mFDR, FDR, and FNR are estimated. These quantities are estimated using $K=50$ samples of $m$ data points. To control for the estimation error introduced by estimating with a finite number of samples, each estimation is repeated $B=100$ times. The estimation proceeds as follows:
	For each repetition $b$, for each sample $k$, $m$ data points are generated. $m_0$ normal data points are generated according to the reference distribution $U([0,1])$. $m_1$ anomalies are generated according to the alternative distribution $U([0,1/\delta])$, where $\delta$ is the degree of atypicity of the anomalies. Two procedures are applied to each of this sequence of $m$ points: $BH_\alpha$ and $mBH_\alpha$. For each procedure, FDP, FNP as well as the number of detections and false positives are calculated. By averaging FDP and FNP over the $K$ samples, FDR and FNR are estimated. By averaging the number of detections and false positives, mFDR can be estimated.
	These steps are then repeated over the different scenarios.
	
\begin{figure}[!h]
		\centering
		\begin{subfigure}[b]{.45\linewidth} 
			\centering
			\includegraphics[width=0.7\linewidth]{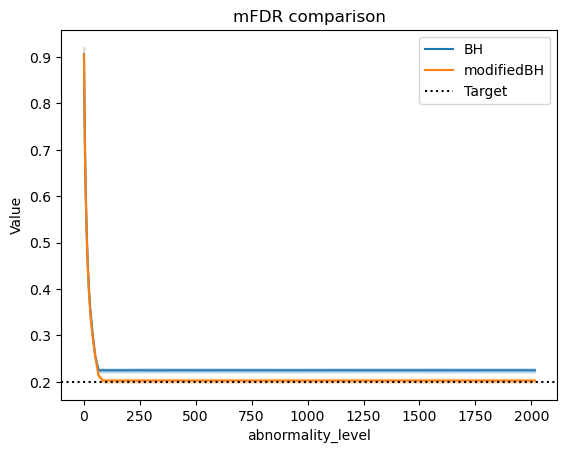}
			\caption[]{mFDR, $\alpha=0.2$, $\pi=0.07$}
			\label{fig:mfdr3}
		\end{subfigure}
		\begin{subfigure}[b]{.45\linewidth} %
			\centering
			\includegraphics[width=0.7\linewidth]{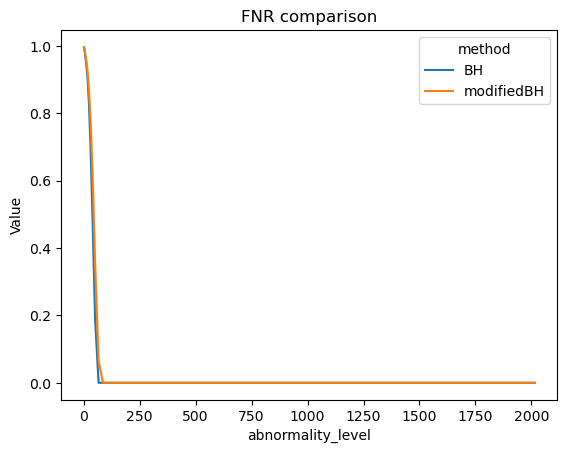}
			\caption[]{FNR, $\alpha=0.2$, $\pi=0.07$}
			\label{fig:fnr3}
		\end{subfigure}
		\begin{subfigure}[b]{.45\linewidth} 
			\centering
			\includegraphics[width=0.7\linewidth]{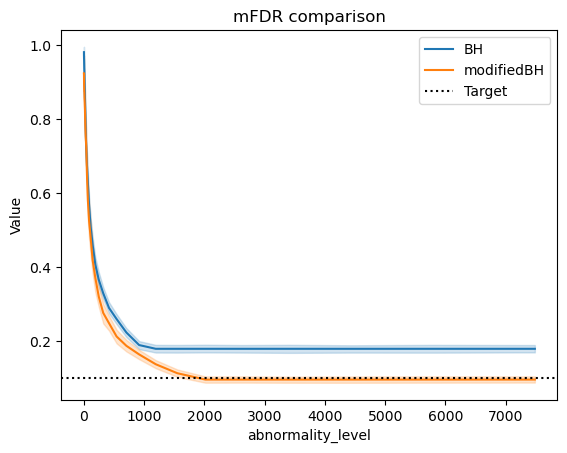}
			\caption[]{mFDR, $\alpha=0.1$, $\pi=0.01$}
			\label{fig:mfdr4}
		\end{subfigure}
		\begin{subfigure}[b]{.45\linewidth} %
			\centering
			\includegraphics[width=0.7\linewidth]{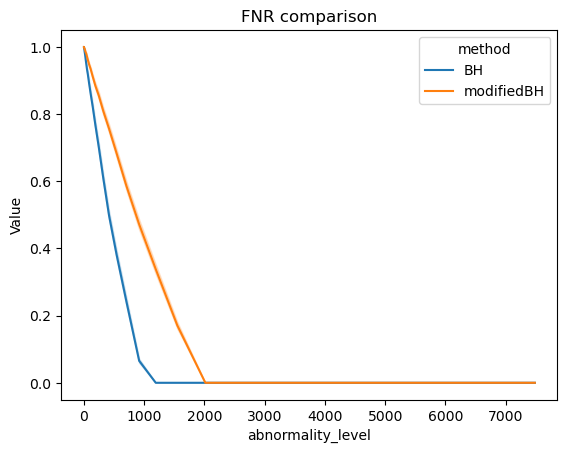}
			\caption[]{FNR, $\alpha=0.1$, $\pi=0.01$}
			\label{fig:fnr4}
		\end{subfigure}
		\caption[]{mFDR and FNR as a function of level of atypicity across different scenarios}
		\label{fig:mfdrcontr}
	\end{figure}

	\paragraph{Results and Analysis}
	TThe results are shown in Figure~\ref{fig:mfdrcontr} by varying $\delta$, $\alpha$, and $m_1$.
	In Figure~\ref{fig:mfdrcontr}, the level of atypicity $\delta$ is shown on the x-axis. The y-axis represents the estimated mFDR (in Figures~\ref{fig:mfdr3} or~\ref{fig:mfdr4}) or FNR (in Figures~\ref{fig:fnr3} or~\ref{fig:fnr4}). Different colors are used to distinguish BH and mBH procedures. 
	For a low level of atypicity $\delta$, FNR and mFDR are high because the anomalies are difficult to detect. By increasing $\delta$, FNR and mFDR decrease.
	As shown in Figure~\ref{fig:fnr3}, for values of $\delta$ around $100$, FNR is equal to $0$, which can also generate a constant mFDR as shown in Figure~\ref{fig:mfdr3}.
	For the mBH-procedure, mFDR is constant and equal to $\alpha$. This is consistent with Theorem~\ref{thm:nor}, which guarantees control at level $\alpha$ if all anomalies are detected.
	Figure~\ref{fig:fnr4} shows that all anomalies are detected for $\delta=2000$. The same result in Figure~\ref{fig:fnr3} with $\delta=100$. This is explained by the different parameters of the experiment. The easier the anomalies are detected, the faster $FNR=0$ is reached for a small $\delta$ and therefore the easier it is to guarantee $mFDR=\alpha$.
	
	\paragraph{Conclusion}
	In order to control mFDR at the desired level $\alpha$ using mBH, FNR has to be equal to 0. The capacity of mBH to control mFDR depends on the difficulty of the problem. When the proportion of anomalies and the level of atypicity are lower, the power of mBH decreases and mFDR is harder to control. The results of this experiment give an idea of the atypicity level that the detector can find.
		
	\subsection{Convergence of false discovery rate control}
	\label{sec:fdr.conv}
	This section studies the convergence rate of FDR over the full time series using $mBH_{\alpha}$.
	\paragraph{Experiment Description}
	The theoretical results obtained in Theorem~\ref{thm:adosw} only guarantee an asymptotic control of FDR on the whole time series. In practice, it is more useful to have a control of FDR at any time, i.e., on subseries of finite size. The question is studied empirically by observing the speed of convergence of FDR to the level $\alpha$.  FDR of the full time series is calculated across different scenarios as a function of time series size. In order to get the distribution of FDR, the experiment is repeated on $B=100$ time series. The maximal time series size explored is $T=10^4$.
	For each time series $b$, for each time step $t$, the data point is generated according to the following mixture:
	\begin{align*}
		A_{b,t} &\sim B(\pi)\\
		p_{b,t} &\sim \begin{cases}
			U([0,1]),& \text{if } A_{b,t}=1\\
			U([0,1/\delta]),              & \text{else}
		\end{cases}
	\end{align*}
	Then the data driven threshold is computed by applying the $\hat \varepsilon_{mBH_\alpha}$ to the subseries $\lbrace p_{t-m},\ldots,p_{t}\rbrace$. The points where $p_t$ is lower than the threshold are detected as anomalies. FDP is measured for each sub-series $\llbracket 0,t\rrbracket$ by comparing the detected anomalies with the true anomalies ($A_t=1$). This allows us to track the evolution of FDP while observing new data points.
	Different scenarios are generated by varying the proportion of anomalies $\pi$ and the atypicity level $\delta$.

\begin{figure}[!h]
		\centering
		\begin{subfigure}[b]{.45\linewidth} 
			\centering
			\includegraphics[width=0.95\linewidth]{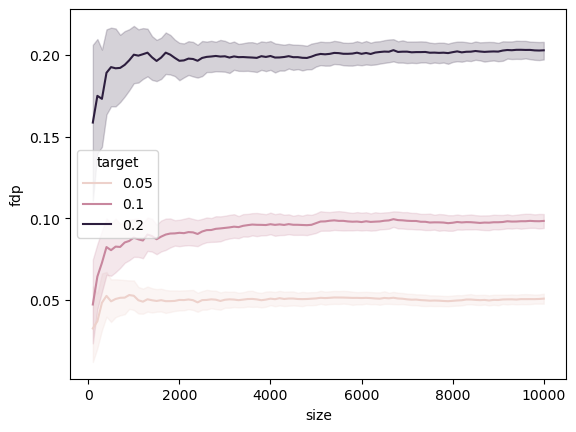}
			\caption[]{$\delta=1000$, $\pi=0.02$}
			\label{fig:convfdr1}
		\end{subfigure}
		\begin{subfigure}[b]{.45\linewidth} %
			\centering
			\includegraphics[width=0.95\linewidth]{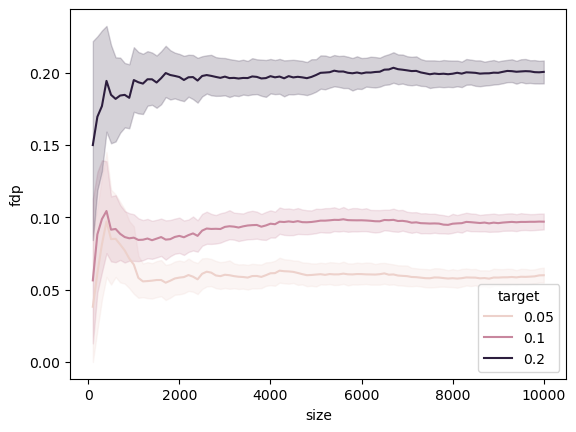}
			\caption[]{$\delta=1000$, $\pi=0.01$}
			\label{fig:convfdr2}
		\end{subfigure}
		\caption[]{FDR over the full time series as a function of the time series size.}
		\label{fig:convfdr}
	\end{figure}

	\paragraph{Results and analysis}
	In Figure~\ref{fig:convfdr}, the false discovery proportion is represented on the y-axis according to the size of the time series given on the x-axis. Different levels of $\alpha$, used to compute the $mBH$ threshold, are tested, with the results of the median FDP and its 95\% band shown in different colors. Different scenarios are represented by varying the proportion of anomalies between the subfigures.
	It can be observed that the convergence is quite fast from a size of 2000 data points, since for a $\alpha$ of 0.05, there is a 95\% chance to have a false positive rate between 0.04 and 0.06, on Figure~\ref{fig:convfdr}.  Thus, the control of the false positive rate can be ensured with a high probability for a series of one data point per minute recorded over a few days.
	
	\paragraph{Conclusion}
	This ensures that control at the desired level $\alpha$ is achieved not only for infinite time series, but also for finite time series, which allows our model to be used in practice.
	
	\subsection{Anomaly detection in presence of time series dependency}
	The theoretical results that are presented in this paper require the time series to be iid. In practice, time series contain dependency. To apply our anomaly detector preprocessing needs to be applied in order to decorrelate the data points. In this section three examples are presented: iid gaussian and non-iid AR(3) and ARCH(3). 
	\begin{itemize}
		\item iid: $\forall t \in \llbracket 1, T\rrbracket, A_t \sim B(\pi)$ and $
		X_t =
		\begin{cases}
			\mathcal N(0,1) & \text{if } A_t = 0 \\
			\Delta & \text{else}
		\end{cases}$
	    \item AR(3): $\forall t \in \llbracket 1, T\rrbracket, A_t \sim B(\pi), \quad
	    z_t =
	    \begin{cases}
	    	\mathcal N(0,1) & \text{if } A_t = 0 \\
	    	\Delta & \text{else}
	    \end{cases}$\\ and $X_t = \phi_1X_{t-1}+\phi_2X_{t-2}+\phi_3X_{t-3} + z_t$
	    \item ARCH(2): $\forall t \in \llbracket 1, T\rrbracket, A_t \sim B(\pi), \quad
	    z_t =
	    \begin{cases}
	    	\mathcal N(0,1) & \text{if } A_t = 0 \\
	    	\Delta & \text{else}
	    \end{cases}$,\\ $\sigma_t^2 = \theta_0 +  \theta_1X^2_{t-1}+\theta_2X^2_{t-2}$ and $X_t = \sigma_t z_t$
	\end{itemize} 
    
    For each of these three examples, 100 time series of 3000 data points are generated. The first half of the time series is used to learn the parameters of the statistical model. For example, in the case of AR(3), the AR(3) model is fitted and $\hat \phi_1$, $\hat \phi_2$ and $\hat \phi_3$ are estimated. This model is then used to obtain the uncorrelated time series: $\hat z_t = X_t - \hat \phi_1X_{t-1}+\hat \phi_2X_{t-2}+\hat\phi_3X_{t-3}$. The anomaly detector is applied to this extracted time series.  The aim of the experiment is to assess whether the estimation of the model is accurate enough to find the anomalies in the series of $\hat z_t$. When the anomaly detector is applied, the $p$-value of $\hat z_t$ is estimated by comparing its value with that of the $n$ points preceding $t$ with an estimated normal status. This estimation method is called the ``Sliding Calibration set''. In doing so, the estimation of the $p$-value may be biased by previous detection errors. An oracle version of our detector is also applied, where the $p$-value $\hat p_t$ is estimated by comparing the value $\hat z_t$ with that of the $n$ points preceding $t$ with a \emph{true} normal status. This estimation method is called ``Sliding Calibration set$-\star$''.
    
    The results are summarized in Table~\ref{table:ad-dependency}, each line represent a different dependency model, with the two $p$-value estimators. FDR and FNR are presented in columns.
    When the ``Sliding Cal.'' estimator is used, the proportion of anomalies is not controlled to the desired level. For example in the AR(3) case FDR is measured at 0.4 instead of 0.2. However, when the ``Sliding Cal.-$\star$'' estimator is used, FDR is controlled to the desired level in all the cases tested. This shows that the lack of control is due to the contamination of the calibration set and not to the dependence between observations
    This issue is discussed more deeply in Section~\ref{sec:res-analysis}.
    
    \begin{table}
    	\begin{subtable}{0.45\linewidth}
    		\centering
    		\begin{adjustbox}{max width=\textwidth}
    		\begin{tabular}{|l|c|c|c|}
    			\hline
    			model & $p$-value & FDR & FNR \\\hline
    			iid   & Sliding Cal.$-\star$   &0.19 &0.01 \\
    			iid   & Sliding Cal.   &0.41 & 0.00\\\hline
    			AR(3)    & Sliding Cal.$-\star$   &0.19 &0.01\\
    			AR(3)    & Sliding Cal.   &0.41&0.00\\\hline
    			ARCH(2)  & Sliding Cal.$-\star$   &0.20&0.01\\
    			ARCH(2)  & Sliding Cal.   &0.44&0.00\\
    			\hline
    		\end{tabular}
        	\end{adjustbox}
    	    \caption{$\alpha=0.2$}
    	\end{subtable}
        \begin{subtable}{0.45\linewidth}
        \begin{adjustbox}{max width=\textwidth}
        \begin{tabular}{|l|c|c|c|}
        	\hline
        	model & $p$-value & FDR & FNR \\\hline
        	iid   & Sliding Cal.$-\star$   & 0.12&0.01 \\
        	iid   & Sliding Cal.   &0.32 & 0.00\\\hline
        	AR(3)    & Sliding Cal.$-\star$   &0.11 &0.01\\
        	AR(3)    & Sliding Cal.   &0.33&0.00\\\hline
        	ARCH(2)  & Sliding Cal.$-\star$   &0.13&0.04\\
        	ARCH(2)  & Sliding Cal.   &0.31&0.00\\\hline
        \end{tabular}
         \end{adjustbox}
		    \caption{$\alpha=0.1$}
		\end{subtable}
       \caption{FDR and FNR results depending on different dependency models.}
       \label{table:ad-dependency}
    \end{table}

    This section shows that it is possible to apply our detector to non-independent time series. The difficulty lies in the possibility of whitening the data.
	
	\section{Comparison against competitor}
	\label{sec:empirical-simulation}
    Corollary~\ref{cor:mbh.contr.fdrg} enables us to conclude that the $mBH_{\alpha}$ procedure can be used to control FDR of the full time series. If the parameter $\alpha'$ of $mBH$ is computed using the heuristics provided in Appendix~\ref{sec.heuristic.arguments}, the assumptions \ref{assume.heuristic} and \ref{assume.power} must be verified. Even though these assumptions are hard to verify theoretically, the experiment in Section~\ref{sec:emp-diswin} shows that $mFDR_1^m$ is controlled for the tested scenario, provided that the anomalies are sufficiently atypical. The experiment in Section~\ref{sec:fdr.conv} shows that control of FDR is possible even when the time series is not infinite, as required by Theorem~\ref{thm:adosw}. These different results provide the conditions for building an anomaly detector that controls FDR of the time series by controlling mFDR on the subseries and the $p$-empirical value.
   Our anomaly detector is compared against Levels based On Recent Discovery (LORD), an online multiple testing procedure introduced in \cite{javanmard2018} to control FDR. 
   The anomaly detectors are evaluated under different scenarios by varying the generated anomalies and the targeted FDR.
   To understand the source of the difficulties that the anomaly detectors may encounter, different sequences of $p$-values with oracle information are introduced.

	\subsection{Data}
	The synthetic data are generated from a Gaussian distribution. With the use of the empirical $p$-value estimator, there is no need to evaluate on other data distributions. Only the anomaly proportion and the distribution shift associated with anomalies affect the performance of the anomaly detector.
Data are generated according to Definition~\ref{def:tsa} with Gaussian reference distribution and anomaly spike as in Section~\ref{sec:empbh-experiment-description}.
The strength of the distribution shift noted by $\Delta$ takes value in $\lbrace 3\sigma, 3.5\sigma, 4\sigma \rbrace$ and the anomaly proportion noted by $\pi$ is equal to $0.01$. Each generated time series contains $T=10^4$ data points. Each experiment is repeated over $100$ time series. For $t$ in $\llbracket 1,T\rrbracket$:
	\begin{align*}
		A_t &\sim B(\pi)\\
		X_t &=
		\begin{cases}
			\mathcal N(0,\sigma^2) & \text{if } A_t = 0 \\
			\Delta\sigma & \text{else}
		\end{cases}
	\end{align*}
	
	The value of $\Delta$ represents the atypicity score of the anomalies. Anomalies with higher $\Delta$ are easier to detect. In this experiment, the standard deviation $\sigma$ is set to 1.
	
	\subsection{Threshold and $p$-value estimators description}
	\paragraph{Our detector mBH on overlapping subseries}
	Using the $p$-value with the empirical estimator, the anomalies are detected by using mBH as the threshold estimator on overlapping subseries in the Algorithm~\ref{alg:fdrcontroad}.
	For each time $t$, the threshold is computed as: $ \hat \varepsilon_{mBH_{\alpha}, t} = f_m(\hat p_{t-m}, \ldots, \hat p_t)$, where $f_m$ is the mBH-procedure.
	To ensure FDR control according to Theorem~\ref{th:empFDR}, the cardinality of the calibration is set equal to $n = \frac{\winsize}{\alpha}-1$. In this experiment, $m$ is equal to $100$ and $\alpha$ takes values $0.1$ and $0.2$ depending on the tested scenario. So the calibration set takes values of $999$ or $1999$.
	\paragraph{LORD}
	LORD introduced in \cite{javanmard2018} is based on alpha-investing rules to define a threshold on $p$-values. For each time $t$, the threshold is computed according to the alpha-investing rules, depending on the previous decisions made by the algorithm. For more precision, see the original article \cite{javanmard2018}.
	The empirical $p$-value given in Eq.~\ref{eq:empv} does not respect the super uniformity property required by LORD to control FDR. The conformal $p$-value defined in Equation~\ref{eq:conf} does.
	\begin{align}
		\hat p_c(s_t, \mathcal S^{cal}_t)= \frac{1}{|\mathcal S_t^{cal}| + 1}\left(1+\sum_{s\in \mathcal S_t^{cal}}\mathbb{1}[s_t>s]\right)\label{eq:conf}
	\end{align}
	Using conformal $p$-values to apply the LORD algorithm leads to a weak power to detect anomalies. The issue is that $\hat p_c \geq \frac{1}{n+1}$ is always checked, and the threshold sequence $\hat\varepsilon_t$ decreases rapidly when no rejections are made. No anomaly can be detected. For these reasons, the empirical $p$ value is used when applying LORD and mBH. This experiment uses LORD3 from \cite{javanmard2018} with the same parameters as in the original paper.

	\paragraph{$p$-value estimation}\label{sec:practical-pvalue-estimators}
	\label{sec:pvalue-estimation}
	Different sequences of $p$-values are used to understand the limitations of our anomaly detector. 
	The true $p$-values are used to evaluate the case where the only limitation comes from the multiple testing procedure. One can thus understand how the estimation of $p$-values affects the detection of anomalies.
	One way to estimate $p$-values in practice is to use the same calibration set for all $p$-values. This is referred to as a fixed calibration set. However, the $p$-values may be biased in that particular calibration set. In practice, the usual way to estimate $p$-values is to use a sliding calibration set. To estimate the $p$-value of a data point $X_t$, $n$ preceding data points are used as a calibration set. To avoid bias in the estimation, the points detected as abnormal cannot be part of the calibration set and are replaced by other data points.
	However, the calibration set may be biased by undetected anomalies. To evaluate this effect, the sliding calibration set -$\star$ is introduced, where the knowledge oracle of the labels is used to construct the calibration set from the previous data points.
	
    The different $p$-value sequences are computed as follows:
	\begin{itemize}
		\item \textbf{Oracle}: The true $p$-value is used instead of the estimated one.
		$$\forall t \in \llbracket 1,T\rrbracket, \hat p_t = \Phi(X_t)  $$
		\item \textbf{Fixed calibration set (Fixed Cal.)}: The $p$-value is estimated using the same calibration set $\lbrace Z_i, i\in[1,n] \rbrace$ for all observations.
		$$\forall t \in \llbracket 1,T\rrbracket, \hat p_t = \frac{1}{n}\sum_{i=1}^n\mathbb{1}[Z_i>X_t] $$
		\item \textbf{Sliding Calibration set-$\star$ (Sliding Cal.-$\star$)}: The $p$-value is estimated using  a calibration that is a sliding windows containing the $n$ previous true normal data.
		$$ \forall t \in \llbracket 1,T\rrbracket, \hat p_t = \frac{1}{n}\sum_{i=1}^n\mathbb{1}[X_{h(t,i)}>X_t] $$
		With $h$ the function that selects observations that respect $\Hzero$. For each $t$ and $i$, $h(t,i)$ returns the $i$-th observation lower than $t$ that satisfies the $\Hzero$ hypothesis.
		\item \textbf{Sliding Calibration set (Sliding Cal.)}: The calibration set is a sliding windows containing the $n$ previous estimated normal data.
		$$ \forall t \in \llbracket 1,T\rrbracket, \hat p_t = \frac{1}{n}\sum_{i=1}^n\mathbb{1}[X_{\hat h(t,i)}>X_t] $$
		With $\hat h$ the function that estimates the function $h$. For each $t$ and $i$, $\hat h(t,i)$ gives the $i$-th observation lower than $t$ and that satisfies $d_{\hat h(t,i)}=0$.
	\end{itemize}

	\subsection{Performance metrics}
	\label{sec:performance-metric}
	The anomaly detectors are evaluated by their ability to control FDR and minimize FNR of the full time series. Therefore, the two applied metrics are FDP and FNP calculated as follows:
	$$FDP = \frac{\sum_{t=1}^T \mathbb{1}[\hat p_t<\hat\varepsilon_t](1-A_t)}{\sum_{t=1}^T\mathbb{1}[\hat p_t<\hat\varepsilon_t]}  $$
	and 
	$$FNP = \frac{\sum_{t=1}^T \mathbb{1}[\hat p_t>\hat\varepsilon_t]A_t}{\sum_{t=1}^T A_t}  $$
	where $\hat\varepsilon_t$ is estimated using mBH or LORD and $\hat p_t$ is estimated using one of the estimators defined in Section~\ref{sec:pvalue-estimation}.

	\subsection{Results}
	\label{sec:illustratino-results}
	The box plots shown in Figures~\ref{fig:fdr01}-\ref{fig:fnr02} represent FDP and FNP distributions for 1000 repetitions. Within each subfigure (a, b, c, d, e,...), the box plots are shown according to the multiple testing procedure mBH or LORD, the $p$-value estimator and the distribution shift between normal data and anomalies.

	Table~\ref{table:fnrfdr} gives a summary using FDR and FNR estimates. It enables an easy comparison between the values obtained with the different strategies combining the multiple testing procedure mBH or LORD, the choice of the level $\alpha$ varying from $0.1$ to $0.2$, the $p$-value estimator and the distribution shift between the normal data and the anomalies.

	\begin{table}[H]
		\footnotesize
		\begin{subtable}{.45\linewidth}
			\centering
			\begin{adjustbox}{max width=\textwidth}
			\begin{tabular}{|l|c|c|c|}
				\hline
				FDR, $\alpha=0.1$ & $\Delta=4\sigma$ & $\Delta=3.5\sigma$ & $\Delta=3\sigma$\\
				\hline
				mBH with Oracle PV & 0.101 & 0.113 & 0.281\\
				mBH with Fixed Cal. & 0.100 & 0.109 & 0.348\\
				mBH with Sliding Cal.-$\star$ & 0.100 & 0.113 & 0.256\\
				mBH with Sliding Cal. & 0.335 & 0.222 & 0.346\\\hline
				LORD with Oracle PV & 0.106 & 0.115 & 0.367\\
				LORD with Fixed Cal. & 0.111 & 0.277 & 0.736\\
				LORD with Sliding Cal.-$\star$ & 0.070 & 0.190 & 0.841\\
				LORD with Sliding Cal. & 0.075 & 0.098 & 0.627\\
				\hline
			\end{tabular}
		   \end{adjustbox}
			\subcaption{FDR, $\alpha=0.1$}
			\label{subtable:fdr1}
		\end{subtable}
		\begin{subtable}{.45\linewidth}
			\centering
			\begin{adjustbox}{max width=\textwidth}
			\begin{tabular}{|l|c|c|c|}
				\hline
				FNR, $\alpha=0.1$ &$\Delta=4\sigma$ & $\Delta=3.5\sigma$ & $\Delta=3\sigma$\\
				\hline
				mBH with Oracle PV&  0.020 & 0.151 & 0.793\\
				mBH with Fixed Cal. & 0.026 & 0.135 & 0.669\\
				mBH with Sliding Cal.-$\star$ & 0.019 & 0.140 & 0.669\\
				mBH with Sliding Cal. & 0.040 & 0.217 & 0.694\\\hline
				LORD with Oracle PV & 0.033 & 0.260 & 0.905\\
				LORD with Fixed Cal. & 0.070 & 0.340 & 0.896\\
				LORD with Sliding Cal.-$\star$ & 0.781 & 0.845 & 0.978\\
				LORD with Sliding Cal.& 0.052 & 0.327 & 0.907\\
				\hline
			\end{tabular}
		    \end{adjustbox}
			\subcaption{FNR, $\alpha=0.1$}
			\label{subtable:fnr1}
		\end{subtable}
		\begin{subtable}{.45\linewidth}
			\centering
			\begin{adjustbox}{max width=\textwidth}
			\begin{tabular}{|l|c|c|c|}
				\hline
				FDR, $\alpha=0.2$ &$\Delta=4\sigma$ & $\Delta=3.5\sigma$ & $\Delta=3\sigma$\\
				\hline
				mBH with Oracle PV& 0.200& 0.208& 0.277\\
				mBH with Fixed Cal.& 0.206& 0.211& 0.301\\
				mBH with Sliding Cal.-$\star$& 0.210& 0.219& 0.283\\
				mBH with Sliding Cal.& 0.833& 0.815& 0.761\\\hline
				LORD with Oracle PV& 0.211& 0.216& 0.290\\
				LORD with Fixed Cal.& 0.210& 0.263& 0.665\\
				LORD with Sliding Cal.-$\star$& 0.061& 0.149& 0.625\\
				LORD with Sliding Cal.& 0.117& 0.133& 0.321\\
				\hline
			\end{tabular}
		    \end{adjustbox}
			\subcaption{FDR, $\alpha=0.2$}
			\label{subtable:fdr2}
		\end{subtable}
		\begin{subtable}{.45\linewidth}
			\centering
			\begin{adjustbox}{max width=\textwidth}
			\begin{tabular}{|l|c|c|c|}
				\hline
				FNR, $\alpha=0.2$ &$\Delta=4\sigma$ & $\Delta=3.5\sigma$ & $\Delta=3\sigma$\\
				\hline
				mBH with Oracle PV& 0.009& 0.062& 0.395\\
				mBH with Fixed Cal.& 0.014& 0.045& 0.355\\
				mBH with Sliding Cal.-$\star$& 0.008& 0.059& 0.339\\
				mBH with Sliding Cal.& 0.003& 0.018& 0.101\\\hline
				LORD with Oracle PV& 0.016& 0.117& 0.610\\
				LORD with Fixed Cal.& 0.04& 0.144& 0.689\\
				LORD  with Sliding Cal.-$\star$& 0.805& 0.835& 0.941\\
				LORD with Sliding Cal.& 0.026& 0.168& 0.692\\
				\hline
			\end{tabular}
		    \end{adjustbox}
			\subcaption{FNR, $\alpha=0.2$}
			\label{subtable:fnr2}
		\end{subtable}
		\caption{Comparison of mBH versus LORD for online anomaly detection in Gaussian white noise with different abnormality levels. \label{table:fnrfdr}}
	\end{table}
	
	\subsection{Analysis}\label{sec:res-analysis}
	\paragraph{Effect of the strength of the distribution shift $\Delta$}
    According to the assumption \ref{assume.power} of the Theorem~\ref{thm:mbh-fdr}, $mBH_\alpha$ allows control of FDR at the level $\alpha$ if all anomalies are detected.
   To test this assertion, the different columns of the Table~\ref{subtable:fdr1}, are compared. In the row ``mBH with Oracle PV'', with $\Delta=4\sigma$ FDR is estimated at $0.101$, which is close to the desired level $\alpha=0.1$. While with $\Delta=3\sigma$ the FDR level is estimated at $0.281$ which is almost three times the desired level $\alpha$. The FNR results in Table~\ref{subtable:fnr1} need to be taken into consideration. When $\Delta=4\sigma$, FNR is close to 0, while when $\Delta=3\sigma$, FNR is $0.793$. Similar results are obtained with other test configurations in Table~\ref{subtable:fdr2} and Table~\ref{subtable:fnr2}. To control FDR at the desired level, FNR needs to be close to 0.
	
	\paragraph{Effect of $p$-value estimation}
	To understand how the $p$-value estimation can affect the control of FDR, the first four rows of Table~\ref{subtable:fdr1} are compared. In the column ``$4\sigma$'', the FDR values for the configurations ``Oracle PV'', ``Fixed Cal.'' and ``Sliding Cal.-$\star$'' are very close to the desired level $\alpha=0.1$. This control is enabled by Theorem~\ref{thm:mbh-fdr}, since the $p$-values verify all hypotheses, in particular all data in the calibration sets are generated according to the reference distribution.  
	However, in the case of ``Sliding Cal.'', FDR increases at a value of $0.335$. 
	For the same configurations, FNR remains low, between $0.019$ and $0.040$, as shown in Table~\ref{subtable:fnr1}. 
	The increase of FDR when using ``Sliding Cal.'' instead of ``Sliding Cal.-$\star$'' is a consequence of calibration set contamination.
	Indeed, according to the procedure used to build the calibration sets described in Section~\ref{sec:pvalue-estimation}, any detected anomalies are removed from the calibration sets used to estimate the next $p$-values. 
	If an observation is incorrectly detected as an anomaly, that data point cannot be part of the calibration set in future steps of online detection. Instead, it is replaced by another data point having a statistically lower atypicity score (false positives have high atypicity scores). As a result, the calibration set contains data points with lower scores than if it had been generated under $\mathcal P_0$. This leads to an underestimation of the $p$-values and an increase in the number of false positives. This illustrates the major drawback of mBH: it is highly sensitive to the lack of robustness of the $p$-value estimator.
	 Figure~\ref{fig:fdr01bh4} shows that using a fixed calibration instead of a sliding calibration-$\star$ gives a larger variance on FDP while FDR is the same. Using a single calibration set for the entire time series means that FDP is highly dependent on the start of the time series. By changing the calibration set at each time step, the statistical fluctuations in FDP are smoothed over the course of the time series analysis.
	
	\paragraph{Comparison with LORD}
	This section compares the results obtained with mBH and those obtained with LORD. As known from the literature, LORD controls FDR of super-uniform $p$-values. In this experiment, the question is the ability of the LORD method to control FDR of empirical $p$-values that have no theoretical guarantees. As shown in Figure~\ref{subtable:fdr1}, LORD is able to ensure control of FDR for all calibration set definitions when anomalies are easier to detect, such as for $\Delta=4\sigma$ or $\Delta=3.5\sigma$. In particular, unlike mBH, LORD is able to control FDR in the case of the sliding calibration set. 
	However, the mBH method has a lower FNR than the LORD method, as shown in \ref{subtable:fdr1} and \ref{subtable:fnr1}. For example, Table~\ref{subtable:fnr1} shows that FNR is equal to $0.019$ with mBH, while it is equal to $0.781$ with LORD, in the case of using the Sliding Calibration set-$\star$ on data with $\Delta=3\sigma$. 
	Nevertheless, in the Sliding Calibration Set case, the LORD method has approximately the same FNR, but with a lower FDR ($0.335$ versus $0.075$). The contamination problem of mBH offsets the superior performance observed in the Sliding Calibration Set-$\star$.
	
	\section{Conclusion}
	A new online anomaly detector has been described that aims at controlling FDR of the full time series at a prescribed level $\alpha$. The main challenges tackled by the new detector are:
	\begin{itemize}
		\item Empirical $p$-values: it clearly identifies and fulfills conditions on the calibration set cardinality (used for computing empirical p-values) ensuring the FDR control. By contrast in situations where empirical p-values are used with no care about these conditions, then FDR is not be controlled at a prescribed level.
		
		\item Online detection: it ensures the FDR control of the full time series (global control) by means of the local control of mFDR of sub-series, using a modified version of the BH-procedure. 
	\end{itemize}

	The theoretical guarantees developed in the present work are mainly established under the independence assumption, where the tightest control is made possible. 
	However some results on the FDR control are also derived under dependence assumptions such as the PRDS property. 
	The reported empirical experiments highlight the ability of the new procedure to control FDR of the full time series, while it also clearly emphasizes the sensitivity of the procedure with respect to its hyper-parameters or the features of the underlying probability distribution of the data.

	\section{Acknowledgments}
	The authors would like to thank Cristian Preda, director of the MODAL team at Inria, for valuable discussions.
	
	\bibliographystyle{imsart-number} 
	\bibliography{new_bibli.bib}

\begin{thebibliography}{31}

\bibitem{ahmed2016survey}
\begin{barticle}[author]
\bauthor{\bsnm{Ahmed},~\bfnm{Mohiuddin}\binits{M.}}, \bauthor{\bsnm{Mahmood},~\bfnm{Abdun~Naser}\binits{A.~N.}} \AND \bauthor{\bsnm{Hu},~\bfnm{Jiankun}\binits{J.}}
(\byear{2016}).
\btitle{A survey of network anomaly detection techniques}.
\bjournal{Journal of Network and Computer Applications}
\bvolume{60}
\bpages{19--31}.
\end{barticle}
\endbibitem

\bibitem{AngelopoulosGentleIntroductionConformal2022}
\begin{barticle}[author]
\bauthor{\bsnm{Angelopoulos},~\bfnm{Anastasios~N}\binits{A.~N.}} \AND \bauthor{\bsnm{Bates},~\bfnm{Stephen}\binits{S.}}
(\byear{2021}).
\btitle{A gentle introduction to conformal prediction and distribution-free uncertainty quantification}.
\bjournal{arXiv preprint arXiv:2107.07511}.
\end{barticle}
\endbibitem

\bibitem{aue2024state}
\begin{barticle}[author]
\bauthor{\bsnm{Aue},~\bfnm{Alexander}\binits{A.}} \AND \bauthor{\bsnm{Kirch},~\bfnm{Claudia}\binits{C.}}
(\byear{2024}).
\btitle{The state of cumulative sum sequential changepoint testing 70 years after Page}.
\bjournal{Biometrika}
\bvolume{111}
\bpages{367--391}.
\end{barticle}
\endbibitem

\bibitem{bates2023testing}
\begin{barticle}[author]
\bauthor{\bsnm{Bates},~\bfnm{Stephen}\binits{S.}}, \bauthor{\bsnm{Cand{\`e}s},~\bfnm{Emmanuel}\binits{E.}}, \bauthor{\bsnm{Lei},~\bfnm{Lihua}\binits{L.}}, \bauthor{\bsnm{Romano},~\bfnm{Yaniv}\binits{Y.}} \AND \bauthor{\bsnm{Sesia},~\bfnm{Matteo}\binits{M.}}
(\byear{2023}).
\btitle{Testing for outliers with conformal p-values}.
\bjournal{The Annals of Statistics}
\bvolume{51}
\bpages{149--178}.
\end{barticle}
\endbibitem

\bibitem{benjamini1995}
\begin{barticle}[author]
\bauthor{\bsnm{Benjamini},~\bfnm{Yoav}\binits{Y.}} \AND \bauthor{\bsnm{Hochberg},~\bfnm{Yosef}\binits{Y.}}
(\byear{1995}).
\btitle{Controlling the false discovery rate: a practical and powerful approach to multiple testing}.
\bjournal{Journal of the Royal statistical society: series B (Methodological)}
\bvolume{57}
\bpages{289--300}.
\end{barticle}
\endbibitem

\bibitem{Benjaminicontrolfalsediscovery2001}
\begin{barticle}[author]
\bauthor{\bsnm{Benjamini},~\bfnm{Yoav}\binits{Y.}} \AND \bauthor{\bsnm{Yekutieli},~\bfnm{Daniel}\binits{D.}}
(\byear{2001}).
\btitle{The control of the false discovery rate in multiple testing under dependency}.
\bjournal{Annals of statistics}
\bpages{1165--1188}.
\end{barticle}
\endbibitem

\bibitem{blazquez-garciaReviewOutlierAnomaly2020}
\begin{barticle}[author]
\bauthor{\bsnm{Bl{\'a}zquez-Garc{\'\i}a},~\bfnm{Ane}\binits{A.}}, \bauthor{\bsnm{Conde},~\bfnm{Angel}\binits{A.}}, \bauthor{\bsnm{Mori},~\bfnm{Usue}\binits{U.}} \AND \bauthor{\bsnm{Lozano},~\bfnm{Jose~A}\binits{J.~A.}}
(\byear{2021}).
\btitle{A review on outlier/anomaly detection in time series data}.
\bjournal{ACM Computing Surveys (CSUR)}
\bvolume{54}
\bpages{1--33}.
\end{barticle}
\endbibitem

\bibitem{blum2010alarms}
\begin{barticle}[author]
\bauthor{\bsnm{Blum},~\bfnm{James~M}\binits{J.~M.}} \AND \bauthor{\bsnm{Tremper},~\bfnm{Kevin~K}\binits{K.~K.}}
(\byear{2010}).
\btitle{Alarms in the intensive care unit: too much of a good thing is dangerous: is it time to add some intelligence to alarms?}
\bjournal{Critical care medicine}
\bvolume{38}
\bpages{702--703}.
\end{barticle}
\endbibitem

\bibitem{braei2020anomaly}
\begin{barticle}[author]
\bauthor{\bsnm{Braei},~\bfnm{Mohammad}\binits{M.}} \AND \bauthor{\bsnm{Wagner},~\bfnm{Sebastian}\binits{S.}}
(\byear{2020}).
\btitle{Anomaly detection in univariate time-series: A survey on the state-of-the-art}.
\bjournal{arXiv preprint arXiv:2004.00433}.
\end{barticle}
\endbibitem

\bibitem{chandolaAnomalyDetectionSurvey2009}
\begin{barticle}[author]
\bauthor{\bsnm{Chandola},~\bfnm{Varun}\binits{V.}}, \bauthor{\bsnm{Banerjee},~\bfnm{Arindam}\binits{A.}} \AND \bauthor{\bsnm{Kumar},~\bfnm{Vipin}\binits{V.}}
(\byear{2009}).
\btitle{Anomaly detection: A survey}.
\bjournal{ACM computing surveys (CSUR)}
\bvolume{41}
\bpages{1--58}.
\end{barticle}
\endbibitem

\bibitem{cvach2012monitor}
\begin{barticle}[author]
\bauthor{\bsnm{Cvach},~\bfnm{Maria}\binits{M.}}
(\byear{2012}).
\btitle{Monitor alarm fatigue: an integrative review}.
\bjournal{Biomedical instrumentation \& technology}
\bvolume{46}
\bpages{268--277}.
\end{barticle}
\endbibitem

\bibitem{fisher1951design}
\begin{barticle}[author]
\bauthor{\bsnm{Fisher},~\bfnm{RA}\binits{R.}}
(\byear{1951}).
\btitle{The Design of Experiments, volume 6th Ed}.
\bjournal{Hafner, New York, NY}.
\end{barticle}
\endbibitem

\bibitem{foster2008alpha}
\begin{barticle}[author]
\bauthor{\bsnm{Foster},~\bfnm{Dean~P}\binits{D.~P.}} \AND \bauthor{\bsnm{Stine},~\bfnm{Robert~A}\binits{R.~A.}}
(\byear{2008}).
\btitle{$\alpha$-investing: a procedure for sequential control of expected false discoveries}.
\bjournal{Journal of the Royal Statistical Society Series B: Statistical Methodology}
\bvolume{70}
\bpages{429--444}.
\end{barticle}
\endbibitem

\bibitem{gretton2012kernel}
\begin{barticle}[author]
\bauthor{\bsnm{Gretton},~\bfnm{Arthur}\binits{A.}}, \bauthor{\bsnm{Borgwardt},~\bfnm{Karsten~M}\binits{K.~M.}}, \bauthor{\bsnm{Rasch},~\bfnm{Malte~J}\binits{M.~J.}}, \bauthor{\bsnm{Sch{\"o}lkopf},~\bfnm{Bernhard}\binits{B.}} \AND \bauthor{\bsnm{Smola},~\bfnm{Alexander}\binits{A.}}
(\byear{2012}).
\btitle{A kernel two-sample test}.
\bjournal{The Journal of Machine Learning Research}
\bvolume{13}
\bpages{723--773}.
\end{barticle}
\endbibitem

\bibitem{huber1992robust}
\begin{bincollection}[author]
\bauthor{\bsnm{Huber},~\bfnm{Peter~J}\binits{P.~J.}}
(\byear{1992}).
\btitle{Robust estimation of a location parameter}.
In \bbooktitle{Breakthroughs in statistics: Methodology and distribution}
\bpages{492--518}.
\bpublisher{Springer}.
\end{bincollection}
\endbibitem

\bibitem{javanmard2018}
\begin{barticle}[author]
\bauthor{\bsnm{Javanmard},~\bfnm{Adel}\binits{A.}} \AND \bauthor{\bsnm{Montanari},~\bfnm{Andrea}\binits{A.}}
(\byear{2018}).
\btitle{Online rules for control of false discovery rate and false discovery exceedance}.
\bjournal{The Annals of statistics}
\bvolume{46}
\bpages{526--554}.
\end{barticle}
\endbibitem

\bibitem{kirch2022sequential}
\begin{barticle}[author]
\bauthor{\bsnm{Kirch},~\bfnm{Claudia}\binits{C.}} \AND \bauthor{\bsnm{Stoehr},~\bfnm{Christina}\binits{C.}}
(\byear{2022}).
\btitle{Sequential change point tests based on U-statistics}.
\bjournal{Scandinavian Journal of Statistics}
\bvolume{49}
\bpages{1184--1214}.
\end{barticle}
\endbibitem

\bibitem{LaxhammarConformalanomalydetection2014}
\begin{bphdthesis}[author]
\bauthor{\bsnm{Laxhammar},~\bfnm{Rikard}\binits{R.}}
(\byear{2014}).
\btitle{Conformal anomaly detection: Detecting abnormal trajectories in surveillance applications},
\btype{PhD thesis},
\bpublisher{University of Sk{\"o}vde}.
\end{bphdthesis}
\endbibitem

\bibitem{lewandowska2023determining}
\begin{barticle}[author]
\bauthor{\bsnm{Lewandowska},~\bfnm{Katarzyna}\binits{K.}}, \bauthor{\bsnm{M{\k{e}}drzycka-D{\k{a}}browska},~\bfnm{Wioletta}\binits{W.}}, \bauthor{\bsnm{Tomaszek},~\bfnm{Lucyna}\binits{L.}} \AND \bauthor{\bsnm{Wujtewicz},~\bfnm{Magdalena}\binits{M.}}
(\byear{2023}).
\btitle{Determining Factors of Alarm Fatigue among Nurses in Intensive Care Units—A Polish Pilot Study}.
\bjournal{Journal of Clinical Medicine}
\bvolume{12}
\bpages{3120}.
\end{barticle}
\endbibitem

\bibitem{li2022intelligent}
\begin{barticle}[author]
\bauthor{\bsnm{Li},~\bfnm{Yichen}\binits{Y.}}, \bauthor{\bsnm{Zhang},~\bfnm{Xu}\binits{X.}}, \bauthor{\bsnm{He},~\bfnm{Shilin}\binits{S.}}, \bauthor{\bsnm{Chen},~\bfnm{Zhuangbin}\binits{Z.}}, \bauthor{\bsnm{Kang},~\bfnm{Yu}\binits{Y.}}, \bauthor{\bsnm{Liu},~\bfnm{Jinyang}\binits{J.}}, \bauthor{\bsnm{Li},~\bfnm{Liqun}\binits{L.}}, \bauthor{\bsnm{Dang},~\bfnm{Yingnong}\binits{Y.}}, \bauthor{\bsnm{Gao},~\bfnm{Feng}\binits{F.}}, \bauthor{\bsnm{Xu},~\bfnm{Zhangwei}\binits{Z.}} \betal{et~al.}
(\byear{2022}).
\btitle{An Intelligent Framework for Timely, Accurate, and Comprehensive Cloud Incident Detection}.
\bjournal{ACM SIGOPS Operating Systems Review}
\bvolume{56}
\bpages{1--7}.
\end{barticle}
\endbibitem

\bibitem{MarandonMachinelearningmeets2022}
\begin{barticle}[author]
\bauthor{\bsnm{Marandon},~\bfnm{Ariane}\binits{A.}}, \bauthor{\bsnm{Lei},~\bfnm{Lihua}\binits{L.}}, \bauthor{\bsnm{Mary},~\bfnm{David}\binits{D.}} \AND \bauthor{\bsnm{Roquain},~\bfnm{Etienne}\binits{E.}}
(\byear{2022}).
\btitle{Machine learning meets false discovery rate}.
\bjournal{arXiv preprint arXiv:2208.06685}.
\end{barticle}
\endbibitem

\bibitem{page1954continuous}
\begin{barticle}[author]
\bauthor{\bsnm{Page},~\bfnm{Ewan~S}\binits{E.~S.}}
(\byear{1954}).
\btitle{Continuous inspection schemes}.
\bjournal{Biometrika}
\bvolume{41}
\bpages{100--115}.
\end{barticle}
\endbibitem

\bibitem{phipson2010permutation}
\begin{barticle}[author]
\bauthor{\bsnm{Phipson},~\bfnm{Belinda}\binits{B.}} \AND \bauthor{\bsnm{Smyth},~\bfnm{Gordon~K}\binits{G.~K.}}
(\byear{2010}).
\btitle{Permutation P-values should never be zero: calculating exact P-values when permutations are randomly drawn}.
\bjournal{Statistical applications in genetics and molecular biology}
\bvolume{9}.
\end{barticle}
\endbibitem

\bibitem{ramdas2018}
\begin{binproceedings}[author]
\bauthor{\bsnm{Ramdas},~\bfnm{Aaditya}\binits{A.}}, \bauthor{\bsnm{Zrnic},~\bfnm{Tijana}\binits{T.}}, \bauthor{\bsnm{Wainwright},~\bfnm{Martin}\binits{M.}} \AND \bauthor{\bsnm{Jordan},~\bfnm{Michael}\binits{M.}}
(\byear{2018}).
\btitle{SAFFRON: an adaptive algorithm for online control of the false discovery rate}.
In \bbooktitle{International conference on machine learning}
\bpages{4286--4294}.
\bpublisher{PMLR}.
\end{binproceedings}
\endbibitem

\bibitem{ruff2021unifying}
\begin{barticle}[author]
\bauthor{\bsnm{Ruff},~\bfnm{Lukas}\binits{L.}}, \bauthor{\bsnm{Kauffmann},~\bfnm{Jacob~R}\binits{J.~R.}}, \bauthor{\bsnm{Vandermeulen},~\bfnm{Robert~A}\binits{R.~A.}}, \bauthor{\bsnm{Montavon},~\bfnm{Gr{\'e}goire}\binits{G.}}, \bauthor{\bsnm{Samek},~\bfnm{Wojciech}\binits{W.}}, \bauthor{\bsnm{Kloft},~\bfnm{Marius}\binits{M.}}, \bauthor{\bsnm{Dietterich},~\bfnm{Thomas~G}\binits{T.~G.}} \AND \bauthor{\bsnm{M{\"u}ller},~\bfnm{Klaus-Robert}\binits{K.-R.}}
(\byear{2021}).
\btitle{A unifying review of deep and shallow anomaly detection}.
\bjournal{Proceedings of the IEEE}
\bvolume{109}
\bpages{756--795}.
\end{barticle}
\endbibitem

\bibitem{solet2012managing}
\begin{barticle}[author]
\bauthor{\bsnm{Solet},~\bfnm{Jo~M}\binits{J.~M.}} \AND \bauthor{\bsnm{Barach},~\bfnm{Paul~R}\binits{P.~R.}}
(\byear{2012}).
\btitle{Managing alarm fatigue in cardiac care}.
\bjournal{Progress in Pediatric Cardiology}
\bvolume{33}
\bpages{85--90}.
\end{barticle}
\endbibitem

\bibitem{staerman2022functional}
\begin{bphdthesis}[author]
\bauthor{\bsnm{Staerman},~\bfnm{Guillaume}\binits{G.}}
(\byear{2022}).
\btitle{Functional anomaly detection and robust estimation},
\btype{PhD thesis},
\bpublisher{Institut polytechnique de Paris}.
\end{bphdthesis}
\endbibitem

\bibitem{WangOnlineFDRControlled2019}
\begin{binproceedings}[author]
\bauthor{\bsnm{Wang},~\bfnm{Weinan}\binits{W.}}, \bauthor{\bsnm{Liu},~\bfnm{Zhengyi}\binits{Z.}}, \bauthor{\bsnm{Shi},~\bfnm{Xiaolin}\binits{X.}} \AND \bauthor{\bsnm{Pierce},~\bfnm{Lucas}\binits{L.}}
(\byear{2019}).
\btitle{Online fdr controlled anomaly detection for streaming time series}.
In \bbooktitle{5th Workshop on Mining and Learning from Time Series (MiLeTS)}.
\end{binproceedings}
\endbibitem

\bibitem{wu2021current}
\begin{barticle}[author]
\bauthor{\bsnm{Wu},~\bfnm{Renjie}\binits{R.}} \AND \bauthor{\bsnm{Keogh},~\bfnm{Eamonn~J}\binits{E.~J.}}
(\byear{2021}).
\btitle{Current time series anomaly detection benchmarks are flawed and are creating the illusion of progress}.
\bjournal{IEEE transactions on knowledge and data engineering}
\bvolume{35}
\bpages{2421--2429}.
\end{barticle}
\endbibitem

\bibitem{wu2007inference}
\begin{bbook}[author]
\bauthor{\bsnm{Wu},~\bfnm{Yanhong}\binits{Y.}}
(\byear{2007}).
\btitle{Inference for change point and post change means after a CUSUM test}
\bvolume{180}.
\bpublisher{Springer Science \& Business Media}.
\end{bbook}
\endbibitem

\bibitem{xu2022dynamic}
\begin{binproceedings}[author]
\bauthor{\bsnm{Xu},~\bfnm{Ziyu}\binits{Z.}} \AND \bauthor{\bsnm{Ramdas},~\bfnm{Aaditya}\binits{A.}}
(\byear{2022}).
\btitle{Dynamic algorithms for online multiple testing}.
In \bbooktitle{Mathematical and Scientific Machine Learning}
\bpages{955--986}.
\bpublisher{PMLR}.
\end{binproceedings}
\endbibitem

\end{thebibliography}
	\appendix
	\section{More precision on the method}
    \subsection{Evaluating the ratio of rejection numbers}\label{sec.evaluate.alphaprim}
    
    \subsubsection{Using heuristic arguments}\label{sec.heuristic.arguments}
    The Section~\ref{sec.mFDR.BH} raises the importance of the ratio $ \mathbb E R_{1,\alpha}^{*,m} / \mathbb{E} R_{1,\alpha}^m$ of rejection numbers. 
    The present section aims at deriving a numeric approximation to this ratio. In a first step, a first result details the value of the denominator. In a second step, an approximation to the numerator is derived based on a heuristic argument and also empirically justified on simulation experiments.

    When $mFDR$ is assumed to equal $\alpha$, the expected number of rejections can be made explicit.
    \begin{proposition}\label{thm:nor}
    	With the previous notation, let $(X_1,\ldots, X_m)$ be given by Definition~\ref{def:tsa}, where $\pi$ denotes the unknown proportion of anomalies, and assume that $mFDR_1^m=\alpha$ and $FNR_1^m=\beta\in[0,1]$.
    	Then		
    	\begin{align}\label{eq.expected.number.rejections}
    		\mathbb E [R_{1,\alpha}^m] = \frac{\winsize\pi(1-\beta)}{1-\alpha} .
    	\end{align}
    \end{proposition}
    The proof is postponed to Appendix~\ref{sec:proof-th-nor}.
    For instance, Eq.~\eqref{eq.expected.number.rejections} establishes that the expected number of rejection output by $BH_\alpha$ increases with $\pi$, the unknown proportion of anomalies along the signal. This makes sens since the more anomalies, the more  expected rejections. 
    The expected number of rejection is also increasing with $\alpha$: the larger $\alpha$, the less restrictive the threshold, and the more rejections should be made.
    However the number of rejection decreases with the FNR value $\beta$. As $\beta$ increases, the proportion of false negatives grows meaning that fewer alarms are raised, which results in a smaller number of rejections.

    In what follows, the assumption is made that anomalies are easy to detect, meaning that the $FNR_1^m$ value $\beta$ is negligible compared to 1. In this context, Proposition~\ref{thm:nor} would yield that
    \begin{align}\label{assume.power}
    	\mathbb E [R_{1,\alpha}^m] \approx \frac{\winsize\pi}{1-\alpha}. \tag{{\bf Power}}
    \end{align}

    An another assumption is also made about the relationship between $\mathbb E[ R_{1,\alpha}^m]$ and $\mathbb E[R_{1,\alpha}^{*,m}]$.
    This assumption is based on a heuristic argument supported by the results of numerical experiments as reported in Table~\ref{table:heuristic}.
    In what follows, it is assumed that
    \begin{align}\label{assume.heuristic}
    	\mathbb E[R_{1,\alpha}^{*,m}] = \mathbb E[ R_{1,\alpha}^m]+1 \tag{{\bf Heuristic}}.
    \end{align}
    No mathematical proof of this statement is given in the present paper. However, Table~\ref{table:heuristic} displays numerical values which empirically support this approximation, whereas further analyzing the connection between these quantities should be necessary. 
    
    \begin{table}[h!]
    	\begin{tabular}{|c|c|c|c|}
    		\hline
    		$BH_\alpha$& $0.05$  & $0.1$ &$0.2$  \\
    		\hline
    		$\mathbb E[R_{1,\alpha}^m]$& 2.14 & 2.32 & 2.78 \\
    		\hline
    		$\mathbb E [R_{1,\alpha}^m(i)]$& 3.18 & 3.44 &  3.99 \\
    		\hline
    	\end{tabular}
    	\caption{Numerical evaluations for different values of $\alpha$ ($10^3$  repetitions)\label{table:heuristic}}
    \end{table}
    Let us emphasize that Table~\ref{table:heuristic} has been obtained with Gaussian data (generated similarly to those detailed in Section~\ref{subsec:emprslt}). 
    For all the three considered values of $\alpha$, one observes that $\mathbb E[R^*]$ remains close to (but also slightly larger than) $\mathbb E[R] + 1$.

   These two assumptions give rise to a strategy for computing the ratio $\frac{\mathbb E[R^*]}{\mathbb E[R]}$. So all ingredient to build a procedure that control mFDR are given. 
    
    \begin{align}
    	\frac{\mathbb E[R^*]}{\mathbb E[R]} &= 1 + \frac{1-\alpha}{m\pi}\\
    	\alpha' &= \alpha \left(1+\frac{1-\alpha}{m\pi}\right)^{-1}\\
    \end{align}
     The value $\alpha'$ can be used with the $mBH$ procedure.
     Under \eqref{assume.heuristic} and \eqref{assume.power}, according to Corollary~\ref{cor:mbh.contr.fdrg}, $mBH$ allows global control of FDR at the level $\alpha$.

    \subsubsection{Estimation on a training set}\label{sec:estimation.training}
    The true proportion of anomalies is usually not known, and estimating the proportion of anomalies is error-prone. Assumptions are also difficult to ensure. The number of detections is known by the user.
    To estimate $\mathbb{E}[R_{\tilde\alpha}]$, the procedure $BH_{\tilde\alpha}$ is applied to each of the subseries of length $m$ from the training set. The average number of detections is computed and noted as $\hat\mu_{R_{\tilde\alpha}}$. However, it is not possible to compute $R_{\tilde\alpha'}^*$, the number of detections after one $p$-value associated with normal data is set to $0$ since the true labels are unknown. But since the proportion of true anomalies $\pi$ is close to $0$, $\mathbb{E}[R_{\tilde\alpha'}^*]$ can be approximate by $\mathbb{E}[R_{\tilde\alpha'}^{**}]$, the number of detections after one $p$-values (possible abnormal) is set to 0.
    Then, to estimate $\mathbb{E}[R_{\tilde\alpha'}^{**}]$, the procedure $BH_{\tilde\alpha}$ is applied again to the same subseries but after a randomly chosen $p$-value is replaced by $0$. Here again, the average number of detections is computed and noted $\hat\mu_{R^{**}_{\tilde\alpha}}$.
    By varying $\tilde\alpha$ it is possible to estimate:
    \begin{align}
    \alpha' = \argmax_{\tilde\alpha}\left\lbrace\frac{\hat\mu_{R^{**}_{\tilde\alpha}}}{\hat\mu_{R_{\tilde\alpha}}}\tilde\alpha\leq\alpha\right\rbrace
    \end{align}
     This estimator of $\alpha'$ can be used for the modified BH procedure. 
     According to Corollary~\ref{cor:mbh.contr.fdrg}, the global control of FDR is ensured by $\hat\varepsilon_{BH_{\alpha'}}$ when the size of the training set goes to infinity.	
	\section{Proofs}
	\subsection{Proof of Theorem \ref{thm:fdr-bh-emp}}\label{sec:proof-thm-empfdr}
		\begin{proof}[Proof of Theorem \ref{thm:fdr-bh-emp}]
		The proof of BH applied to true $p$-values is reproduced (see Section~1.1 of the Supplementary Materials), but in the case of empirical $p$-values. The only modification is that
		$\mathbb P(\hat p_i \leq \frac{\alpha k}{\winsize})$ is not equal to $\frac{\alpha k}{\winsize}$ since $\hat p_i$ now follows the discrete uniform distribution
		$$\mathbb P(\hat p_i \leq \frac{\alpha k}{\winsize})= \sum_{\ell=0}^{\lfloor n \alpha k/\winsize \rfloor}\mathbb P(n\hat p_i = \ell) = \frac{\lfloor\frac{\alpha kn}{\winsize}\rfloor+1}{n+1}.$$
		Plugging this in the FDR expression, it gives
		$$FDR_1^m(\hat \varepsilon_{{BH}_{\alpha}}, \hat p) = \winsize_0\sum_{k=1}^\winsize \frac{\frac{\lfloor \frac{\alpha k n}{\winsize}\rfloor+1}{n+1}}{k}\mathbb{P}(R(i)=k).$$
		Recall that $\hat p_i$ follows $U(0,1/n,2/n,\ldots,1)$ entails that $n\hat p_i$ follows $U(0,1,2,\ldots,n)$.
	\end{proof}
	\subsection{PRDS property for $p$-values having overlapping calibration set}
	\label{sec:proof-prds}
	The following construction is used to describe a family of $p$-values with overlapping calibration set.
	Let $Z$ the vector that combine all calibration set, the $Z_i$ are i.i.d. with marginal probability $\mathcal P_0$. The set of the $n$ indices defining the elements of the calibration set related to $\hat p_i$ in $Z$ is noted $\mathcal{D}_i$. The calibration related to $X_1$ is noted $Z_{\mathcal D_1}=(Z_{i_1},\ldots,Z_{i_n})$.
	For all $i$ in $\llbracket 1, m \rrbracket$: $\hat p_i = p\mbox{-value}(X_i, Z_{\mathcal{D}_i})$.
	
	To proof that $p$-values with overlapping calibration set are PRDS as described in Definition~\ref{def:prds}, the methodology used in \cite{bates2023testing} to be extended in the case of overlapping calibration set.
%
%
	For $i$ in $\llbracket 1, m \rrbracket$ the calibration set associated to $X_i$ is noted $Z_{\mathcal D_i}$. The law of total probabilities gives:
	
	\begin{align*}
		\mathbb{P}\left[ \hat p_1^m \in A|\hat p_i=u\right] &= \int \mathbb{P}\left[ \hat p_1^m \in A|\hat p_i=u| Z_{\mathcal D_i}=z\right]\mathbb{P}\left[ Z_{\mathcal D_i}=z\right|]dz\\
		&=\mathbb{E}_{Z_{\mathcal D_i}|\hat p_i=u} \mathbb{P}\left[ \hat p_1^m \in A|\hat p_i=u| Z_{\mathcal D_i}=z\right]
	\end{align*}

   If these two lemma are suppose to be true, the PRDS property is verified.  
    
    \begin{lemma}\label{lemma:pz}
    	For non-decreasing set $A$ and vectors $z$, $z'$ such that $z \succeq z'$, then
    	\begin{equation}
    		\mathbb{P}\left[ \hat p_1^m \in A| Z_{\mathcal D_i}=z\right] \geq  \mathbb{P}\left[ \hat p_1^m \in A| Z_{\mathcal D_i}=z'\right]
    	\end{equation}
    \end{lemma}

    \begin{lemma}\label{lemma:z1z2}
    	For $u \geq u'$, if $i$ belongs to the set of inliers, the exists $Z_{{\mathcal D_i},1} \sim Z_{\mathcal D_i}|\hat p_i=u$ and $Z_{{\mathcal D_i},2}\sim Z_{\mathcal D_i}|\hat p_i=u'$ such that $\mathbb{P}[Z_{{\mathcal D_i},1} ]  \succeq\mathbb{P}[Z_{{\mathcal D_i},2}] $
    \end{lemma}

    Indeed, take $i\in \llbracket 1, m\rrbracket$ and $u \geq u'$ and define $Z_{{\mathcal D_i},1}$ and $Z_{{\mathcal D_i},2}$ as in the statement of Lemma \ref{lemma:z1z2}. 
    
    \begin{align*}
    	\mathbb{P}[\hat p_1^m \in A|p_i=u] &= \mathbb{E}_{Z_{{\mathcal D_i},1}}[\mathbb{P}[\hat p_1^m \in A|Z_{\mathcal D_i}=Z_{{\mathcal D_i},1}]] \text{   (Lemma \ref{lemma:z1z2})}\\
    	&\geq \mathbb{E}_{Z_{{\mathcal D_i},2}}[\mathbb{P}[\hat p_1^m \in A|Z_{\mathcal D_i}=Z_{{\mathcal D_i},2}]]\text{   (Lemma \ref{lemma:pz})}\\
    	&\geq\mathbb{P}[\hat p_1^m \in A|p_i=u']\text{   (Lemma \ref{lemma:z1z2})} 
    \end{align*}
     It shows that, when $u\geq u'$ then $\mathbb{P}[\hat p_1^m \in A|p_i=u] \geq \mathbb{P}[\hat p_1^m \in A|p_i=u']$, which means $\mathbb{P}[\hat p_1^m \in A|p_i=u]$ is increasing in $u$. The PRDS property is satisfied.
     To complete the proof, the introduced lemmas are proven.
    
    \begin{proof}[Proof of Lemma \ref{lemma:pz}] 
    	
    	Let be $i$ in  $\llbracket 1, m\rrbracket$ and  vectors $z$, $z'$ and $\overline z$ vectors such that $z  \succeq z'$. The vectors $z$, $z'$ are used to define the calibration set related to the $p$-values $\hat p_i$ and $\overline z$ is used to define elements of calibrations sets that are not in the calibration set of $\hat p_i$.
    	By conditioning on the calibration sets defined by $(z, \overline z)$ and $(z', \overline z)$ it gives:
    	\begin{align}\label{eq:prds.decomp}
    		\mathbb{P}\left[ \hat p_1^m \in A| Z_{\mathcal D_i}=z, Z_{\overline{\mathcal D_i}}=\overline{z}\right] \geq \mathbb{P}\left[ \hat p_1^m \in A| Z_{\mathcal D_i}=z', Z_{\overline{\mathcal D_i}}=\overline{z}\right]
    	\end{align}
    
    This result comes from the decomposition the following decomposition, for all $j$ in $\llbracket 1, m\rrbracket$
    \begin{align*}
    	\hat p_j &= \frac{1}{n}\sum_{k \in \mathcal{D}_j}\mathbb{1}[a(Z_{k}) \geq a(X_j)]\\
    	&=  \frac{1}{n}\left(\sum_{k \in \mathcal{D}_j \cap  \mathcal{D}_i}\mathbb{1}[a(Z_{k}) \geq a(X_j)] + \sum_{k \in \mathcal{D}_j\setminus \mathcal{D}_i}\mathbb{1}[a(Z_{k}) \geq a(X_j)]\right)
    \end{align*}
    The conclusion comes from $Z_{\mathcal{D}_i} \succeq Z'_{\mathcal{D}_i}$ which implies $Z_{\mathcal{D}_i\cap \mathcal{D}_j} \succeq Z'_{\mathcal{D}_i\cap \mathcal{D}_j}$.
    
    Since $Z_{\mathcal{D}_j\setminus \mathcal{D}_i} \perp Z_{\mathcal{D}_i}$, Eq.~\ref{eq:prds.decomp} can be integrated over $Z_{\overline{\mathcal D_i}}$ to give:

	\begin{align}
		\mathbb{P}\left[ \hat p_1^m \in A| Z_{\mathcal D_i}=z\right] \geq \mathbb{P}\left[ \hat p_1^m \in A| Z_{\mathcal D_i}=z'\right]
	\end{align}
    \end{proof}

    \begin{proof}[Proof of Lemma \ref{lemma:z1z2}]
    	Let $S'_{i,(1)}\leq S_{i,(2)} \leq \ldots \leq S_{i,(n)}$ the order statistics of $(a(Z_{{\mathcal D_i},1}),\ldots, a(Z_{{\mathcal D_i},n}))$.
    	Let $S'_{i,(1)}\leq S'_{i,(2)}\leq\ldots \leq S'_{i,(n+1)}$ the order statistics of $(a(Z_{{\mathcal D_i},1}),\ldots, a(Z_{{\mathcal D_i},n}), a(X_i))$. And $R_i$ the rank of $a(X_i)$ among these.
    	\begin{equation}
    		\left\lbrace (S_{(1)}, \ldots, S_{(n)}) | R_i = k, S'_{i, (1)},\ldots, S'_{i, (n+1)}  \right\rbrace = (S'_{(1)}, \ldots, S'_{(k-1)}, S'_{(k+1)}, \ldots, S'_{(n+1)})
    	\end{equation}
    Using that $R_i$ is independent of $S'_{i, (1)},\ldots, S'_{i, (n+1)}$:
    
        \begin{equation}
        	\left\lbrace (S_{(1)}, \ldots, S_{(n)}) | R_i = k  \right\rbrace = (S'_{(1)}, \ldots, S'_{(k-1)}, S'_{(k+1)}, \ldots, S'_{(n+1)})
        \end{equation}
    The right-hand side is not increasing with $k$ and $\hat p_i = \frac{R_i - 1}{n}$
    \end{proof}

\subsection{Proof of Corollary~\ref{th:empFDR}}\label{proof:empFDR}
	\begin{proof}[Proof of Corollary~\ref{th:empFDR}]
		To get a deeper understanding of the FDR expression obtained in Theorem~\ref{thm:fdr-bh-emp}, $q_{n, k}$ the fractional part of $\frac{\alpha kn}{\winsize}$ is introduced:
	\begin{align*}
		q_{n, k} = \frac{\alpha kn}{\winsize} - \left\lfloor \frac{\alpha k n}{\winsize}\right\rfloor 
	\end{align*}
    When plugged into the FDR expression, it gives:
    \begin{align}
		&FDR =  \winsize_0\sum_{k=1}^\winsize \frac{\frac{ \frac{\alpha k n}{\winsize} +1 - q_{n, k}}{n+1}}{k}\mathbb{P}(R(1)=k)\nonumber\\
		&FDR =  \frac{\winsize_0\alpha}{\winsize}\frac{n}{n+1} + \frac{\winsize_0}{n+1}\sum_{k=1}^\winsize  \frac{1 - q_{n,k}}{k}\mathbb{P}(R(1)=k)\label{formul1}
	\end{align}
	
%
	

	In order to get lower and upper bounds of FDR, the value of $q_{n,k}$ should be expressed as a function of $\alpha$, $k$, $n$ and $m$.
	
	For the next part of the proof, it is useful to express the relation between $q_{n,k}$ and $q_{n+1,k}$. It gives the effect of increasing the cardinality of the calibration by one. Using the definition of the fractional part:
	 \begin{align*}
	 	q_{n+1,k} - q_{n,k} &= \frac{\alpha k(n+1)}{\winsize} - \left\lfloor \frac{\alpha k (n+1)}{\winsize}\right\rfloor - \frac{\alpha kn}{\winsize} + \left\lfloor \frac{\alpha k n}{\winsize}\right\rfloor\\
	 	q_{n+1,k} - q_{n,k} &= \frac{\alpha k}{\winsize} - \left\lfloor \frac{\alpha k (n+1)}{\winsize}\right\rfloor  + \left\lfloor \frac{\alpha k n}{\winsize}\right\rfloor\\
	 \end{align*}
	

     Which can be expressed as a congruence relation:
     \begin{align}
		q_{n+1,k} - q_{n,k} \equiv \frac{\alpha k}{\winsize} \pmod 1\label{eq:cong}
	\end{align}

    Two cases are studied:
    \begin{enumerate}
    	\item Particular case: there exists an integer $1\leq \nu$ such that $\frac{\nu m}{\alpha}$ is an integer. the notation $n_\nu=\frac{\nu m}{\alpha}$ is introduced.
    	Since: $\frac{\alpha kn_\nu}{\winsize} =  \frac{\alpha k\nu m/\alpha}{\winsize} = k\nu$ is an integer, then the fractional part is null:
    	$$q_{n_\nu, k} = 0 $$
    	If the calibration set cardinality $n$ is equal to $n = n_\nu-1=\frac{\nu m}{\alpha} - 1 $. Then, the congruence relation in Eq.~\ref{eq:cong} gives:
    	\begin{align*}
    		q_{n_\nu -1,k} \equiv q_{n_l,k} - \alpha k/\winsize \pmod 1\\
    		q_{n_\nu -1,k} \equiv 0 - \alpha k/\winsize \pmod 1\\
    	\end{align*}
    	
    	Using the fact that fractional part of a number belongs to  $[0,1[$, the only possible value to $q_{n_\nu -1,k}$ is:
    	$$q_{n_\nu -1,k} = 1 - \alpha k/\winsize $$
    	Plugging the value of $q_{n_\nu -1,k}$ into Eq.~\ref{formul1}, it gives:
    	\begin{align*}
    		FDR &= \frac{\winsize_0 \alpha n}{\winsize(n+1)} + \frac{\winsize_0}{n+1}\sum_{k=1}^\winsize  \frac{\alpha k}{k \winsize}\mathbb{P}(R(1)=k)
    	\end{align*}
        Simplifying by $k$ and using that $\sum_{k=1}^m\mathbb{P}(R(i)=k)=1$, the  result is obtained:
        \begin{align*}
    		FDR &= \frac{\winsize_0 \alpha n}{\winsize(n+1)} + \frac{\winsize_0\alpha}{(n+1)\winsize}\\
    		FDR &= \frac{\winsize_0 \alpha }{\winsize}
    	\end{align*}
        \item General case: With $\alpha \in ]0,1]$, for each $\nu$ the notation $n_\nu=\left\lceil\frac{\nu m}{\alpha}\right\rceil$ is introduced. Notice that this definition is consistent with the particular case.
        The ceiling function definition gives:
        \begin{align*}
        	&\left\lceil \frac{\nu m}{\alpha}\right\rceil - 1 < \frac{\nu m}{\alpha} \leq \left\lceil \frac{\nu m}{\alpha}\right\rceil\\
        \end{align*}
    Multiplying by $\alpha k$ on each side and the $n_\nu$ notation: 
        \begin{align*}
        	&	\frac{\alpha k (n_\nu-1)}{m} < k\nu \leq \frac{\alpha k (n_\nu)}{m} \\
        \end{align*}
        
        It implies that $\lfloor \frac{\alpha k (n_{\nu}-1)}{m}\rfloor < \lfloor \frac{\alpha k (n_{\nu})}{m}\rfloor$. Also, Eq.~\ref{eq:cong} is expressed as $q_{n_\nu,k} - q_{n_\nu-1,k} \equiv \frac{\alpha k}{m} \pmod{1}$: 
        \begin{align}
        	 1-\frac{\alpha k}{m} \leq q_{n_\nu-1, k} < 1\label{eq:qn.upper.lower}
        \end{align}
        Indeed, the fractional part of a number as to be larger than $1-\alpha k/m$ so that adding $\alpha k/m$ increase the integer part.
        
        By plugin the bounds of $q_{n_\nu-1, k}$ into Eq.~\ref{formul1}, it can gives the bounds of FDR. At first, to compute the upper bound of  FDR the lower bound of $q_{n_\nu-1, k}$ is used:
        \begin{align*}
        	 FDR \leq \frac{m_0 (n_{\nu}-1)\alpha}{m n_\nu} + \frac{\winsize_0}{n+1}\sum_{k=1}^\winsize  \frac{\alpha k}{k \winsize}\mathbb{P}(R(1)=k)\\
        \end{align*}
         With the same calculations as for the ``Particular case'', it gives:
         \begin{align*}
         	FDR \leq \frac{m_0 \alpha}{m }
         \end{align*}

         Similarly, the lower bound of FDR can be obtained using the $q_{n,k}$  upper bound from Eq.~\ref{eq:qn.upper.lower} plugged into Eq.~\ref{formul1}:
        \begin{align*}
        	&\frac{m_0 (n_{\nu}-1)\alpha}{m n_\nu} + \frac{\winsize_0}{n+1}\sum_{k=1}^\winsize \frac{(1-1)}{k }\mathbb{P}(R_1=k) < FDR \\
        	& \frac{m_0 (n_{\nu}-1)\alpha}{m n_\nu} < FDR \\
        \end{align*} 
    \end{enumerate}

\end{proof}

\subsection{Proof of Theorem \ref{thm:adosw}}\label{sec:proof-thm-adosw}
\begin{proof}[Proof of Theorem \ref{thm:adosw}]
	Let us start with the FDP expression for a time series of length $T$.
	\begin{align*}
		FDP_{t=1}^T & = \frac{\sum_{t=1}^T\mathbb{1}[\hat p_t < \hat\varepsilon_{o,t}](1-A_t)}{\sum_{t=1}^T \mathbb{1}[\hat p_t < \hat\varepsilon_{o,t}]}\\
		& = \frac{\frac{1}{T}\sum_{t=1}^T \mathbb{1}[\hat p_t <f_m(\hat{\mathbf{P}}_t)](1-A_t)}{\frac{1}{T}\sum_{t=1}^T \mathbb{1}[\hat p_t <f_m(\hat{\mathbf{P}}_t)]},
	\end{align*}
	The decision process $(\mathbb{1}[\hat p_t < \hat\varepsilon_{o,t}])_t$ and the false positives process $(\mathbb{1}[\hat p_t < \hat\varepsilon_{o,t}]A_t)_t$ are not independent, therefore it is not possible to use the Law of Large Numbers directly. 
	The alternative strategy consists first in splitting the numerator and denominator into several disjoint subseries corresponding to independent and identically distributed processes.
	Then partitioning the times series of length $T=T^\prime (n+m)$ into $T^\prime$ subseries, each of length $n+m$, it results that 
	\begin{align}
		& \frac{1}{T}\sum_{t=1}^T \mathbb{1}[\hat p_t <f_m(\hat{\mathbf{P}}_{t})](1-A_t) \notag \\
		&= \frac{1}{n+m}\sum_{k=1}^{n+m}\left(\frac{1}{T'}\sum_{t=0}^{T'-1} \mathbb{1}[\hat p_{t(n+m)+k} <f_m(\hat{\mathbf{P}}_{t(n+m)+k})](1-A_{t(n+m)+k})\right). \label{lign:cut-ts}
	\end{align}
	Interestingly for each $k$ from 1 to $m+n$, the summands within the brackets do all belong to different subseries, which makes the sum over $t$ a sum of independent and identically distributed random variables.
	%
	%
	It results that, for each $1\leq k \leq n+m$, the average within the brackets is converging to its expectation by the LLN theorem. 
	
	Since the limit of a (finite) sum is equal to the sum of the limits, the average in Eq.~\eqref{lign:cut-ts} is converging and
	\begin{align}
		\lim_{T \rightarrow \infty} \frac{1}{T}\sum_{t=1}^{T} \mathbb{1}[\hat p_t < f_m(\hat{\mathbf{P}}_t)](1-A_{t}) 	& = \sum_{k=1}^m\mathbb E \left[\mathbb{1}[\hat p_k < f_m(\hat{\mathbf{P}}_k)](1-A_k) \right]\mbox{   a.s.}\\
		& = m\mathbb E \left[\mathbb{1}[\hat p_m < f_m(\hat{\mathbf{P}}_m)](1-A_m) \right]\mbox{   a.s.}.\label{eq:conv-fp}
	\end{align}

    Furthermore,
    \begin{align*}
    	FP_1^m(\hat \varepsilon_o, \hat{\mathbf{P}}_m) = \sum_{k=1}^m \mathbb{1}[\hat{\mathbf{P}}_{m,k}<f_m(\hat{\mathbf{P}}_m)](1-A_k) 
    \end{align*}
    Using the exchangeability of $\hat{\mathbf{P}}_m$ and the permutation invariance of $f_m$ is comes:
     \begin{align*}
    	\mathbb{E}FP_1^m(\hat \varepsilon_o, \hat{\mathbf{P}}_1) = m \times \mathbb{E}[ \mathbb{1}[\hat{\mathbf{P}}_{m,m}<f_m(\hat{\mathbf{P}}_m)](1-A_m) ]
    \end{align*}
     Finally with $\hat{\mathbf{P}}_{m,m}=\hat p_m$ and Eq.~\ref{eq:conv-fp} it gives:
     \begin{align*}
     \lim_{T \rightarrow \infty} \frac{1}{T}\sum_{t=1}^{T} \mathbb{1}[\hat p_t < f_m(\hat{\mathbf{P}}_m)](1-A_{t}) = \mathbb{E}FP_1^m(\hat \varepsilon_o, \hat{\mathbf{P}}_m) 
     \end{align*}

	Then after applying the same reasoning on the denominator, it gives:
	$$FDP_1^\infty(\hat \varepsilon_{o}, \hat p) = \frac{\mathbb{E}FP_1^m(\hat \varepsilon_o, \hat{\mathbf{P}}_m) }{\mathbb{E}R_1^m(\hat \varepsilon_o, \hat{\mathbf{P}}_m) } $$

\end{proof}

\subsection{Proof of Proposition \ref{thm:mfr-bh}}\label{sec:proof-thm-mfr-bh}

\begin{lemma}\label{lemma:proba-fp}
	Let $(p_i)_{1\leq i \leq m}$ be a sequence of $m$ $p$-values with $m_0$ true null hypothesis.
	Suppose $i$ belong to the set of true negative $\mathcal H_0$, and $D_i$ the random variable equal to 1 if $i$ is detected by the $BH_\alpha$ procedure.
	\begin{itemize}
		\item If the $p$-values are independent and verify that: $\forall k \in \llbracket 1, m \rrbracket, \mathbb{P}(p_i\leq \frac{\alpha k}{m})$, then:
		\begin{align}
			\mathbb{P}(D_i) = \frac{\alpha \mathbb{E}[R^*]}{m}
		\end{align}
	    Where $R$ is the number of hypotheses rejected by $BH_\alpha$ and $R^*$ is the number of hypotheses rejected after $p_i$ is set to $0$.
	    \item If the $p$-values are empirical $p$-values using a unique calibration set with cardinality $\nu \frac{m}{\alpha}-1$, then:
	    \begin{align}
	    	\mathbb{P}(D_i) = \frac{\alpha \mathbb{E}[\tilde R^*]}{m}
	    \end{align}
        Where $\tilde R^*$ is the number of rejected hypothesis by applying $BH_\alpha$ after on $(p_j')_{1 \leq j \leq m}$, where $p'_i=0$, and for $j\neq i, p'_j=p_j - \frac{1}{n}\mathbb{1}[p_j<p_i]$.
	\end{itemize}
\end{lemma}

\begin{proof}[Proof of Lemma~\ref{lemma:proba-fp}]
	\textbf{Proof of the first statement:}\\
	Using, the random variable $R$ representing the number of rejection of $BH_\alpha$, $i$ is rejected if $p_i$ is below the threshold $\frac{\alpha R}{m}$.
	\begin{align}
		\mathbb{P}[D_i] = \mathbb{E}[\mathbb{1}[p_i \leq \frac{\alpha R}{m} ]]
	\end{align}
    
    Let $(p_j')_{1\leq j\leq m}$ be defined by $p_i'=0$ and $p'_j=p_j$. The conditions of Lemma D6 from \cite{MarandonMachinelearningmeets2022} are satisfied, it follows:
    \begin{align}
    	\mathbb{P}[D_i] =   \mathbb{E}[\mathbb{1}[p_i \leq \frac{\alpha R(i)}{m} ]]
    \end{align}
     Where $R(i)$ is the number detection when applying $BH_\alpha$ on $(p_j')_{1\leq j\leq m}$.
     
     According to the law of total expectation:
     \begin{align}
     	\mathbb{P}[D_i] &=  \mathbb{E}[\mathbb{E}[\mathbb{1}[p_i \leq \frac{\alpha R(i)}{m} ]|p\backslash p_i]].
     \end{align}
     Since $R(i)$ is measurable is $p\backslash p_i$ and $p_i$ is independent from $p\backslash p_i$, it gives:
     \begin{align}
     	\mathbb{P}[D_i] &= \mathbb{E}[\mathbb P_{p_i}(p_i\leq \frac{\alpha R(i)}{m})]
     \end{align}
    By hypothesis $\mathbb P_{p_i}(p_i\leq \frac{\alpha R(i)}{m}) = \frac{\alpha R(i)}{m}$, then:
    \begin{align}
    	\mathbb{P}[D_i] &= \frac{\alpha}{m}\mathbb{E}[R(i)]
    \end{align}
    which conclude the proof of the first statement. 
    
    \textbf{Proof of the second statement:}\\
	Let $W_i$, $C_{i,j}$ defined as:
	\begin{align}
		W_i = ( \lbrace s_1, ...., s_n, s_{n+i}\rbrace, (s_i, i \in \mathcal{H}_0, i\neq j),(s_i, i \in \mathcal{H}_1))
	\end{align}

    \begin{align}
    	C_{i,j} = \frac{1}{n}\left(\sum_{s \in \lbrace s_1, \ldots, s_n, s_{n+i}} \mathbb{1}[s>s_{n+j}] - 1\right)
    \end{align}
    \begin{enumerate}
    	\item $p_j = C_{i,j} + \frac{1}{n}\mathbb{1}[s_{n+j}>s_{n+i}]$
    	\item $p_i$ independent of $W_i$
    	\item $p_i$ follow uniform distribution in $\lbrace 0, \frac{1}{n}, \ldots, 1\rbrace$
    \end{enumerate}

    Lemma D.6 from \cite{MarandonMachinelearningmeets2022} is applied. Let $(p_j')_{1 \leq j \leq m}$ be defined by $p'_i=0$ and for $j\neq i$, $p'_j=C_{i,j}$. For all $j$, $p_j'\leq p_j$ and if $p_j>p_i$ then $p_j'=p_j$. Thus, the conditions of the Lemma D.6 from \cite{MarandonMachinelearningmeets2022} are verified, which gives:
    \begin{align}
    	\mathbb{1}[p_i \leq \frac{\alpha R}{m}] &= \mathbb{1}[p_i \leq \frac{\alpha \tilde R(i)}{m}]
    \end{align}

	\begin{align}
		\mathbb{P}[D_i] &= \mathbb{E}[\mathbb{E}[p_i \leq \frac{\alpha \tilde R(i)}{m}| W_i]]
	\end{align}
Since, $\tilde R(i)$ in measurable in $W_i$ and $p_i$ is independent of $W_i$.
 
 \begin{align}
 	\mathbb{P}[D_i] &= \mathbb{E}[\mathbb{P}_{p_i}(p_i\leq \frac{\alpha \tilde R(i)}{m})]
 \end{align}
By hypothesis, $p_i$ is a empirical $p$-value with calibration set verified that there exist an integer $\nu$ such that $n=\nu\frac{m}{\alpha}-1$. So according to Corollary~\ref{th:empFDR}, $\mathbb{P}_{p_i}(p_i\leq \frac{\alpha \tilde R(i)}{m}) = \frac{\alpha \tilde R(i)}{m}$.

Finally, the second statement is verified with:
 \begin{align}
	\mathbb{P}[D_i] &= \frac{\alpha}{m}\mathbb{E}[\tilde R(i)]
\end{align}

\end{proof}

\begin{lemma}\label{lemma:negcov}
	Let $X$ and $Y$ be two random variables. Suppose that $k \mapsto \mathbb{E}[X|Y=k]$ is decreasing, then:
	\begin{align}
		\mathbb{E}[XY] \leq \mathbb{E}[X]\mathbb{E}[Y].
	\end{align}
\end{lemma}
\begin{proof}[Proof of Lemma~\ref{lemma:negcov}]
	Let $Z$ be a random variable that follows the same law than $Y$ but is independent.
	Since $k \mapsto \mathbb{E}[X|Y=k]$ is decreasing:
	\begin{align}
		(Y - Z)(\mathbb{E}[X|Y] - \mathbb{E}[X|Z]) &\leq 0\\
		\mathbb E\left[(Y - Z)(\mathbb{E}[X|Y] - \mathbb{E}[X|Z])\right] &\leq 0\\
	\end{align}
    By distributing the product and using that $Y$ and $Z$ follow the same law, this gives:
    \begin{align}
    	2\mathbb E[Y\mathbb{E}[X|Y]] - 2\mathbb E[Y\mathbb{E}[X|Z]]\leq 0\\
    \end{align}
    Finally using $\mathbb E[Y\mathbb{E}[X|Y]] = \mathbb E[XY]$ and independence of $Y$ and $Z$:
    \begin{align}
    	\mathbb E[XY]\leq \mathbb E[X]\mathbb E[Y]\\
    \end{align}
\end{proof}

	\begin{proof}[Proof of Proposition \ref{thm:mfr-bh}]
	The mFDR formula is given by
	$$mFDR_1^m(p) = \frac{\mathbb{E}[FP_{1,\alpha}^m(p)]}{\mathbb{E}[R_{1,\alpha}^m(p)]}. $$
	
	Let us compute the numerator $\mathbb{E}[FP_1^m(p)]$ value after applying the $BH_\alpha$. 
	Keep in mind that here the family of true null hypotheses is random, generated by $(A_i)_{1 \leq i \leq m}$. In order to meet the conditions of Lemma~\ref{lemma:proba-fp}, it is possible to condition with respect to $\mathcal H_0$.
	Using $D_i= \mathbb{1}[p_i\leq \frac{R}{m}\alpha]$, it appears that
	\begin{align*}
		FP_1^m(p) & = \sum_{i \in \mathcal H_0} D_i.
	\end{align*}
	Lemma~\ref{lemma:proba-fp} allows to calculate its conditional expectation:
	\begin{align*}
		\mathbb  E[FP_{1,\alpha}^m|\mathcal{H}_0]
		   & = \sum_{i\in \mathcal{H}_0} \frac{\alpha}{m}\mathbb{E}[R(i)|\mathcal{H}_0]\\
		& =  \frac{\alpha m_0}{m}\mathbb{E}[R^*|\mathcal{H}_0] . 
	\end{align*}
With $R^*$ the number of rejection where one $p$-values is set to 0.
Integrating with respect to $\mathcal H_0$. Then using the fact that $m_0=|\mathcal H_0|$ is measurable with respect to $\mathcal H_0$.
\begin{align*}
	\mathbb  E[FP_{1,\alpha}^m] &=  \mathbb E\left[\frac{\alpha m_0}{m}\mathbb{E}[R^*|\mathcal{H}_0]\right]\\
		&=   \alpha  \mathbb E\left[\frac{m_0}{m}R^*\right]\\
\end{align*}

Finally, if $\mathbb E[R^*| m_0]$ is decreasing, Lemma~\ref{lemma:negcov} gives,
\begin{align*}
	 \mathbb  E[FP_{1,\alpha}^m] \leq \alpha (1-\pi) \mathbb{E}[R^*] 
\end{align*}
with $1-\pi = \mathbb E\frac{m_0}{m}$, the proportion of data generated by the reference distribution.

\end{proof}

\subsection{Proof of Proposition~\ref{thm:nor}}\label{sec:proof-th-nor}
  \begin{proof}[Proof of Proposition \ref{thm:nor}]    
  	By definition $mFDR_1^m=\frac{\mathbb{E}[FP_1^m]}{\EE R_1^m}$, and $R_1^m = FP_1^m + TP_1^m$. With hypothesis the $mFDR$ is equal to $\alpha$, this gives:
  	\begin{align*}
  		\alpha &= \frac{\mathbb{E}[FP_1^m]}{\EE R_1^m}\\
  		\alpha &= \frac{\mathbb{E}[FP_1^m]}{\EE [FP_1^m + TP_1^m]}\\
  		\alpha (\mathbb E[FP_1^m] + \mathbb E[TP_1^m]) &= \mathbb E[FP_1^m] \\
  		(\alpha - 1) \mathbb E[FP_1^m] &= - \alpha \mathbb E[TP_1^m] \\
  		\mathbb E[FP_1^m] &= \frac{\alpha}{1-\alpha}\mathbb E[TP_1^m]\\
  	\end{align*}
  	Then, the expectation of true positives is expressed using the proportion of false negatives $\beta$, the proportion of anomaly $\pi$ in the $m$ observations, $A_i$ the random variable equal to $1$ if the observation $X_i$ is an anomaly and $D_i$ the random variable equal to $1$ if the observation $X_i$ is detected as anomaly :
  	\begin{align*}
  		\mathbb{E}[TP_1^m] 
  		&= \sum_{i=1}^m \mathbb{P}[A_i=1\mbox{ and } D_i=1]\\
  		&= \sum_{i=1}^m \mathbb{P}[A_i=1]\mathbb P[D_i=1|A_i=1]\\
  		&=  m\pi (1-\beta)\\
  	\end{align*}
  	
  	Therefore, the $\EE [FP_1^m]$ can be expressed as:
  	\begin{equation*}
  		\mathbb E[FP_1^m] = \frac{\alpha  m\pi (1-\beta)}{1-\alpha}
  	\end{equation*}
  	
  	So the $\EE [R_1^m]$ is expressed as follows:
  	\begin{align*}
  		\mathbb E[R_1^m] &= \frac{\alpha  m\pi (1-\beta)}{1-\alpha} +  m\pi (1-\beta)\\
  		&=\frac{m\pi (1-\beta)}{1-\alpha}
  	\end{align*}
  \end{proof}

\subsection{Proof of Corollary~\ref{cor:mbh.contr.fdrg}}\label{sec:proof-cor-mbh.contr.fdrg}
\begin{proof}[Proof of Corollary~\ref{cor:mbh.contr.fdrg}]
	All conditions being satisfied Theorem~\ref{thm:adosw} gives that:
	\begin{align}\label{eq:fdrmfdr}
		FDR_1^\infty(\hat \varepsilon_{BH_{\alpha'}}, \hat p) = mFDR_1^m(\hat \varepsilon_{BH_{\alpha'}}, \hat{\textbf{P}}_m)
	\end{align}
	According to Proposition~\ref{thm:mfr-bh}, if one of the 3 statements is true:
	\begin{align*}
		 mFDR_1^{m}(\hat \varepsilon_{BH_{\alpha'}}, \hat p) \leq (1-\pi)\alpha'\frac{\mathbb{E}[R^*_{\alpha'}]}{\mathbb{E}[R_{\alpha'}]}
	\end{align*}

	By hypothesis $\alpha'\frac{\mathbb{E}[R^*_{\alpha'}]}{\mathbb{E}[R_{\alpha'}]}=\alpha$ which allows to conclude.
	\begin{align*}
		mFDR_1^{m}(\hat \varepsilon_{BH_{\alpha'}}, \hat p) \leq (1-\pi)\alpha
	\end{align*}
\end{proof}

\subsection{Proof of Theorem~\ref{thm:mbh-fdr}}\label{sec:proof-thm-mbh-fdr}
\begin{proof}[Proof of Theorem~\ref{thm:mbh-fdr}]
	The Corollary~\ref{cor:mbh.contr.fdrg} gives the two properties that the $p$-values families has to verify to control FDR of the time series:
	\begin{itemize}
		\item The $\hat{\textbf{P}}_t$ are identically distributed and independent when time distance is larger than $n+m$.
		\item For each  $t$, $\hat{\textbf{P}}_t$ are either true $p$-values, or empirical $p$-values with independent or unique calibration set.
	\end{itemize}
	In the following, these properties are verified for the different $p$-values.
	\begin{enumerate}
		\item The sequence of true $p$-value $\mathbb{P}_{X\sim\mathcal P_0}(a(X)>a(X_t))$ is i.i.d., because the time series mixture is i.i.d. Then $\hat{\textbf{P}}_t = (\hat p_{t-m+1}, \ldots, p_{t})$ are independent for a time distance larger than $m$.
		 Using the first statement of Corollary~\ref{cor:mbh.contr.fdrg} FDR of the whole time series is controlled at level $(1-\pi)\alpha$.
		\item The sequence of empirical $p$-value is i.i.d., because the time series mixture i and the calibration sets are i.i.d. Then $\hat{\textbf{P}}_t = (\hat p_{t-m+1}, \ldots, p_{t})$ are independent for a time distance larger than $m$. Using the second statement of Corollary~\ref{cor:mbh.contr.fdrg} FDR of the whole time series is controlled at level $(1-\pi)\alpha$.
		\item This $p$-value family is not i.i.d. However, because the calibration are build using a sliding window of size $n$, two $p$-values subseries of length $m$, $\hat{\textbf{P}}_{t_1}$ and $\hat{\textbf{P}}_{t_2}$, are independent when $|t_1 - t_2|>m+n$. Then the third statement of Corollary~\ref{cor:mbh.contr.fdrg} ensure that FDR of the whole time series is controlled at level $(1-\pi)\alpha$.
	\end{enumerate}	 
\end{proof}

\section{Supplementary experiments}
\subsection{Overlapping Calibration set and FDR control}\label{sec:experiment-overlapping-calibrationset}
 The following experiments aims at drawing a comparison between the FDR values in three scenarios: independent calibration sets, partially overlapping calibration sets with an overlap size driven by the value of $sn$ (size of the shift), and the same calibration set for all empirical $p$-values.
To be more specific, the calibration sets (and corresponding empirical $p$-values) were generated according to the following scheme. Each calibration set is of cardinality $n$. When moving from one calibration set to the next one, the shift size is equal to $sn$, where $s$ in $[0,1]$ is the proportion of independent data between calibration sets, resulting in an overlap of cardinality $(1-s)n$. Therefore an overlap occurs as long as $s < 1$. All these ways to build the calibration sets are called ``calibration sets strategies''.
\begin{enumerate}
	\item The independent $p$-values (iid Cal.) are generated according to
	\begin{equation}
		\forall i \in \llbracket 1, m \rrbracket, \quad \textbf{Z}_i \sim \mathcal P_0^n, \quad  \hat p_{1,i} = \hat p\mbox{-value}(X_i, \textbf{Z}_i).
	\end{equation}
	\item The $p$-values with the same calibration set (Same Cal.) are generated by
	\begin{equation}
		\forall i \in \llbracket 1, m\rrbracket, \quad	\textbf{Z} \sim \mathcal P_0^n,  \quad \hat p_{2,i} = \hat p\mbox{-value}(X_i, \textbf{Z}) .
	\end{equation}
	\item The $p$-values with overlapping calibration sets (Over. Cal.) are generated given, for $0<s<n$, by
	\begin{align}
		\forall i \in \llbracket 1,m \rrbracket, \quad \textbf{Z}_i & = \{Z_{\lfloor isn\rfloor+1},\ldots, Z_{\lfloor isn\rfloor+n}\}, \quad \hat p_{3,i} = \hat p\mbox{-value}(X_i, \textbf{Z}_i), \notag \\
		\mbox{and }\qquad &  \{Z_{s+1},\ldots, Z_{\lfloor 2sn\rfloor+1},\ldots,Z_{\lfloor msn\rfloor+1},\ldots, Z_{\lfloor msn\rfloor+n}\} \sim \mathcal P_0^{ms+n}.
	\end{align}
\end{enumerate}
According to these three scenarios, as $s$ increases, the overlap cardinality decreases, which results in more and more (almost) independent calibration sets. 
This is illustrated by the empirical results collected in Table~\ref{table:fdr-dependence-calibration}. For each calibration set strategy, presented in row, and for each calibration set cardinality in column, the estimated FDR is shown. In this experiment, the reference distribution $\mathcal P_0$ is the Gaussian $\mathcal N(0,1)$ and the anomalies are equal to $\Delta=4$. The number of tested $p$-values, noted $m$, is equal to $100$ and $m_1$, the number of anomalies, is equal to $1$. On each sample, BH-procedure is applied with $\alpha=0.1$ and the $FDP$ is computed. Each FDR is estimated over $10^3$ repetitions.

\begin{table}[H]
	\caption{FDR results with overlapping calibration sets  \label{table:fdr-dependence-calibration}}
	\begin{tabular}{l|rr|rr|rr|rr}
		\toprule
		n &  249  &  250  &  499  &  500  &  749  &  750  &  999  &  1000  \\
		\midrule
		Same Cal.   & 0.164 & 0.175 & 0.112 & 0.183 & 0.137 & 0.131 & 0.097 & 0.154\\
		Over Cal. (s=0.1\%)   & 0.167 & 0.174 & 0.100 & 0.156 & 0.138 & 0.125 & 0.093 & 0.140\\
		Over Cal. (s=0.2\%)   & 0.162 & 0.176 & 0.095 & 0.170 & 0.124 & 0.127 & 0.109 & 0.143\\
		Over Cal. (s=0.5\%)   & 0.163 & 0.166 & 0.110 & 0.170 & 0.116 & 0.132 & 0.111 & 0.149\\
		Over Cal. (s=1\%)   & 0.151 & 0.180 & 0.094 & 0.177 & 0.127 & 0.128 & 0.099 & 0.143\\
		Over Cal. (s=2\%)   & 0.164 & 0.180 & 0.108 & 0.179 & 0.133 & 0.140 & 0.097 & 0.143\\
		Over Cal. (s=5\%)   & 0.168 & 0.172 & 0.108 & 0.169 & 0.125 & 0.130 & 0.096 & 0.144\\
		Over Cal. (s=10\%)  & 0.165 & 0.181 & 0.104 & 0.185 & 0.122 & 0.140 & 0.105 & 0.146\\
		Over Cal. (s=20\%)  & 0.173 & 0.207 & 0.109 & 0.171 & 0.136 & 0.149 & 0.101 & 0.140\\
		Over Cal. (s=50\%)  & 0.180 & 0.187 & 0.103 & 0.183 & 0.121 & 0.128 & 0.094 & 0.143\\
		iid Cal. & 0.171 & 0.188 & 0.115 & 0.174 & 0.138 & 0.143 & 0.104 & 0.132\\
		\bottomrule
	\end{tabular}
\end{table}

The values of $n$ in the columns of Table~\ref{table:fdr-dependence-calibration} are chosen such that, for each pair of columns, the FDR value is smaller for the left column and larger for the right column (see Figure~\ref{fig:motivexp} for a visual illustration of this phenomenon). 
Table~\ref{table:fdr-dependence-calibration} illustrates that, in the context of the present numerical experiments, the FDR estimation is not too strongly impacted by the value of $s$ (proportion of the overlap). To assert that the observed differences between FDR estimations in each column  are not significant, permutation tests \citep{fisher1951design, phipson2010permutation} are performed. Under $\mathcal H_{0n}$ hypothesis, the $FDR$ are the same across all calibration set strategies for the calibration set cardinality $n$. Under $\mathcal H_{1n}$ there are at least two calibration strategies leading to different $FDR$. The $FDP$ samples that have been used to estimate the $FDR$ are reused. The maximal gap between sample means is used as statistic. The test is performed using the function ``permutation\_test'' from the Python library called Scipy. The significance level is fixed at $0.05$. Since multiple tests are performed over the different cardinalities, the threshold for rejecting a hypothesis is $0.00625$, according Bonferroni correction.
The results are display in Table~\ref{table:fdr-dependence-calibration-permtest}. All tested hypotheses have a $p$-values greater than the threshold $0.00625$. There are no significant difference in the resulting FDR between the different proportions of overlapping in calibration sets. This would suggest that considering overlapping calibration sets should not worsen too much the control of false positives and negatives. 

\begin{table}[H]
	\caption{$p$-values resulting from permutations test \label{table:fdr-dependence-calibration-permtest}}
	\begin{tabular}{l|rr|rr|rr|rr}
		\toprule
		n &  249  &  250  &  499  &  500  &  749  &  750  &  999  &  1000 \\
		\midrule
		$p$-value of the test &0.300& 0.0326& 0.572& 0.313& 0.588& 0.435& 0.735& 0.690\\
		\bottomrule
	\end{tabular}
\end{table}

\section{Figures}
\subsection{Figures related to the experiment of Section \ref{sec:illustratino-results}}
\begin{figure}[H]
	\centering
	\begin{subfigure}[b]{.45\linewidth} 
		\centering
		\includegraphics[width=0.95\linewidth]{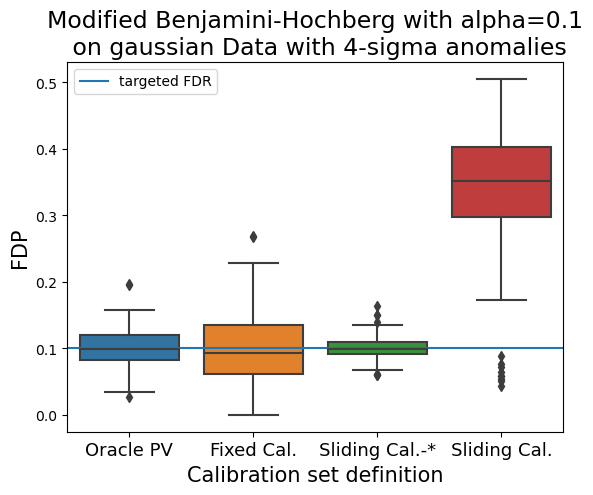}
		\caption[]{mBH, $4-$sigma}
		\label{fig:fdr01bh4}
	\end{subfigure}
	\begin{subfigure}[b]{.45\linewidth} %
		\centering
		\includegraphics[width=0.95\linewidth]{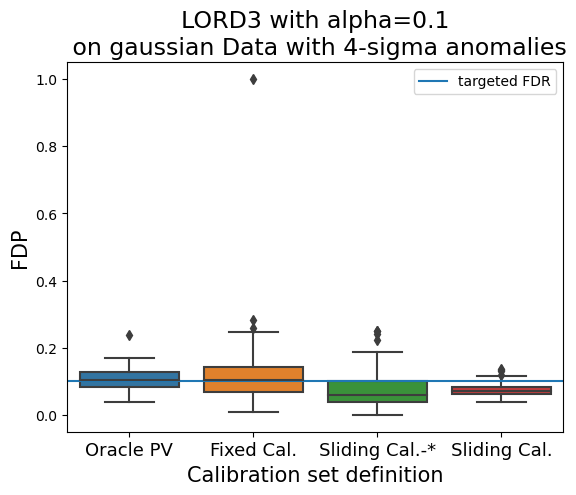}
		\caption[]{LORD, $4-$sigma}
		\label{fig:fdr01lord4}
	\end{subfigure}
	\begin{subfigure}[b]{.45\linewidth} 
		\centering
		\includegraphics[width=0.95\linewidth]{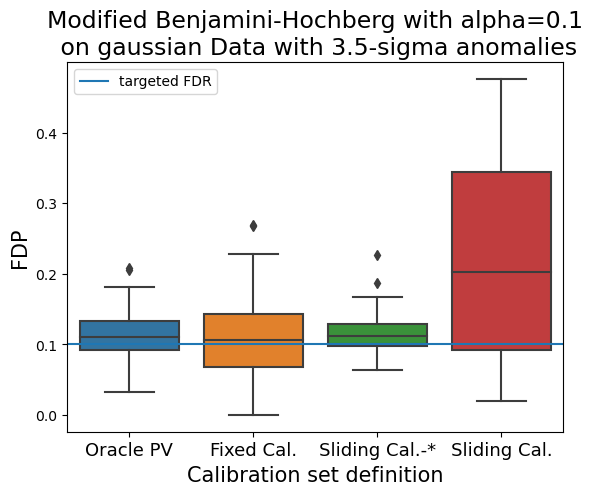}
		\caption[]{mBH, $3.5-$sigma}
		\label{fig:fdr01bh35}
	\end{subfigure}
	\begin{subfigure}[b]{.45\linewidth} %
		\centering
		\includegraphics[width=0.95\linewidth]{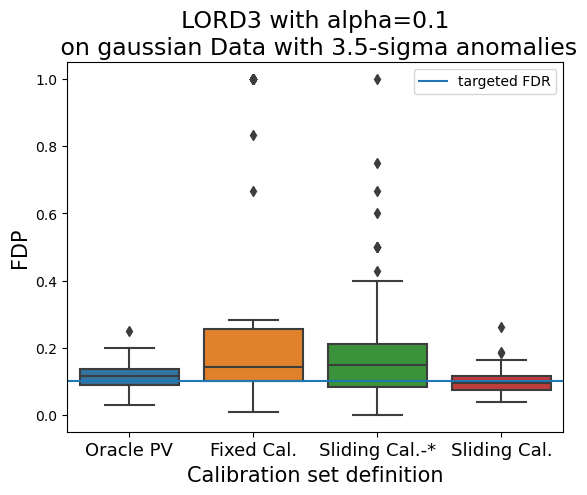}
		\caption[]{LORD, $3.5-$sigma}
		\label{fig:fdr01lord35}
	\end{subfigure}
	\begin{subfigure}[b]{.45\linewidth} 
		\centering
		\includegraphics[width=0.95\linewidth]{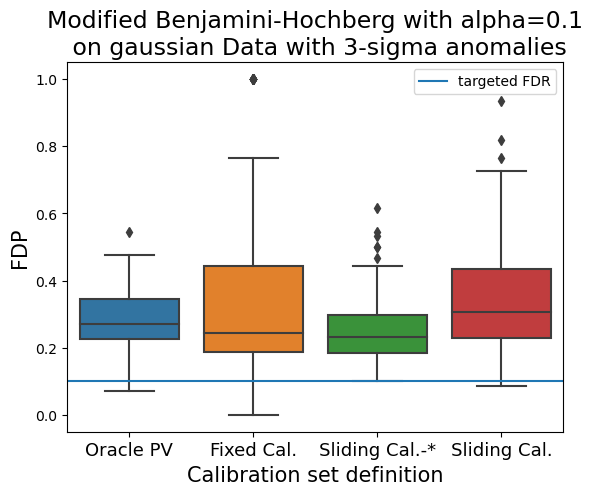}
		\caption[]{mBH, $3-$sigma}
		\label{fig:fdr01bh3}
	\end{subfigure}
	\begin{subfigure}[b]{.45\linewidth} %
		\centering
		\includegraphics[width=0.95\linewidth]{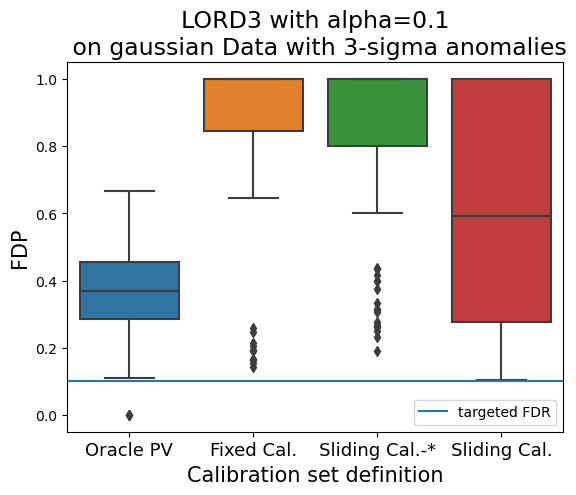}
		\caption[]{LORD, $3-$sigma}
		\label{fig:fdr01lord3}
	\end{subfigure}
	\caption[]{Comparison of FDPs acquired from using different multiple testing procedures, mBH or LORD, and from the way the $p$-values are calculated, in the case $\alpha=0.1$.}
	\label{fig:fdr01}
\end{figure}

\begin{figure}[H]
	\centering
	\begin{subfigure}[b]{.45\linewidth} 
		\centering
		\includegraphics[width=0.95\linewidth]{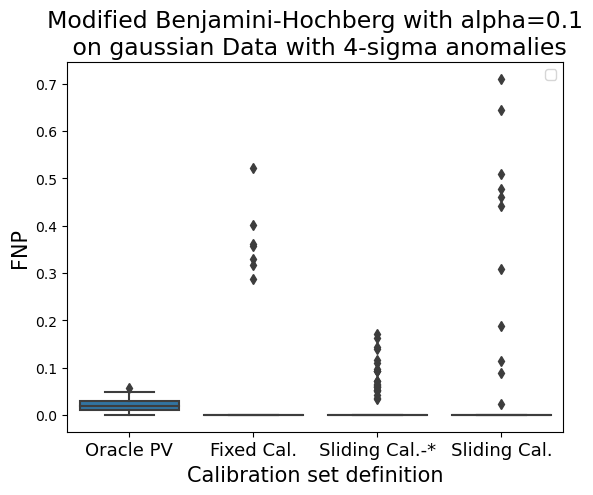}
		\caption[]{mBH, $4-$sigma}
		\label{fig:fnr01bh4}
	\end{subfigure}
	\begin{subfigure}[b]{.45\linewidth} %
		\centering
		\includegraphics[width=0.95\linewidth]{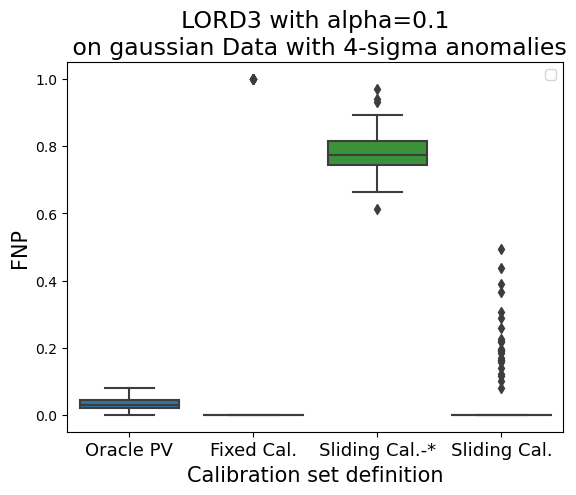}
		\caption[]{LORD, $4-$sigma}
		\label{fig:fnr01lord4}
	\end{subfigure}
	\begin{subfigure}[b]{.45\linewidth} 
		\centering
		\includegraphics[width=0.95\linewidth]{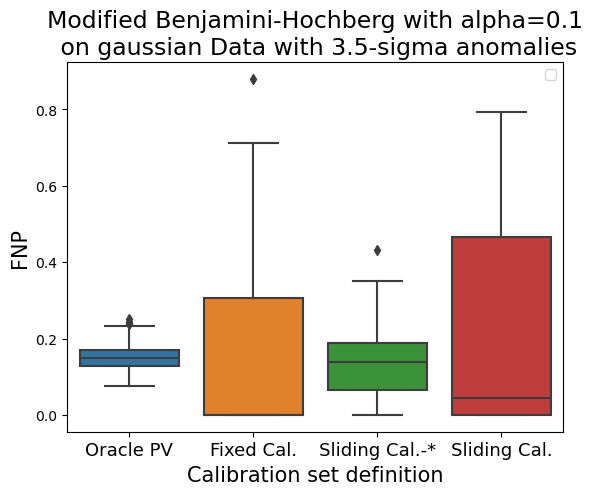}
		\caption[]{mBH, $3.5-$sigma}
		\label{fig:fnr01bh35}
	\end{subfigure}
	\begin{subfigure}[b]{.45\linewidth} %
		\centering
		\includegraphics[width=0.95\linewidth]{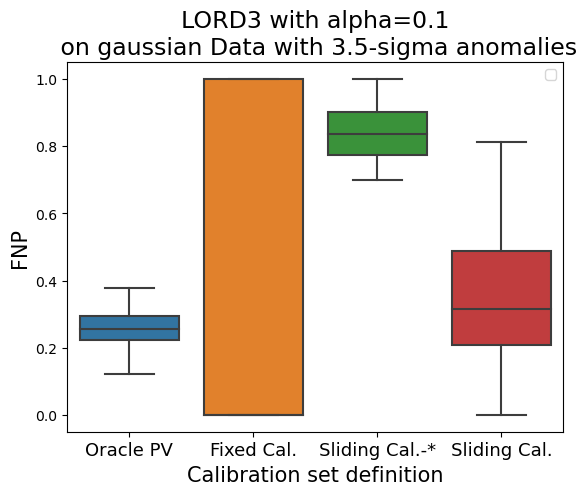}
		\caption[]{LORD, $3.5-$sigma}
		\label{fig:fnr01lord35}
	\end{subfigure}
	\begin{subfigure}[b]{.45\linewidth} 
		\centering
		\includegraphics[width=0.95\linewidth]{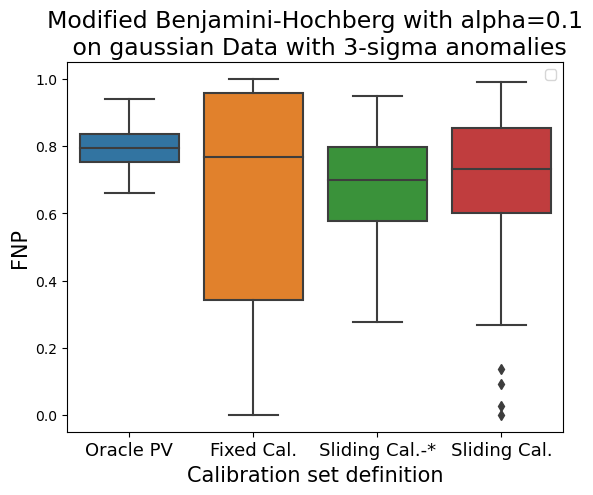}
		\caption[]{mBH, $3-$sigma}
		\label{fig:fnr01bh3}
	\end{subfigure}
	\begin{subfigure}[b]{.45\linewidth} %
		\centering
		\includegraphics[width=0.95\linewidth]{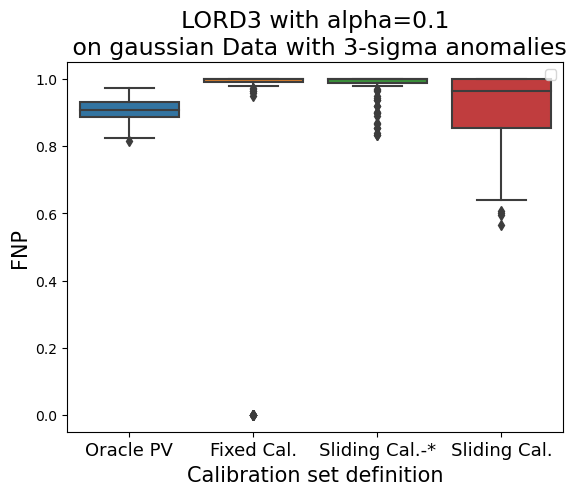}
		\caption[]{LORD, $3-$sigma}
		\label{fig:fnr01lord3}
	\end{subfigure}
	\caption[]{Comparison of FNPs acquired from using different multiple testing procedures, mBH or LORD, and from the way the $p$-values are calculated, in the case $\alpha=0.1$.}
	\label{fig:fnr01}
\end{figure}

\begin{figure}[H]
	\centering
	\begin{subfigure}[b]{.45\linewidth} 
		\centering
		\includegraphics[width=0.95\linewidth]{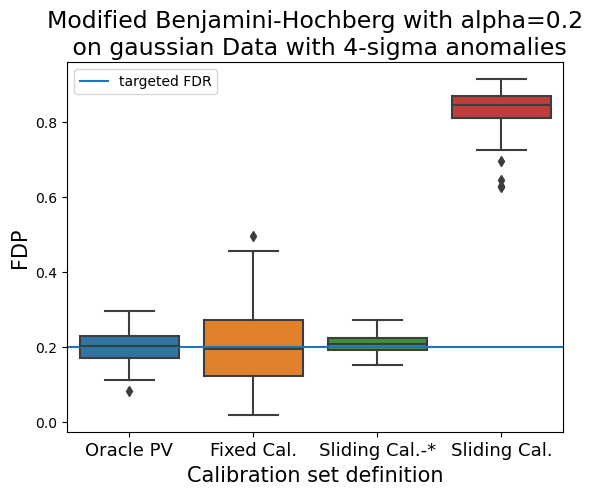}
		\caption[]{mBH, $4-$sigma}
		\label{fig:fdr02bh4}
	\end{subfigure}
	\begin{subfigure}[b]{.45\linewidth} %
		\centering
		\includegraphics[width=0.95\linewidth]{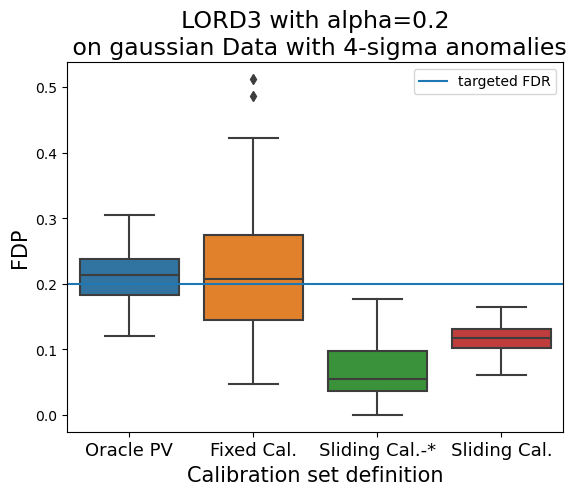}
		\caption[]{LORD, $4-$sigma}
		\label{fig:fdr02lord4}
	\end{subfigure}
	\begin{subfigure}[b]{.45\linewidth} 
		\centering
		\includegraphics[width=0.95\linewidth]{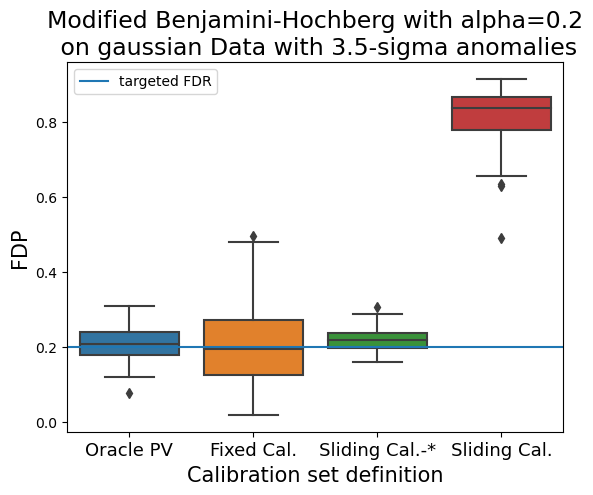}
		\caption[]{mBH, $3.5-$sigma}
		\label{fig:fdr02bh35}
	\end{subfigure}
	\begin{subfigure}[b]{.45\linewidth} %
		\centering
		\includegraphics[width=0.95\linewidth]{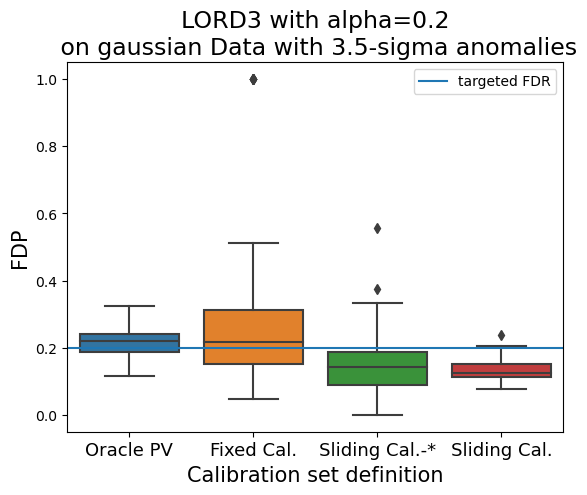}
		\caption[]{LORD, $3.5-$sigma}
		\label{fig:fdr02lord35}
	\end{subfigure}
	\begin{subfigure}[b]{.45\linewidth} 
		\centering
		\includegraphics[width=0.95\linewidth]{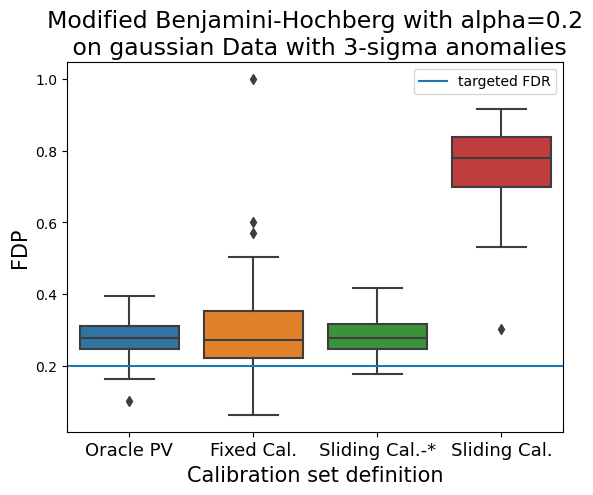}
		\caption[]{mBH, $3-$sigma}
		\label{fig:fdr02bh3}
	\end{subfigure}
	\begin{subfigure}[b]{.45\linewidth} %
		\centering
		\includegraphics[width=0.95\linewidth]{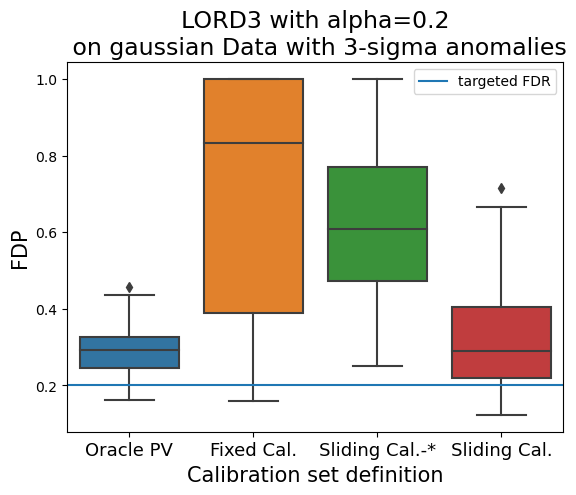}
		\caption[]{LORD, $3-$sigma}
		\label{fig:fdr02lord3}
	\end{subfigure}
	\caption[]{Comparison of FDPs acquired from using different multiple testing procedures, mBH or LORD, and from the way the $p$-values are calculated, in the case $\alpha=0.2$}
	\label{fig:fdr02}
\end{figure}

\begin{figure}[H]
	\centering
	\begin{subfigure}[b]{.45\linewidth} 
		\centering
		\includegraphics[width=0.95\linewidth]{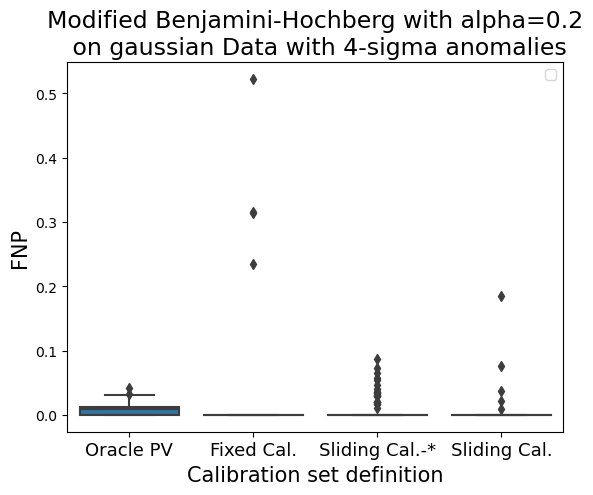}
		\caption[]{mBH, $4-$sigma}
		\label{fig:fnr02bh4}
	\end{subfigure}
	\begin{subfigure}[b]{.45\linewidth} %
		\centering
		\includegraphics[width=0.95\linewidth]{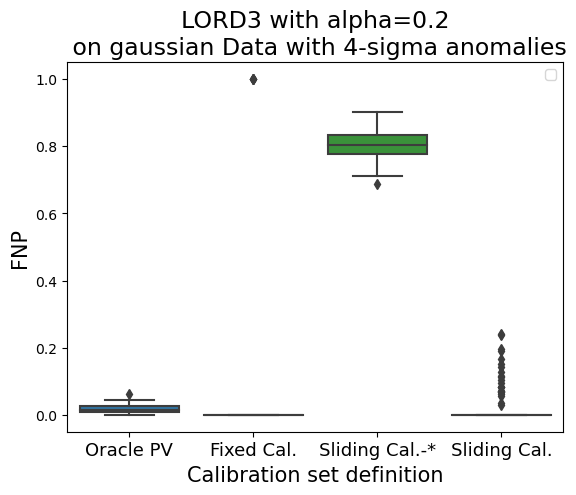}
		\caption[]{LORD, $4-$sigma}
		\label{fig:fnr02lord4}
	\end{subfigure}
	\begin{subfigure}[b]{.45\linewidth} 
		\centering
		\includegraphics[width=0.95\linewidth]{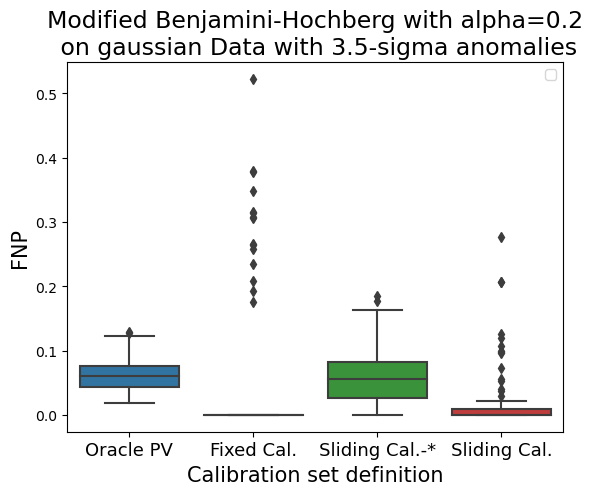}
		\caption[]{mBH, $3.5-$sigma}
		\label{fig:fnr02bh35}
	\end{subfigure}
	\begin{subfigure}[b]{.45\linewidth} %
		\centering
		\includegraphics[width=0.95\linewidth]{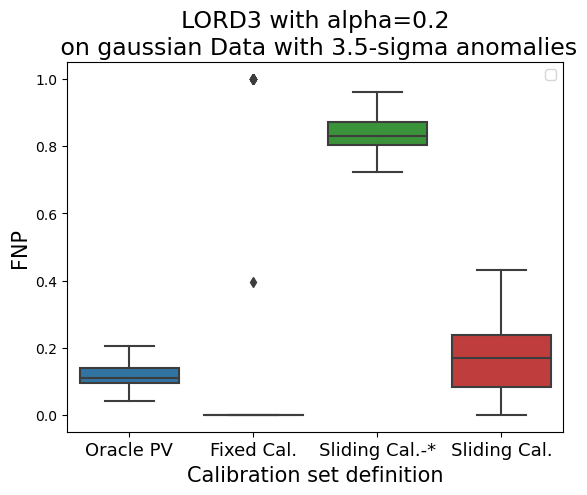}
		\caption[]{LORD, $3.5-$sigma}
		\label{fig:fnr02lord35}
	\end{subfigure}
	\begin{subfigure}[b]{.45\linewidth} 
		\centering
		\includegraphics[width=0.95\linewidth]{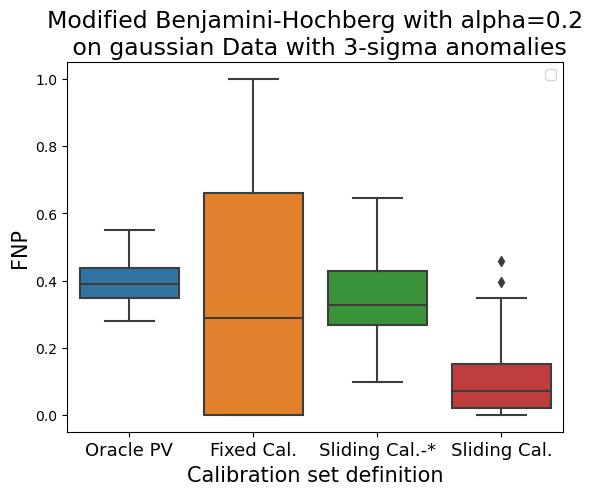}
		\caption[]{mBH, $3-$sigma}
		\label{fig:fnr02bh3}
	\end{subfigure}
	\begin{subfigure}[b]{.45\linewidth} %
		\centering
		\includegraphics[width=0.95\linewidth]{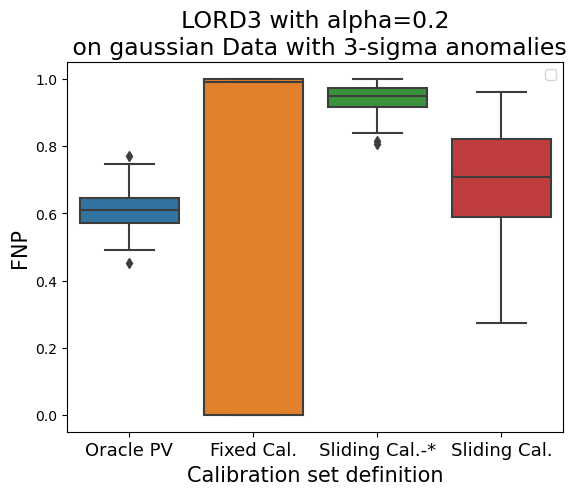}
		\caption[]{LORD, $3-$sigma}
		\label{fig:fnr02lord3}
	\end{subfigure}
	\caption[]{Comparison of FNPs acquired from using different multiple testing procedures, mBH or LORD, and from the way the $p$-values are calculated, in the case $\alpha=0.2$.}
	\label{fig:fnr02}
\end{figure}

\end{document}